\documentclass[iop,numberedappendix,appendixfloats]{emulateapj}
\usepackage{lscape}

\shorttitle{THE STELLAR MASS ASSEMBLY OF FOSSIL GALAXIES}
\shortauthors{HARRISON ET AL.}

\begin{document}

\title{THE \textit{XMM} CLUSTER SURVEY: THE STELLAR MASS ASSEMBLY OF
  FOSSIL GALAXIES}

\author{Craig D.\ Harrison$^1$, Christopher J.\ Miller$^1$, Joseph
  W.\ Richards$^2$, E.\ J.\ Lloyd-Davies$^3$, Ben Hoyle$^{4,5,6}$, A.\ Kathy
  Romer$^3$, Nicola Mehrtens$^3$, Matt Hilton$^{7,8}$, John P.\
  Stott$^{9,10}$, Diego Capozzi$^9$, Chris A.\ Collins$^9$, Paul-James
  Deadman $^2$, Andrew R.\ Liddle$^3$, Martin Sahl{\'e}n$^{11}$, S.\ Adam
  Stanford$^{12,13}$, and Pedro T.\ P.\ Viana$^{14,15}$}
\affiliation{$^1$ Department of Astronomy, University of Michigan,
  Ann Arbor, MI 48109, USA}
\affiliation{$^2$ Center for Time Domain Informatics, University of
  California, Berkeley, CA 94720, USA}
\affiliation{$^3$ Astronomy Centre, University of Sussex, Falmer,
  Brighton, BN1 9QH, UK}
\affiliation{$^4$ Institute of Sciences of the Cosmos (ICCUB) and
  IEEC, Physics Department, University of Barcelona, Barcelona 08024,
  Spain}
\affiliation{$^5$ CSIC, Consejo Superior de Investigaciones
  Cientificas, Serrano 117, Madrid, 28006, Spain}
\affiliation{$^6$ Helsinki Institute of Physics, P.O. Box 64,
  FIN-00014 University of Helsinki, Helsinki, Finland}
\affiliation{$^7$ Astrophysics \& Cosmology Research Unit, School of
  Mathematical Sciences, University of KwaZulu-Natal, Private Bag
  X54001, Durban 4000, South Africa}
\affiliation{$^8$ School of Physics \& Astronomy, University of
  Nottingham, Nottingham, NG7 2RD, UK}
\affiliation{$^9$ Astrophysics Research Institute, Liverpool John Moores
  University, Twelve Quays House, Egerton Wharf, Birkenhead, CH41 1LD, UK}
\affiliation{$^{10}$ Department of Physics, Institute for Computational
  Cosmology, Durham University, South Road, Durham DH1 3LE, UK}
\affiliation{$^{11}$ The Oskar Klein Centre for Cosmoparticle Physics,
  Department of Physics, Stockholm University, AlbaNova, SE-106 91
  Stockholm, Sweden}
\affiliation{$^{12}$ Physics Department, University of California, Davis,
  CA 95616, USA}
\affiliation{$^{13}$ Institute of Geophysics and Planetary Physics, Lawrence
  Livermore National Laboratory, Livermore, CA 94551, USA}
\affiliation{$^{14}$ Centro de Astrof\'{\i}sica da Universidade do
  Porto, Rua das Estrelas, 4150-762, Porto, Portugal} 
\affiliation{$^{15}$ Departamento de F\'{\i}sica e Astronomia da
  Faculdade de Ci\^{e}ncias da Universidade do Porto, Rua do Campo Alegre,
  687, 4169-007 Porto, Portugal}
\email{craigha@umich.edu}

\begin{abstract}
This paper presents both the result of a search for fossil systems
(FSs) within the \textit{XMM} Cluster Survey and the Sloan Digital Sky
Survey and the results of a study of the stellar mass assembly and
stellar populations of their fossil galaxies. In total, 17 groups and
clusters are identified at $z < 0.25$ with large magnitude gaps
between the first and fourth brightest galaxies. All the information
necessary to classify these systems as fossils is provided. For both
groups and clusters, the total and fractional luminosity of the
brightest galaxy is positively correlated with the magnitude gap. The
brightest galaxies in FSs (called fossil galaxies) have stellar
populations and star formation histories which are similar to normal
brightest cluster galaxies (BCGs). However, at fixed group/cluster
mass, the stellar masses of the fossil galaxies are larger compared to
normal BCGs, a fact that holds true over a wide range of group/cluster
masses. Moreover, the fossil galaxies are found to contain a
significant fraction of the total optical luminosity of the
group/cluster within $0.5R_{200}$, as much as 85\%, compared to the
non-fossils, which can have as little as 10\%. Our results suggest
that FSs formed early and in the highest density regions of
the universe and that fossil galaxies represent the end products of
galaxy mergers in groups and clusters. The online FS catalog can be
found at
http://www.astro.ljmu.ac.uk/$\sim$xcs/Harrison2012/XCSFSCat.html.
\end{abstract}


\section{Introduction}\label{introduction}

Hierarchical models of structure formation predict that galaxy groups
and clusters (hereafter referred to collectively as galaxy systems)
form via the gravitational infall of field galaxies and smaller
systems \citep{white78}. The effects of dark energy will eventually
halt the infall \citep{nagamine03}, at which time the system can be
considered fully assembled. Dynamical friction will cause some of
these galaxies to merge, resulting in the formation of a massive
galaxy near the base of the gravitational potential
\citep[e.g.,][]{barnes89, dubinski98}. A magnitude gap can then
develop as the timescale for orbital decay by dynamical friction is
inversely proportional to galaxy mass and so the more massive galaxies
will tend to merge first \citep{d'onghia05, dariush10}. The cooling
timescale of the intergalactic medium is longer, so the system will be
surrounded by a halo of X-ray-emitting gas. Such systems have come to
be referred to as ``fossil groups''.

The most commonly accepted definition of a fossil group
\citep{jones03} requires (1) an $R$-band magnitude gap of 2 or greater
between the two brightest galaxies located within half the virial
radius of the system and (2) extended X-ray emission with a bolometric
X-ray luminosity $L_{X,\mathrm{bol}}\gtrsim 5\times
10^{41}\,h_{70}^{-2}$ erg s$^{-1}$. The X-ray criterion guarantees the
existence of at least a group-size halo while the optical criterion
ensures (approximately) that there are no $L^*$ galaxies inside the
radius for orbital decay by dynamical friction.

Originally, the term ``fossil group'' (or ``fossil galaxy group'')
referred to an apparently isolated galaxy, which due to a surrounding
halo of X-ray-emitting gas, was assumed to have previously resided in
a group \citep{ponman94}. The word ``fossil'' was used because the
merger history of the group was thought to be contained within this
single galaxy. Since then, fossil groups (i.e., galaxies) have been
found within actual galaxy groups and this has lead to a degree of
confusion and a gradual shift from ``fossil group'' referring to the
individual galaxy to the group as a whole. Additionally, since the
definition of a fossil group that has developed only sets lower limits
on both the size of the magnitude gap and the X-ray luminosity
($L_X$), it is possible to find fossil groups with properties that are
more consistent with those of galaxy clusters, as we do in this work
\citep[but see also][]{mendesdeoliveira06, mendesdeoliveira09,
khosroshahi06b}. Therefore, we replace the word ``group'' with the
word ``system'', so that ``fossil system'' (FS) refers to a galaxy
group or cluster that satisfies the above two criteria and ``fossil
galaxy'' (FG) refers to the most-luminous galaxy in an FS. This is the
nomenclature that we shall use throughout the remainder of this paper.

The initial interpretation of FGs was that they represent the end
product of galaxy merging in groups or clusters. The magnitude gap in
an FS was accounted for by an early formation epoch that allowed time
for sufficient mergers. The observed regular, symmetric X-ray emission
\citep{khosroshahi07} and highly concentrated mass profiles
\citep{jones03, khosroshahi04} of FSs support this idea. Likewise,
simulations have shown that FSs form a large fraction of their mass at
high redshift, and that they form earlier than non-FSs
\citep{d'onghia05, dariush07, vonbendabeckmann08}. Additionally,
\citet{d'onghia05} found a correlation between the magnitude gap at
$z=0$ and the formation time of the FS in the sense that early-forming
systems have larger gaps \citep[see also][]{dariush10}. Note, however,
that infall can result in a smaller magnitude gap without the system
necessarily being late-forming.

Simulations by \citet{vonbendabeckmann08} suggest that the formation
of FSs is primarily driven by the relatively early in-fall of massive
satellites with the magnitude gap arising after the system has built
up half of its final mass. The formation appears particularly
efficient if the in-fall occurs along filaments with small impact
parameters \citep{d'onghia05, sommerlarsen06, vonbendabeckmann08}.

Studies of the FS luminosity function \citep{mendesdeoliveira06,
  cypriano06, zibetti09, aguerri11} have found parameters that are
consistent with the universal luminosity function of clusters derived
by \citet{popesso05}, although they are deficient in $\sim L^*$
galaxies. It is possible that FSs formed with a deficit of $\sim L^*$
galaxies \citep{mulchaey99}. This alternate formation scenario
interprets FSs as `failed groups' in which the majority of the
available gas was initially used up in a single luminous galaxy rather
than in several. Similarly, \citet{proctor11} found that FSs have low
richnesses but masses comparable to those of clusters resulting in
high dynamical mass-to-light ratios. As with their masses, FS X-ray
scaling relations are more consistent with clusters than groups
\citep{mendesdeoliveira06, mendesdeoliveira09, khosroshahi06b},
leading to the suggestion that FSs are merely representative of an
evolutionary phase \citep{vonbendabeckmann08}--after 4 Gyr $\sim
90\%$ of FSs in simulations are found to have become non-fossil
\citep{dariush07}.

In one of the only studies of its kind, \citet{diaz-gimenez08} looked
at the merger history of FGs in simulated FSs within the Millennium
Simulation Galaxy Catalogue. They find that, like brightest cluster
galaxies (BCGs), FGs are mainly formed by gas-poor mergers but that
FGs are formed later than BCGs, i.e., they undergo mergers at lower
redshifts, despite the fact that FSs assembled most of their virial
mass at higher redshifts in comparison with non-FSs. No age
differences were found between the stellar populations in FSs and
bright field ellipticals \citep{labarbera09}, which is consistent with
a later formation if the mergers were gas-poor. Numerous studies have
been made of the growth of BCGs and the results are
conflicting. Simulations performed in a hierarchical context predict
that BCGs are the result of multiple mergers and therefore should
continue to grow until $z \sim 0.5$
\citep[e.g.,][]{delucia07}. Observationally, however, BCGs are found
to evolve passively since $z\sim1$ \citep{brough02, brough07, stott08,
stott10, whiley08, collins09}.

Defining FS samples is difficult due to the need for both good quality
X-ray and optical data, and so most of our understanding has come from
simulations. The initial observational studies focused on small
samples or single objects and so the number of known FSs remains small
\citep[see][for a good summary]{mendesdeoliveira06}. Enabled by the
large quantity of data produced by surveys such as the Sloan Digital
Sky Survey \citep[SDSS;][]{york00}, more recent studies have focused
on defining and analyzing larger samples of FSs
\citep[e.g.,][]{santos07, labarbera09, voevodkin10, miller11}, but in
most cases these samples lack the high-quality X-ray data necessary to
ensure that the definition of an FS is met and often the criteria have
been relaxed to such an extent that the possibility of substantial
contamination is high.

The purpose of this study is two-fold. First, we want to take the high
quality X-ray data from the \textit{XMM} Cluster Survey
\citep[XCS;][]{romer01,lloyddavies11, mehrtens11} and combine it with
the optical data from the SDSS Data Release 7 (DR7) to produce a
secure sample of FSs; we choose purity over quantity. Second, we want
to compare the stellar mass assembly and the stellar populations of
FGs to various other samples, and to examine the X-ray scaling
relations of FSs.

The XCS is a serendipitous search for galaxy clusters using all
publicly available data in the \textit{XMM} Science Archive. Its main
aims are to measure cosmological parameters and trace the evolution of
X-ray scaling relations. The first data release from XCS
\citep[XCS-DR1][hereafter M11]{mehrtens11} contains 503 clusters of
which 402 have measured X-ray temperatures ($T_X$) and
luminosities. The serendipitous nature of XCS is a big advantage
because it allows the detection of smaller systems, which have a
higher probability of satisfying the optical criterion in the FS
definition \citep{cui11}.

The layout of the remainder of the paper is as follows. We outline the
samples used in Section \ref{samples}, including a review of the
definition of an FS and our methodology for searching for them. In
Section \ref{data}, we describe the various data sets utilized in this
study. The results of our study into the stellar masses and the
stellar populations of fossil galaxies are presented in Section
\ref{stellar populations}, while Section \ref{assem} presents the
results of our study of the X-ray scaling relations and stellar mass
assembly of FGs and BCGs. A discussion of all these results can be
found in Section \ref{discussion} and our conclusions are presented in
Section \ref{conclusions}.

We assume a $\Lambda$CDM cosmology with Hubble parameter $H_0=70$ km
s$^{-1}$ Mpc$^{-1}$, dark matter density parameter
$\Omega_\Lambda=0.73$, and matter density parameter $\Omega_M=0.27$.

\section{Samples}\label{samples}

In this section, we outline our methodology for searching for FSs
within the XCS and the SDSS and present our final sample of 17. We
also discuss two samples of BCGs, one optically selected and one X-ray
selected, which we use as comparison samples. We begin by making a few
comments on how FSs are defined.

\subsection{The Definition of a Fossil System}

The most commonly used definition of an FS is that found in
\citet{jones03}, which is reproduced in the introduction. This
definition is motivated by physical arguments but is not beyond
modification. The X-ray criterion was set to only select galaxies that
reside in at least a group-sized halo and so does not have much scope
for modification, the adopted magnitude gap, on the other hand, does.

This is because studies of the magnitude gap in both clusters
\citep{milosavljevic06} and groups \citep{tavasoli11} have shown a
wide range of gaps. The average magnitude gap between a $0.1<z<0.2$
BCG and the second brightest cluster member is found to be 0.5
\citep[][see also Loh \& Strauss 2006]{pipino11}, with poor systems
exhibiting a larger average gap. Statistically, it is easier to get
large magnitude gaps when the number of galaxies in the system is low
\citep{dariush07, cui11}. In simulations, \citet{dariush07} found that
the strongest X-ray FS candidates are those with the highest X-ray
luminosity as these systems are not expected to have a large
luminosity gap entirely by chance.

\citet{dariush10} have suggested, based on numerical simulations, that
a magnitude gap of 2.5 between the brightest and fourth-brightest
galaxies is a better indicator of an FS. They find that the 2 mag gap
is better at finding high-mass systems but the 2.5 mag gap finds 50\%
more early-forming systems and those that are in the fossil phase
longer.

The definition of an FS that we adopt here is a combination of the
X-ray criterion of \citeauthor[]{jones03} and the optical criterion of
\citet{dariush10}. Therefore, for a system to be classified as a
fossil by us it needs to have $L_X\gtrsim 5\times
10^{41}\,h_{70}^{-2}$ erg s$^{-1}$ and a magnitude gap of 2.5 in the
$r$ band between the brightest and the fourth brightest galaxies
located within half the virial radius, which we denote as $\Delta
m_{14}$. The virial radius is approximated with $R_{200}$, the radius
at which the average density is equal to 200 times the critical
density of the universe, and $L_X$ is the bolometric X-ray luminosity
inside this radius. In Appendix \ref{robust} we investigate how our
results change if the standard definition of an FS is adopted, i.e.,
replacing the \citeauthor{dariush10} magnitude gap with that of
\citeauthor{jones03}.

\subsection{The Fossil System Sample}\label{fs sample}

Previous searches for FSs began by looking for galaxies that satisfied
the magnitude gap criterion and then tried to match them to an X-ray
detection. Here, we take the opposite approach and examine the
galaxies associated with an extended X-ray source to determine the
magnitude gap.

We begin with the $Zoo^{\mathrm{DR7}}$ and $Zoo^{\mathrm{S82}}$
candidate catalogs as described in M11. In addition, we also allow for
\textit{XMM} PI-targeted cluster observations to be included in our
final sample (such systems are noted in Table \ref{fs_details}). Thus,
our catalog of FSs contains serendipitously discovered FSs, as well as
previously known groups/clusters which have been re-classified as
FSs. Since we are studying the stellar populations of FGs, we also
required that the brightest galaxy near the center of each candidate
have a measured SDSS spectrum and we used the redshift of that galaxy
to define the redshift of the system. These constraints complicate the
FS selection, and so for this work we avoid making any conclusions
which require a known selection function (e.g., constraining the
number density of FSs).

We then identified FSs by examining the color--magnitude relation
around each candidate. We used a system similar to the {\sc XCS-Zoo}
that is described in detail in M11. The main difference being the
addition of color--magnitude diagrams (CMDs) to help with the
estimation of the magnitude gap. Specifically, we visually examined
X-ray images and optical color images from the SDSS overlaid with
X-ray contours. These X-ray and optical images were created over
$3\arcmin\times3\arcmin$, $6\arcmin\times6\arcmin$, and
$12\arcmin\times12\arcmin$ fields of view. At the same time, we
compared the images to the $r$ versus $r-i$ CMDs using SDSS DR7 imaging
data.

A system was either designated an FS candidate or not based on the
inspection of these images and CMDs. To be classified as an FS
candidate the system must satisfy the following criteria.
\begin{itemize}
\item \textit{Optical.} A bright elliptical galaxy at or near the
  location of the X-ray source.
\item \textit{X-ray.} Obviously an extended source, e.g., rather than a blend
  of point sources (see M11).
\item \textit{CMD.} $\Delta m_{14}>2.0$ in the
$3\arcmin\times3\arcmin$ image.
\end{itemize}
 
It has recently been noted that the SDSS photometry systematically
underestimates the luminosities of nearby BCGs \citep{bernardi07,
lauer07} with the discrepancy being a function of BCG radius. This
problem affects DR7, which we use, and is best described in the
``Imaging Caveats'' page of
DR8\footnote{http://www.sdss3.org/dr8/imaging/caveats.php}. Therefore,
we use a magnitude gap of only 2.0 when selecting the FS candidates,
instead of the adopted gap of 2.5, to allow for the correction of this
problem \citep{vonderlinden07}. All sources were examined by at least
three coauthors and those that were classified as an FS candidate at
least twice ($\sim 60$ in total) were then examined more closely.

Defining magnitude gaps in clusters requires accurate measures of
$R_{200}$, $L_X$, and the magnitudes of the galaxies, as well the
ability to reject foreground/background galaxies. Therefore, the FS
candidates that passed the above initial criteria were then put
through the XCS analysis pipelines in order to measure $T_X$ and $L_X$
(see Section \ref{xcs} for more details). After doing so, we
re-calculated the CMDs using the estimated $R_{200}$ values
(specifically, out to $0.5R_{200}$). We discuss the robustness of our
sample and results to changes to $R_{200}$ in Appendix \ref{robust}.

We also applied background corrections to the magnitudes of the large
central galaxies using the algorithm defined in \citet{vonderlinden07}
and tested in \citet{voevodkin08}. The average size of this correction
is 0.3 mag but could be as large as 1.0 mag. We note that many
existing FS samples that rely on SDSS data \citep[e.g.,][]{santos07,
labarbera09, miller11} have not made this important correction. This
would result in incorrect FG luminosities (which, as we shall see in
Section \ref{assem}, is a significant fraction of the total system
luminosity), smaller magnitude gaps, and lower levels of completeness.

Only then were we able to measure the magnitude gap from the brightest
central galaxy out to some physical distance. In estimating the
magnitude gap we only considered galaxies that were within $\pm 0.2$
from the $r-i$ color of the central galaxy. This is reasonably
generous since the scatter in the red sequence is found to be less
than half this amount. We also discarded galaxies from the CMDs that
had SDSS spectroscopic redshifts $>2000$ km s$^{-1}$ away from the
central galaxy. For those galaxies with a photometric redshift
\citep{csabai07}, we discarded galaxies beyond $\Delta
z_\mathrm{phot}=0.1$. This cut is conservative since the average
photometric redshift error for galaxies with $m_r\sim 19.7$ is
0.04. No difference in the number of systems classified as fossils was
found when this photometric redshift cut was doubled to $\Delta
z_\mathrm{phot}=0.2$. This resulted in a clean (but possibly
incomplete) sample of 17 FS. This iterative procedure is important and
it allowed us to rigorously check which of the 60 FS candidates were
in fact true FSs.

\begin{figure}
\includegraphics[width=0.5\textwidth]{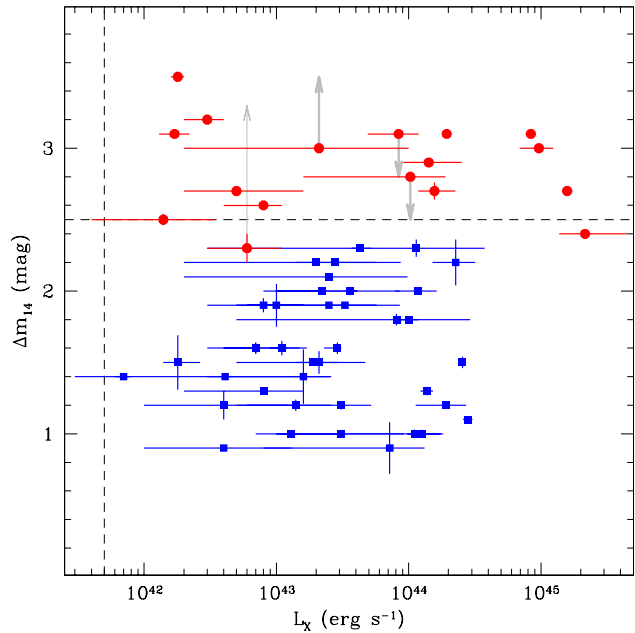}
\caption{Distribution of our sample (red circles) and the XCS
groups/clusters (blues squares) in the parameter space used to
classify FSs. Points that lie above and to the right of the dashed
lines are classified as FSs. Two systems that have $\Delta m_{12}<2.5$
are classified as FSs. The system with the arrow is
XMMXCS~J124425.9+164758.0, which has a highly uncertain
sky-subtraction correction that we elect not to apply. If we apply the
correction (of 1.0 mag) then this results in a magnitude gap of 3.3,
as denoted by the thin arrow. The other system is
XMMXCS~J172010.0+263724.7, which has a magnitude gap $\Delta
m_{14}=2.45$ and is close enough to 2.5 that we accept it as an
FS. The three thick arrows indicate how the magnitude gap changes when
the error on $R_{200}$ is taken into account. For all other systems
the magnitude gap is unaffected by the errors on $R_{200}$.}
\label{lx_gap}
\end{figure}

The positions of these systems in the parameter space used to define
FSs are shown in Figure \ref{lx_gap}. The red circles are our sample
while the blue squares are the XCS-DR1 groups/clusters of M11 (see
Section \ref{bcgs} for more details). The horizontal and vertical
dashed lines denote the optical and X-ray criterion respectively;
points that lie above and to the right of these lines are classified
as FSs. There are two systems that have $\Delta m_{14}<2.5$, which we
accept as FSs. The system with the arrow is XMMXCS~J124425.9+164758.0
(hereafter referred to simply as J124425.9), which has an FG with a
double nucleus. This makes the sky-subtraction correction highly
uncertain and so we elect not to apply it. If we were to apply the
estimated correction of 1.0 mag then $\Delta m_{14}$ changes from 2.3
to 3.3, which is denoted by the thin arrow in Figure
\ref{lx_gap}. Even if we only apply the average sky-subtraction
correction of the other 16 FSs (0.3 mag) then $\Delta m_{14}=2.6$ and
so we accept J124425.9 as an FS. The other system is
XMMXCS~J172010.0+263724.7, which actually has a magnitude gap $\Delta
m_{14}=2.45$ and is close enough to 2.5 that we accept it as an FS
too. Our results do not change if these two FSs are excluded. We note
that four FSs were targeted for observation by \textit{XMM} and are
thus not a part of the XCS-DR1, which includes only serendipitously
discovered clusters. These targeted systems are labeled in Table
\ref{fs_details} and our conclusions are unaffected if we exclude them
from our analyses.

For three FSs, there are galaxies near $0.5R_{200}$, such that a
slightly smaller $R_{200}$ excludes one or more of them as members or
a slightly larger $R_{200}$ includes one or more of them as
members. If the galaxy is among the four brightest in the system then
this can have the effect of increasing or decreasing the magnitude
gap, respectively. We show the magnitude of this systematic effect in
Figure \ref{lx_gap} (thick arrows) after changing $R_{200}$ by one
standard deviation of its statistical error (see Section
\ref{errors}). The magnitude gaps for the other 14 FSs are unaffected
by the statistical uncertainties on $R_{200}$.

Details of the FSs are given in Table \ref{fs_details} and details of
the FGs are given in Table \ref{fg_details}. Descriptions of the
individual systems, along with CMDs and SDSS images with XCS contours
overlaid are given in Appendix \ref{notes}.

\setcounter{table}{1}
\begin{table*}
\caption{The details of the fossil galaxies in our sample. The ID
matches a galaxy to a system in Table
\ref{fs_details}. $L_\mathrm{gal}$ is the luminosity of the FG in
$10^{11}\,L_\odot$, $M_*$ is the stellar mass of the FG in
$10^{12}\,M_\odot$, the age of the FG is in Gyr, $Z$ is the
metallicity of the FG as a percentage, and SSFR is the specific
star-formation rate in yr$^{-1}$. Luminosities and $z$ are taken from
the SDSS; $M_*$, age, and $Z$ are from {\sc starlight}; and SSFR are
taken from the MPA-JHU database.}\label{fg_details}
\begin{center}
\begin{tabular}{lccccccc}
\hline
ID & SDSS Name & $L_\mathrm{gal}$ & $z$ & $M_*$ & Age & $Z$ & log(SSFR)\\
\hline
1  & J015315.24+010220.6 & $1.81\pm 0.06$ & 0.0597 & 0.47 & 10.4 & 0.030 & $-12.23$\\
2  & J030658.71+000833.2 & $0.69\pm 0.03$ & 0.0751 & 0.29 & 10.5 & 0.030 & $-12.20$\\ 
3  & J073422.21+265144.9 & $1.88\pm 0.05$ & 0.0796 & 0.82 & 8.1  & 0.032 & $-12.37$\\ 
4  & J083454.90+553421.1 & $6.69\pm 0.28$ & 0.2412 & $\cdots$  & $\cdots$  & $\cdots$   & $\cdots$     \\
5  & J092539.05+362705.5 & $2.24\pm 0.06$ & 0.1121 & 0.75 & 11.3 & 0.030 & $-12.31$\\ 
6  & J101703.63+390249.4 & $7.08\pm 0.21$ & 0.2056 & 1.56 & 10.8 & 0.031 & $-12.70$\\ 
7  & J104044.49+395711.2 & $4.95\pm 0.17$ & 0.1381 & $\cdots$  & $\cdots$  & $\cdots$   & $\cdots$     \\ 
8  & J123024.67+111122.8 & $1.27\pm 0.05$ & 0.1169 & 0.37 & 10.9 & 0.021 & $-11.96$\\ 
9  & J123337.74+374122.0 & $0.81\pm 0.02$ & 0.1023 & 0.33 & 11.8 & 0.026 & $-12.30$\\ 
10 & J124425.43+164756.9 & $3.64\pm 0.35$ & 0.2346 & 0.52 & 11.8 & 0.027 & $-12.29$\\ 
11 & J130749.23+292548.2 & $6.71\pm 0.22$ & 0.2406 & 1.36 & 10.9 & 0.028 & $-12.13$\\ 
12 & J131146.19+220137.2 & $4.36\pm 0.26$ & 0.1715 & 0.97 & 9.5  & 0.027 & $-12.13$\\ 
13 & J134825.78+580018.7 & $2.35\pm 0.07$ & 0.1274 & 0.59 & 8.5  & 0.033 & $-11.77$\\ 
14 & J141627.37+231522.5 & $3.53\pm 0.10$ & 0.1382 & 1.08 & 9.9  & 0.032 & $-12.29$\\ 
15 & J141657.46+231242.5 & $1.27\pm 0.03$ & 0.1159 & 0.42 & 8.7  & 0.033 & $-11.26$\\ 
16 & J160129.75+083850.6 & $1.71\pm 0.06$ & 0.1875 & 1.02 & 11.3 & 0.028 & $-12.50$\\ 
17 & J172010.04+263732.0 & $5.40\pm 0.18$ & 0.1596 & 0.92 & 5.6  & 0.031 & $-11.44$\\ 
\hline
\end{tabular}
\end{center}
\end{table*}

\subsection{The BCG Samples}\label{bcgs}

One of the main aims of this study is to characterize the stellar
populations and the stellar mass assembly of FGs and to compare them
with those of BCGs in non-FSs. For this purpose we use two samples of
BCGs, one optically selected and one X-ray selected. For a fuller
description of the various catalogs on which the samples were based
the reader is referred to the references below.

The first sample of BCGs consists of optically selected galaxies drawn
from two catalogs: the C4 BCG catalog \citep{vonderlinden07} and the
maxBCG catalog \citep{koester07a}. We use both these catalogs because,
combined, they fully cover the redshift range of our FGs. The second
sample of BCGs is drawn from the XCS (M11). We will refer to the
former sample as the optical BCGs, the latter as the XCS BCGs, and,
when combined, simply as BCGs. There are considerably less XCS BCGs
than optical BCGs, but they are drawn from an X-ray-selected cluster
sample and therefore are useful in assessing any bias that may result
from comparing an optically selected sample (the optical BCGs) to an
X-ray selected sample (the FGs).

The \citeauthor{vonderlinden07} C4 catalog is based on the
\citet{miller05} C4 catalog but utilizes a better algorithm for
identifying the BCG. It is derived using the SDSS DR4 spectroscopic
sample and it identifies clusters in position, redshift, and color
parameter space assuming that a fraction of the cluster galaxies form
a red sequence. The C4 catalog contains 625 BCGs at $z<0.1$.

The maxBCG algorithm for cluster detection relies on two
characteristics of galaxy clusters: the brightest galaxies in a
cluster occupy a narrow region of color--magnitude space (the red
sequence) and the brightest galaxy in the cluster is located near the
center of the galaxy distribution. Galaxies are designated as BCGs
based on the product of two likelihoods. The first is the likelihood
that the galaxy is spatially located in an overdensity of galaxies
with similar $g-r$ and $r-i$ colors. The second is the likelihood that
it has the color and magnitude properties typical of BCGs. The
algorithm can be run on any catalog of galaxies and doing so on the
SDSS database results in a catalog consisting of 13823 BCGs with
redshifts $0.1<z<0.3$, which is over 85\% complete for halos with
masses above $1 \times 10^{14}\,h^{-1}\,M_\odot$. To match our FS
sample, we restrict the maxBCG catalog to $z\le 0.25$.

XCS-DR1 (M11) consists of 503 optically confirmed X-ray clusters at
$z<1.46$, 402 of which have published X-ray temperatures and
luminosities. From these we draw a sample of clusters that are within
the SDSS footprint, have an observed spectrum, and have $z<0.25$ (the
redshift of our highest-redshift FS), for use as a comparison
sample. For each cluster a BCG was identified from an SDSS image and
in the cases where the BCG was not obvious the brightest galaxy
closest to the X-ray position was selected. The positions of these
clusters in the parameter space used to define FSs are shown in Figure
\ref{lx_gap}. Note in defining the magnitude gap for the XCS BCGs we
used the same method that was used in defining the FS magnitude gaps.

There are 39 XCS clusters that satisfy the above three criteria and
these clusters have a flat distribution in redshift space very similar
to that of the FSs. The optical BCGs have a distribution that peaks at
$z\sim 0.25$, therefore, to minimize any evolutionary effects, we
randomly resampled the original sample of 14448 BCGs (C4 and maxBCG
combined) to create a sample that has a flat distribution over the
same redshift range as our FSs, resulting in a final sample of 2687
optical BCGs.

\section{Data}\label{data}

In this section, we describe the various data sets (X-ray, optical,
and stellar population parameters) that were used in this study. We
conclude the section with a discussion of the errors on the various
quantities.

\subsection{X-Ray Data}\label{xcs}

The X-ray ($T_X$ and $L_X$) and $R_{200}$ data used in this paper come
from the XCS-DR1 (M11). The XCS is a serendipitous search for galaxy
clusters using all publicly available data in the \textit{XMM} Science
Archive \citep{romer01}. The procedures used for measuring $T_X$ and
$L_X$, and for estimating $R_{200}$ are outlined in Section 4 of
\citet[][hereafter LD11]{lloyddavies11}\footnote{Actually, the process
for estimating $L_{X,500}$ and $R_{500}$ is described, but it is
similar to that for estimating $L_{X,200}$ and $R_{200}$.}, so we give
only a brief description here. $T_X$ was measured by performing
spectral fitting to background-subtracted spectra with {\sc
xspec}. The best fit was determined using the maximum likelihood Cash
statistic \citep{cash79}. $L_X$ was measured by extrapolating surface
brightness fits to a simple one-dimensional, spherically symmetric,
$\beta$-profile model \citep{cavaliere76}. $R_{200}$ is then estimated
from the \citet{arnaud05} $T_X$--$R_{200}$ relation.

\subsection{Optical Data}

All galaxy magnitudes are SDSS DR7 Petrosian magnitudes that have been
extinction corrected, $k$-corrected \citep[kcorrect
v4.1.4;][]{blanton07}, and sky-subtraction corrected
\citep{vonderlinden07}. Total optical luminosities of groups/clusters
($L_\mathrm{tot}$) are estimated from all galaxies that are: (1) less
than 0.2 redder and 0.2 bluer than the $r-i$ color of the BCG and
(2) dimmer than the BCG and within a radius of $0.5 R_{200}$ to an
absolute $r$-band magnitude of $-19.67$ (or 20.8 at
$z=0.25$). Galaxies with spectra are restricted to those within 2000
km s$^{-1}$ of the cluster, while those without are restricted to
$\Delta z_\mathrm{phot}=0.1$, where we use the $z_\mathrm{phot}$ from
the SDSS DR7. These selection criteria are identical to those used in
defining the magnitude gaps of the systems (see Section \ref{fs
sample}). The radius of $0.5 R_{200}$ was chosen as it matches the
radius used to define an FS and changing it to $R_{200}$ has no effect
on the results. In addition to the optical data, spectroscopic
redshifts ($z$) of the FGs/BCGs are also obtained from the SDSS and
used to define the redshift of the system.

\subsection{Stellar Population Data}\label{stellar pop}

The SDSS has provided galaxy spectra in unprecedented numbers, which
has allowed for statistical studies of the star formation histories
and stellar populations of various galaxy samples to be performed. The
by-products of these efforts have been numerous new methods for
estimating the associated parameters, such as ages, metallicities, and
elemental abundances, etc. Commonly, these data are stored in public
databases for general use or the method is made public to allow it to
be applied to different samples.

Each of these methods approach what is essentially the same task in a
variety of ways. Thus, each method has it own set of strengths and
weaknesses. In this study, we utilize two different methods/databases:
{\sc starlight} \citep{cidfernandes05} and MPA-JHU \citep{kauffmann03,
brinchmann04, tremonti04, salim07}. Brief outlines of both are given
below but for more details readers are referred to the references
provided. We obtain stellar masses ($M_*$), ages, and metallicities
($Z$) from {\sc starlight} and specific star formation rates (SSFRs)
from MPA-JHU, which also provides stellar masses that we use to test
the robustness of some of our results. We chose to use the {\sc
Starlight} data over the MPA-JHU data for two reasons: the diffusion
map method of selecting a population basis used in \citet{richards09}
has been shown to reduce the age/metallicity degeneracy; and we note
that the MPA-JHU did not release ages nor metallicities beyond DR4,
which is a much smaller sample than DR7.

\subsubsection{\sc Starlight}

The population synthesis code {\sc starlight} \citep{cidfernandes04,
cidfernandes05} fits an observed spectrum with a linear combination of
single theoretical stellar populations (coeval and chemically
homogeneous) computed with the \citet{bruzual03} evolutionary
synthesis models. \citet{richards09} show that the accuracy of this
technique is highly dependent on the choice of input basis of simple
stellar population (SSP) spectra. Thus, we use the
\citeauthor{richards09} diffusion map methods to choose appropriate
bases of prototype SSP spectra. These methods allow us to approximate
the continuous grid of age and metallicity of SSPs of which galaxies
are truly composed. As shown in detail in \citeauthor{richards09}, we
obtain robust stellar ages, stellar metallicities, and stellar masses
and at the same time significantly reduce the degeneracy between age
and metallicity. We correct the stellar masses for the fact that they
are observed within a 3\arcsec\ diameter fiber using the same
correction as applied to the MPA-JHU stellar masses (see Section
\ref{mpa}). In Table \ref{fg_details}, we provide the results of these
fits for stellar mass, luminosity-weighted ages, and stellar
mass-weighted metallicities.

\subsubsection{MPA-JHU}\label{mpa}

The MPA-JHU team has publicly released catalogs of derived physical
properties for a sample of SDSS galaxies, but the only property that
we make use of is the SSFR \citep{brinchmann04}, although we make use
of their stellar masses to estimate an average error for the {\sc
starlight} stellar masses. The catalog is being recreated for the SDSS
DR7, for which there are properties for 818,333 galaxies.

The method of \citet{brinchmann04} to estimate star formation rates is
built on the methodology of \citet{charlot02} and the modeling of
emission lines by \citet{charlot01}, which combine the
\citet{bruzual93} galaxy evolution models with the emission line
modeling from {\sc cloudy} \citep{ferland96}. Dust attenuation is the
major source of uncertainty in estimating star formation rates and
\citeauthor{brinchmann04} follow the dust treatment outlined in
\citet{charlot00}, making an initial estimate of dust attenuation
based on H$\alpha$/H$\beta$. A grid of model H$\alpha$ line strengths
(with a particular dust attenuation applied) is compared with the
observed spectrum and the best match selected using a Bayesian
approach similar to \citet{kauffmann03}. H$\alpha$ luminosity is then
converted to an SSFR following \citet{charlot01}. The estimated SSFR
of each FG is given in Table \ref{fg_details}.

Stellar masses are estimated by multiplying the dust-corrected
luminosity of the galaxy by the stellar mass-to-light ratio predicted
by their best-fit model using the Bayesian approach mentioned above
\citep{kauffmann03}. They extrapolate the mass-to-light ratio and the
dust attenuation values estimated within the SDSS fiber to obtain
total stellar masses.

\subsection{Errors}\label{errors}

Derivation of the X-ray data errors is described in LD11 and we will
only give a summary here. For $L_X$ and $T_X$ the parameter is stepped
(both in the negative and positive directions) from its best-fit value
until the fit statistic increases by the amount required for the
confidence region needed (i.e., 68\%). The $R_{200}$ errors were
calculated by propagating the $T_X$ errors through the
\citet{arnaud05} $T_X$--$R_{200}$ relation.

Errors on $L_\mathrm{gal}$, $L_\mathrm{tot}$, $\Sigma_{24}$, and the
magnitude gaps are from the SDSS photometry and when converting to
solar luminosities both the error on the absolute $g$-band magnitude
(0.02 mag) and the solar $g-r$ color \citep[0.02 mag;][]{bilir05} are
propagated through. In addition to these statistical errors we also
derive errors on $\Delta m_{14}$ due to the errors on $R_{200}$. This
only affects the magnitude gap in three out of the 17 FSs and in none
of the three does it negate the system's classification as an
FS. These systems have this error provided in parentheses after the
statistical error in Table \ref{fs_details}. In some cases, the errors
on $R_{200}$ also induce additional errors on the total optical
luminosities. In half of our sample, the error on $R_{200}$ does not
add or subtract any galaxies into the total optical luminosity. In the
other half, the luminosities do change and these have the largest
errors in Table 2. These systematic luminosity errors are still small
when compared to the total optical luminosity of the system which is
dominated by a single very bright galaxy. However, the fractional
error on the $\Sigma_{24}$ can be quite large and one-sided.

There are no errors on age and metallicity because {\sc Starlight}
does not provide any. MPA-JHU does provide errors on their SSFR but we
do not use them since, as for age and metallicity, we are comparing
the SSFR distributions and not individual systems. We estimate average
stellar mass error from the scatter in the differences between the
{\sc Starlight} and MPA-JHU stellar mass estimates, which encodes the
scatter introduced by the use of two distinct statistical techniques
and two different SSP models. We use these average errors (0.047 dex
for the FSs and 0.079 dex for the XCS BCGs) in relevant plots but do
not provide them in Table \ref{fg_details}.

\section{Stellar Populations}\label{stellar populations}

If FGs formed at a different epoch to BCGs or followed a different
evolutionary path then evidence of this may be found within their
stellar populations. In this section, we use the stellar masses, ages,
and metallicities from {\sc starlight}, and the SSFRs from the MPA-JHU
database to compare the stellar populations of the FGs with those of
the BCGs and investigate this issue.

\begin{figure}
\includegraphics[width=0.5\textwidth]{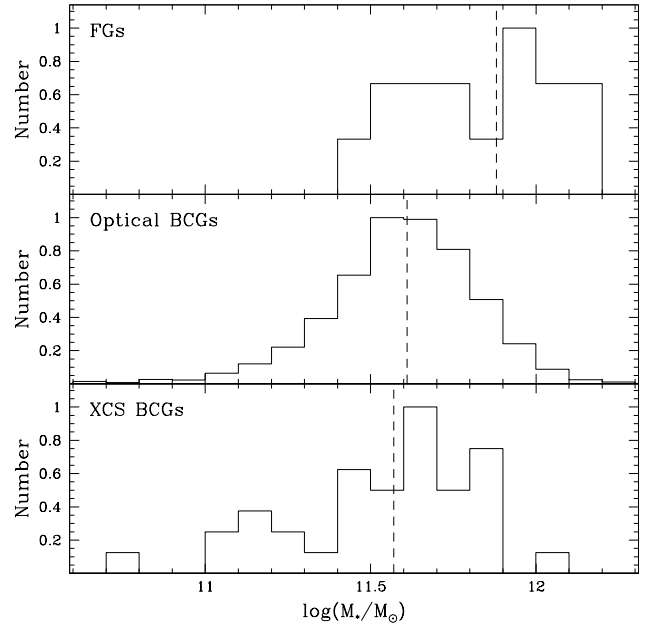}
\caption{{\sc starlight} stellar mass distributions for the various
samples showing that FGs are among the most massive galaxies in the
universe. The dashed lines mark the median mass for each of the
samples.}
\label{star_mass_hist}
\end{figure}

We start by simply comparing the stellar masses of the three samples,
the distributions of which are shown in Figure
\ref{star_mass_hist}. The dashed lines mark the median mass for each
of the samples. From this plot it can be clearly seen that FGs are
among the most massive galaxies in the universe as, on average, they
are more massive than both samples of BCGs. In fact, $\sim 80\%$ of
the FGs are more massive than the average BCG. The median stellar mass
of the FGs is $\langle M_*\rangle=(7.6\pm 1.2)\times
10^{11}\,M_\odot$, while optical BCGs have $\langle M_*\rangle=(4.1\pm
1.9)\times 10^{11}\,M_\odot$ and XCS BCGs have $\langle
M_*\rangle=(3.7\pm 0.1)\times 10^{11}\,M_\odot$, i.e., FGs are roughly
twice as massive as BCGs. We note that the same results are found when
the MPA-JHU photometric-based stellar masses are substituted for the
{\sc starlight} spectroscopically based stellar masses.

Is this difference between the FG and the BCG stellar masses
significant? A simple statistic to use would be the Students $t$-test,
however, this test compares the samples individually and assumes that
the stellar masses are drawn from a normal distribution. The Welch
variation of the $t$-test is applicable if the sample variances are
unequal or the sample sizes are different. If the distributions are
not normal but they are similar (homoscedastic), then the
Mann--Whitney $U$-test is appropriate for individual
comparisons. Unfortunately, our data are not homoscedastic and we need
to make multiple tests simultaneously, therefore none of these
standard techniques are desirable.

We are interested in a joint comparison of the masses from
distributions which are not Gaussian and from samples of widely
varying sizes (see Figure \ref{star_mass_hist}). Thus we use multiple
hypothesis testing (or simultaneous inference, see \citet{shaffer95}
for an excellent review on this subject), since if not accounted for,
the multiplicity can result in an overestimation of statistical
significance. Since we are comparing samples that have different sizes
and distributions (i.e., the random variables are not
heteroscedastic), the $t$-test family is not appropriate. Instead, we
take the most robust statistical approach and control the family-wise
error rates for all pairwise comparisons of group differences via the
max $t$-test using a heteroscedastic consistent covariance estimation
\citep[see][]{herberich10, richardson11}. On a technical note, we
utilize single-step procedures to adjust the $p$-values for their
multiplicity, meaning that the order of the tests is not important.

We find that the stellar masses of the FGs are significantly higher
than those of the optical BCGs and the XCS BCGs. The probabilities
that the FGs and the optical (XCS) BCGs have the same stellar masses
is $<0.005$ (0.003) at the 95\% confidence level. We also find that
there is no difference between the stellar masses of the XCS and
optical BCGs. We note again that these statistics are robust, in that
no assumptions regarding the distribution, sample sizes, or variance
homogeneity have been made. Had we not accounted for the
heteroscedasticity of the samples, our reported probabilities would
have been more than 10 times smaller (and our significance much
higher). Had we used the Mann--Whitney $U$-test (which assumes that
the distributions of the masses are the same in the different sets),
we would have also reported 10 times smaller probabilities (and higher
significances).

Performing the same statistical test on the stellar population
parameters as the stellar masses we find that there are no differences
between any of the three samples in their distributions of SSFRs,
ages, and metallicities, results which are confirmed by K-S
testing. Similar results were found by \citet{labarbera09}, although
when comparing FGs to a sample of bright field ellipticals. No
differences were reported between the ages and metallicities of both
these samples.

\begin{figure}
\begin{center}
\includegraphics[width=0.47\textwidth]{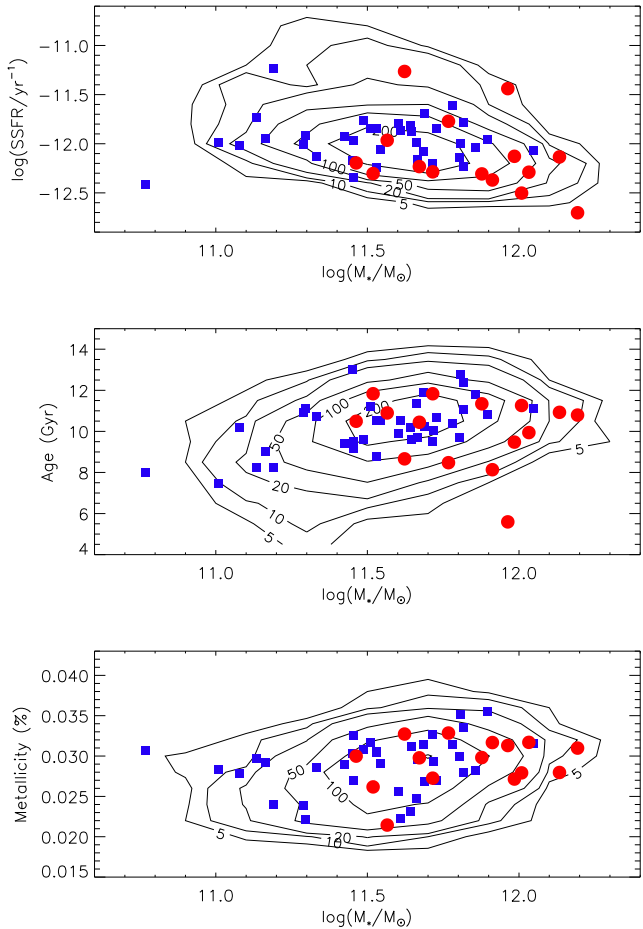}
\caption{Scaling relations of SSFR (top), age (middle), and
 metallicity (bottom). The red circles are the FGs, the blue squares
 are the XCS BCGs, and the density contours are for the optical
 BCGs. The density contour levels are 5, 10, 20, 50, 100, and 200
 galaxies per bin.  The stellar populations of the FGs are found to be
 consistent with those of the BCGs.}
\label{star_mass_trends}
\end{center}
\end{figure}

We show the distribution of the three stellar population parameters as
a function of stellar mass for all three samples in
Figure~\ref{star_mass_trends}, where red circles are the FGs, the blue
squares are the XCS BCGs, and the density contours are for the optical
BCGs. This is for the sake of clarity since there are $\sim 2700$
optical BCGs. The density contour levels are 5, 10, 20, 50, 100, and
200 galaxies per bin. The fact that FGs are among the most massive
galaxies in the universe is highlighted by these distributions where
they mostly hug the massive edge of the BCG distributions.

Using a two-dimensional, two-sample K-S test to compare the optical
BCGs to the XCS BCGs (i.e., X-ray selected) we find no difference
between any of the stellar population parameters, which allows us to
compare the X-ray-selected FGs to the optically selected BCGs and take
advantage of their larger sample size. Comparing the stellar
populations in FGs with this sample we find that the probability that
their SSFRs and ages were drawn from the same distribution is $<0.005$
and $<0.01$ for their metallicities. Given that one-dimensional test
above found no differences between the stellar population parameters
of all three samples, we suggest that this confirms the earlier result
that, on average, FGs are significantly more massive than BCGs and
that the differences found in the two-dimensional testing are driven
by these differences in mass.

In summary, we have performed a careful statistical analysis of the
differences between the FG stellar masses and stellar populations, and
those of the BCGs. Our analysis takes into account the differences in
the sample distributions and sizes, as well as the multiplicity of the
tests. We find that the FGs have significantly higher stellar masses
than the BCGs but similar stellar populations. In the following
sections, we explore the cause of this stellar mass growth in the FG
with respect to the magnitude gap in the FS.

\section{Stellar Mass assembly}\label{assem}

\begin{figure}
\includegraphics[width=0.45\textwidth]{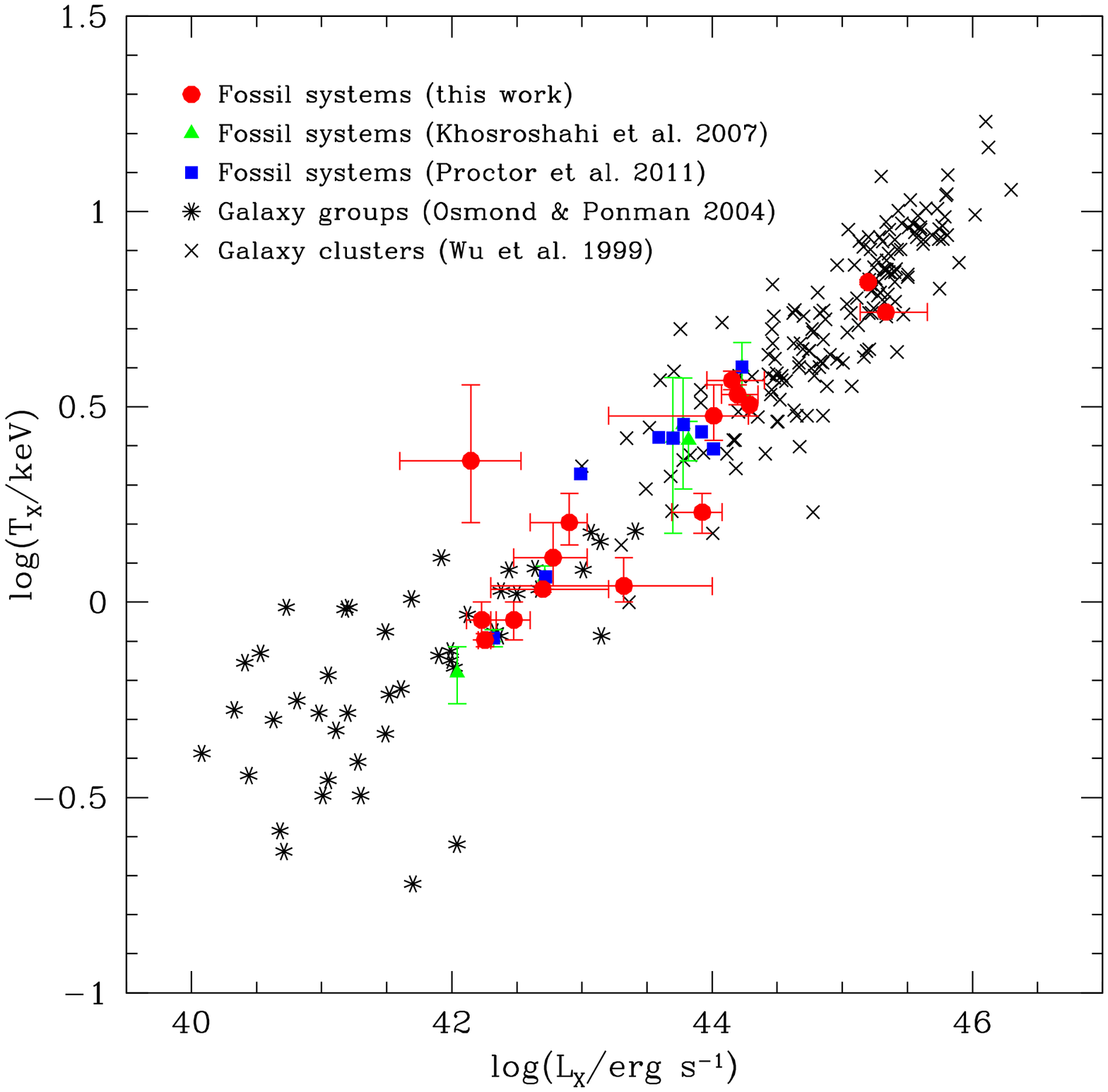}
\caption{$L_X-T_X$ relation for the FSs (red circles). Also plotted
  are the FS data from \citet[][green triangles]{khosroshahi07} and
  \citet[][blue squares]{proctor11}, the \citet{wu99} cluster data
  (black crosses), and the \citet{osmond04} group data (black
  stars). FSs are found at all $L_X$ and $T_X$ except those of the
  most massive clusters. The absence of FSs at low $L_X$ and $T_X$ is
  due to the $L_X$ cutoff used in the definition of an FS.}
\label{lx-tx}
\end{figure}

We begin this section by looking at the X-ray scaling relations of the
FSs and comparing them to those of the XCS BCGs. In Figure
\ref{lx-tx}, we show the $L_X-T_X$ relation of the FSs (red
circles). Also plotted are the FSs from \citet[][green
triangles]{khosroshahi07} and \citet[][blue squares]{proctor11}, the
galaxy group data from \citet[][black stars]{osmond04}, and the galaxy
cluster data from \citet[][black crosses]{wu99}. From this figure we
see that FSs fall on the same relation that both galaxy groups and
clusters do. In fact they are found at all $L_X$, ranging over three
orders of magnitude, except those of the most massive clusters. There
is an absence of FSs at low $L_X$, which is due to a selection effect,
i.e., the $L_X$ cutoff used in the definition of an FS. The
lowest-$L_X$ FS, which has too large a $T_X$ for its $L_X$, is XMMXCS
J030659.8+000824.9 (hereafter J030659.8) and we note that its
classification as an FS is uncertain for reasons given in Appendix
\ref{notes}.

In Figure \ref{lx_ltot}, we examine the optical luminosity of the
galaxy systems within $0.5R_{200}$ ($L_\mathrm{tot}$) as a function of
$L_X$. Red circles are FSs and blue squares are XCS clusters. In a
comprehensive study of the X-ray scaling relations of FSs by
\citet{khosroshahi07}, the authors found evidence that FSs were
boosted in $L_X$ for a given optical luminosity \citep[see
also][]{santos07}, which is also predicted by in the $N$-body
simulations of \citet{d'onghia05}. However, other studies have found
no difference \citep{aguerri11} and have concluded that there is a
possible systematic difference between the studies or that the
difference is real but only for less massive systems. We find no
evidence for boosted $L_X$ in our data, rather we find that FSs, over
more than three orders of magnitude in $L_X$, populate this diagram in
the same way that non-FSs do.

\begin{figure}
\includegraphics[width=0.48\textwidth]{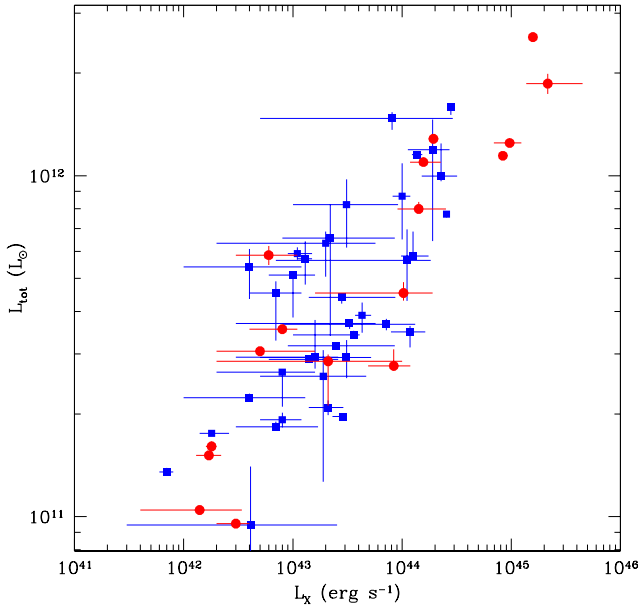}
\caption{Optical luminosity of the galaxy system within $0.5R_{200}$
  as a function of $L_X$. There is no difference between the FS
  relation and that of the XCS clusters. Red circles are FSs and blue
  squares are XCS clusters.}
\label{lx_ltot}
\end{figure}

\begin{figure}
\begin{center}
\includegraphics[width=0.47\textwidth]{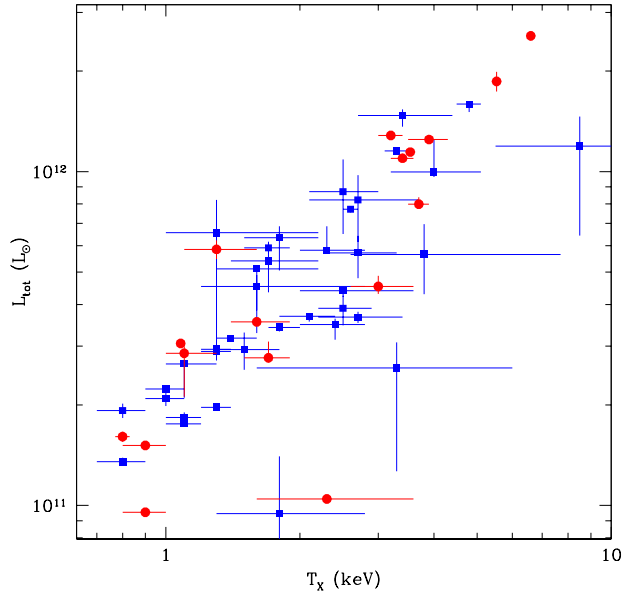}
\caption{Optical luminosity of the galaxy system within $0.5R_{200}$
  as a function of $T_X$. The intrinsic scatter about the FS relation
  is greatly reduced both when compared to the FS relation with $L_X$
  and the XCS cluster relations with $L_X$ and $T_X$. Red circles are
  FSs and blue squares are XCS clusters.}
\label{tx_ltot}
\end{center}
\end{figure}

The optical luminosity of the FSs is strongly correlated with X-ray
luminosity ($r_S=0.92$, $p<1e-7$) and the intrinsic scatter in
log($L_\mathrm{tot}$) is only $0.14\pm 0.02$ dex. Similar results are
found for the XCS clusters, however, the intrinsic scatter is slightly
larger. Substituting $T_X$ for $L_X$ (Figure \ref{tx_ltot}) produces
an even tighter relation, the intrinsic scatter in
log($L_\mathrm{tot}$) is reduced to $0.10\pm 0.01$ dex, although the
correlation is slightly weakened ($r_S=0.85$, $p<0.0001$) due to a
larger number of outliers. Again similar results are found for the XCS
clusters.

The X-ray temperatures are used to infer the radii within which we
calculate the total optical luminosity of the systems. So it is
reasonable to expect there to be some induced correlation between
$T_X$ and $L_\mathrm{tot}$. For example, as we scatter upward in X-ray
temperature, we would of course infer a larger radius to include the
optical light from member galaxies (but never less). We tested the
amplitude of this correlated scatter by examining the change in the
fraction of light as we change $R_{200}$ based on the $T_X$ errors
($\pm 1\sigma$). We find that half of the FSs have optical
luminosities that are identical after changing the radius. The other
half have scatter in their $L_\mathrm{tot}$ that correlates with
scatter in $T_X$. These eight systems have the largest
$L_\mathrm{tot}$ errors in Table 2. However, the slope of the fit to
the fractional scatter in $L_\mathrm{tot}$ versus $T_X$ in these
systems is only 0.3. For a given system, if the $T_X$ scatters up by
100\%, the optical luminosity scatters up by $\sim 30\%$. In Figure
\ref{tx_ltot}, we see that the relationship between total optical
luminosity and X-ray temperature is much steeper than this correlated
scatter. For this reason and the fact that half the FS have
uncorrelated scatter between $T_X$ and $L_\mathrm{tot}$, we conclude
that the real scatter in Figure \ref{tx_ltot} is not much larger than
our measured scatter (see also Appendix \ref{robust}).

\begin{figure}
\includegraphics[width=0.5\textwidth]{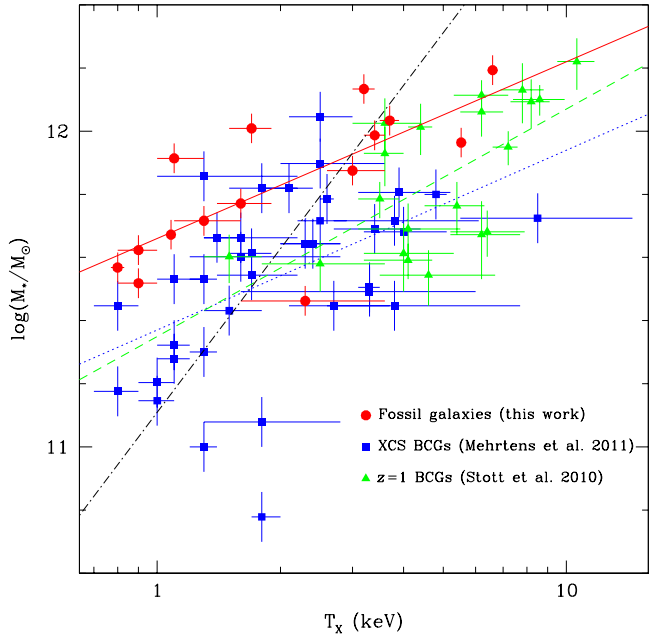}
\caption{Variation of the stellar mass of the dominant galaxy with
  $T_X$ showing that, for a given $T_X$, FGs are among the most
  massive galaxies in the universe. The red circles are the FGs, the
  blue squares are the XCS BCGs and the green triangles are the $z\sim
  1$ BCGs of \citet{stott10}. The red solid line is the fit to the FG
  data, the blue dotted line is the fit to the XCS BCGs, and the
  dashed green line is the fit to the \citeauthor{stott10} BCG
  data. The black dot-dashed line shows the slope (1.72) of the
  power-law relation between $M_{500}$ and $T_X$ from \citet{stott10},
  which has been normalized to the FG fit at $T_X=3.0$ keV.}
\label{tx_mass}
\end{figure}

\begin{figure}
\includegraphics[width=0.48\textwidth]{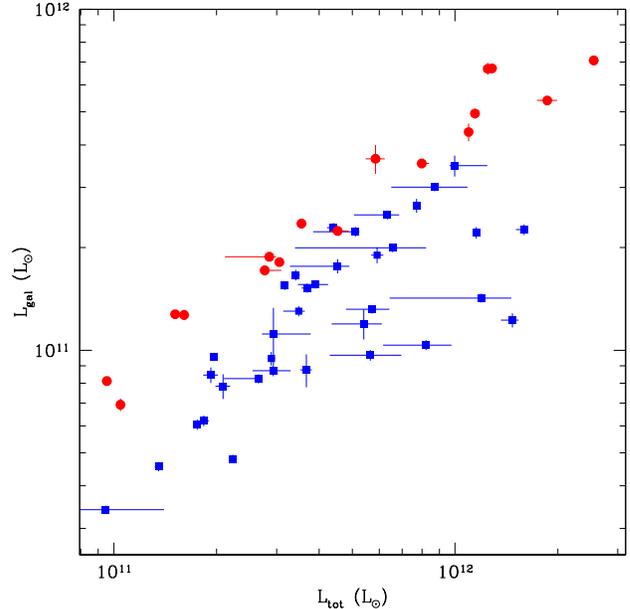}
\caption{Same as Figure \ref{tx_mass} but now we
  show the variation of the optical luminosity of the dominant galaxy
  with the total optical luminosity of the system. As in Figure
  \ref{tx_mass} the FGs lie at the upper edge of the BCG
  distribution. Red circles are FGs and blue squares are BCGs.}
\label{ltot_lr}
\end{figure}

In Figure \ref{tx_mass}, we show the stellar mass of the dominant
galaxy (i.e., an FG in an FS or a BCG in a non-FS) as a function of
$T_X$.  The red circles are our FGs, the blue squares are the $z\leq
0.25$ XCS BCGs, and the green triangles are the $z\sim 1$ BCGs from
\citet{stott10}. While the fits to the two BCG samples are consistent
with each other (the green dashed and blue dotted lines) the fit to
the FGs is offset to higher mass for a given $T_X$ (the red solid
line). In Figure \ref{ltot_lr} we present the same information as in
Figure \ref{tx_mass}, except we represent the stellar mass of the
dominant galaxy by its optical luminosity and the system mass by the
total optical luminosity (see Section \ref{stellar pop}). Since we do
not have stellar masses for all our FGs nor BCGs we make use of the
$r$-band luminosity, which we expect to be a good proxy for stellar
mass, to maximize our sample sizes. Just as in Figure \ref{tx_mass},
the FGs sample the upper edge of the BCG distribution and similar to
Figures \ref{lx_ltot} and \ref{tx_ltot} the scatter in the FGs is
remarkably small.

We find that there is a strong correlation between $T_X$ and $M_*$
($r_S=0.76$, $p<0.001$) but that none of the stellar population
parameters are found to be correlated. These results show that, while
the mass of the system hosting the FG determines how massive it can
become, it has little effect on the formation and evolution of the
stars that compose it. Similar results are found when substituting
$T_X$ with $L_X$ and the significance of the result is unaffected
after excluding the two systems with $\Delta m_{14}<2.5$.

There are two results in Figure \ref{tx_mass}. The first is that FGs
are more massive than XCS BCGs for any given $T_X$, and so the mean
stellar mass of the ensemble of FGs has to be higher than
non-FGs. This explains our results in Figure \ref{star_mass_hist} and
in Section \ref{stellar populations}, where we found that the mean
stellar mass of the FG sample is significantly higher than the non-FS
samples.

Figure \ref{tx_mass} shows that, at any given $T_X$, FGs have the
highest stellar mass. There are, however, a few XCS BCGs that have
large stellar masses and are not classified as fossils. We note that
these exceptions all have $\Delta m_{14}$ of $\sim 2$, which is close
to the FS classification threshold. There is one FG that lies well
below the rest but we have reason to doubt the X-ray emission in this
system and we refer the reader to Appendix \ref{J0306} for a
discussion.

The second result from Figure \ref{tx_mass} is that the $z\sim 0.1$
FGs are $\sim 100\%$ ($\sim 40\%$) more massive than the $z\sim 1$
BCGs at 1 keV (10 keV). The fits to the low- and high-redshift BCG
samples are consistent with each other. Therefore, if the $z\sim 1$
BCGs are the progenitors of the XCS BCGs, which are at $z\leq 0.25$,
then this figure shows that BCGs do not have any significant growth at
$z \lesssim 1$, consistent with previous studies
\citep[e.g.,][]{stott10}. However, if BCGs from either sample are the
progenitors of FGs, then the BCGs need to grow by $\sim 40\%$--$100\%$
to reach the average mass of an FG at $z\sim 0.1$. Semi-analytic
models of early-type galaxy formation in clusters
\citep[e.g.,][]{delucia06} suggest that FGs form via gas-poor mergers
even at low redshifts \citep{diaz-gimenez08}, despite the fact that
FSs assemble most of their virial mass at higher redshifts. There is
some observational evidence that suggests BCGs continue to grow
through dissipationless mergers even today \citep{brough05, brough11}
and simulations predict that a large fraction of the BCG mass today
was accrued in the last 5 Gyr \citep{delucia07, ruszkowski09}. Our
results suggest that neither the high- or low-redshift BCGs have
evolved and that neither are the progenitors of FGs, unless
significant major merging events by galaxies well outside the systems
have yet to occur \citep[e.g.,][]{brough11}.

The black dashed line shows how the mass of the underlying system
(i.e., the group or cluster mass) scales with temperature via the
$M$--$T_X$ relation (core-excised) from Table 1 in
\citet{maughan07}. Both the stellar masses of the dominant galaxies
and the total masses of the groups/clusters scale with $T_X$, but
their relationships are decoupled.

\begin{figure}
\includegraphics[width=0.48\textwidth]{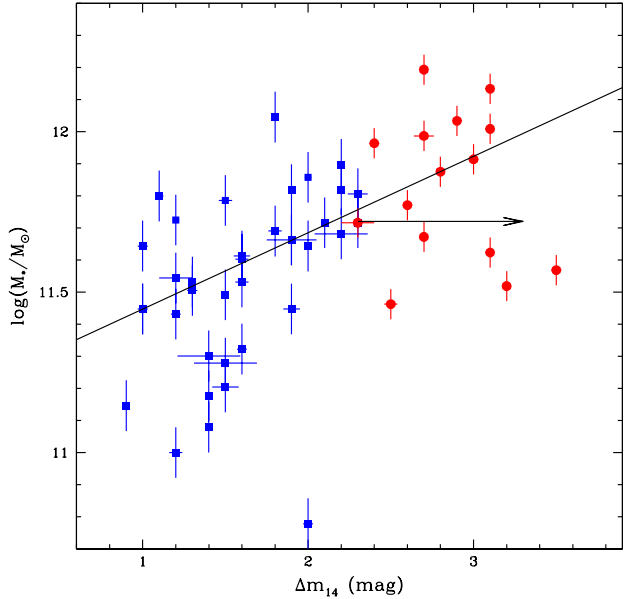}
\caption{Highly significant correlation between $\Delta m_{14}$ and
  the stellar mass of the dominant galaxy. Red circles are FGs and
  blue squares are XCS BCGs. The solid line is the fit to the data,
  both the FGs and the XCS BCGs. The data point with the arrow is
  J124425.9 (see the text for details).}
\label{gap_mass}
\end{figure}

We can see from Figure \ref{gap_mass} that the mass of the dominant
galaxy is correlated with the size of the magnitude gap. The question
is whether the magnitude gap and the stellar mass of the galaxy are
part of a cause and effect relationship: are the FG stellar masses
large because the magnitude gap is large? These two quantities are
found to be positively correlated with high significance ($r_S=0.58$,
$p<0.00001$). The solid line is the fit to the data, both the FGs and
the XCS BCGs. The data point with the arrow in this plot (and the
following plots) is J124425.9. The FG in this system has a double
nucleus and so we elect not to apply the sky-subtraction correction;
the arrow denotes the position of this system in this plot if we did
apply the correction.

\begin{figure}
\includegraphics[width=0.48\textwidth]{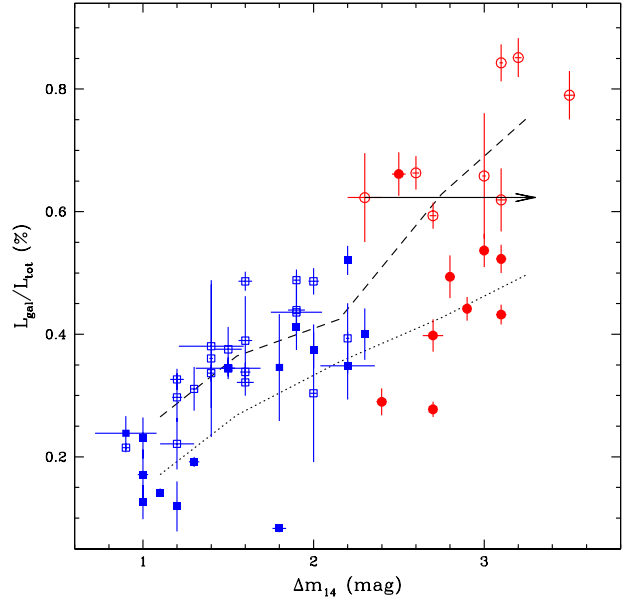}
\caption{Fraction of total cluster light contained within the dominant
galaxy as a function of $\Delta m_{14}$. The FGs can contain a
significant fraction of the total light of a system (up to 85\%) while
for the XCS BCGs this fraction is generally lower. Red circles are FGs
and blue squares are XCS BCGs. The dashed line shows the median
fraction for low-mass systems (open symbols; $T_X<2$ keV) and the
dotted line is for high-mass systems (filled symbols; $T_X>2$
keV). The data point with the arrow is J124425.9 (see the text for
details).}
\label{frac_opt_gap}
\end{figure}

In Figure \ref{frac_opt_gap} we examine the fraction of total optical
luminosity contained within the dominant galaxy as a function of the
magnitude gap $\Delta m_{14}$. Red circles are FGs and blue squares
are XCS BCGs. The trends seen in this figure are still present after
we split the sample into high-mass systems (with $T_X > 2$
keV---dashed line and open symbols) and low-mass systems ($T_X < 2$
keV---dotted line and filled symbols). The magnitude gap bins, within
which the medians were calculated, were chosen to contain
approximately equal numbers and are as follows: 0.9--1.29, 1.3--1.79,
1.8--2.49, 2.5--2.99, and 3.0--3.5. The data point with the arrow is
J124425.9 and it is not use to calculate the median in this or any
following plots.

\begin{figure}
\includegraphics[width=0.48\textwidth]{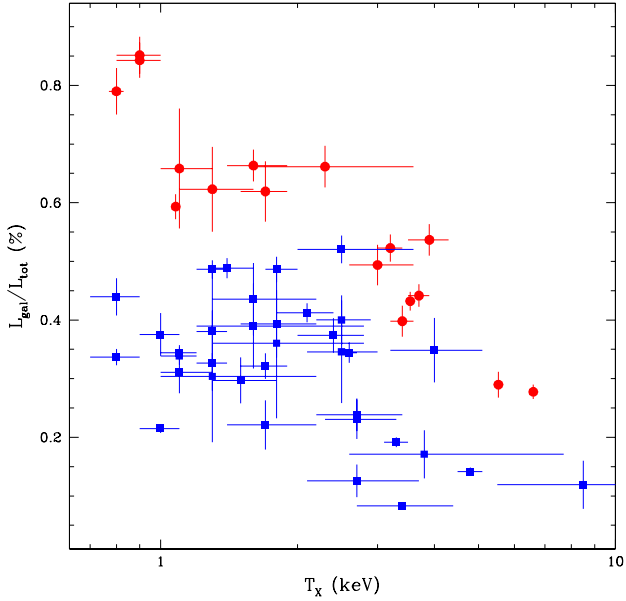}
\caption{Fraction of the total optical luminosity in the system
  contained within the dominant galaxy as a function of $T_X$. Again,
  the FGs lie at the upper edge of the BCG distribution and form a
  tight sequence. The red circles are the FGs and the blue squares are
  the XCS BCGs.}
\label{frac_opt_tx}
\end{figure}

The fraction of total optical luminosity and the magnitude gap are
highly correlated ($r_S=0.80$, $p=0$ (actually $2e-12$)). There is a
large spread in the fraction of light contained within the dominant
galaxy, especially for the FGs. The values range from 30\% all the way
up to 85\%, consistent with the values that have been reported in the
literature (\citeauthor{aguerri11} 2011 found a value of 15\% and
\citeauthor{jones00} 2000 a value of 70\%). This spread is caused by
the strong correlation that exists between the dominant galaxy light
fraction and the mass of the system \citep{lin04}, which we show in
Figure \ref{frac_opt_tx} using $T_X$ as a proxy for system mass, and
the lack of a correlation between $\Delta m_{14}$ and system mass. In
this figure red circles are FGs and blue squares are XCS BCGs. Similar
to Figures \ref{tx_mass} and \ref{ltot_lr}, the FGs sample the upper
edge of the BCG distribution and have a much smaller scatter.

Looking at the trends with system mass, in Figure \ref{frac_opt_gap},
it appears that there is a bi-modality at high magnitude gaps with a
cleaner separation between the high-mass and low-mass systems. FGs in
low-mass systems contain a larger fraction of the total optical luminosity
of the system compared to the FGs in low-mass systems. However, our
data are insufficient to be able to make any definite claims. The
high-mass outlier is J030659.8, which was also an outlier in Figures
\ref{lx-tx} and \ref{tx_mass}.

\begin{figure}
\includegraphics[width=0.48\textwidth]{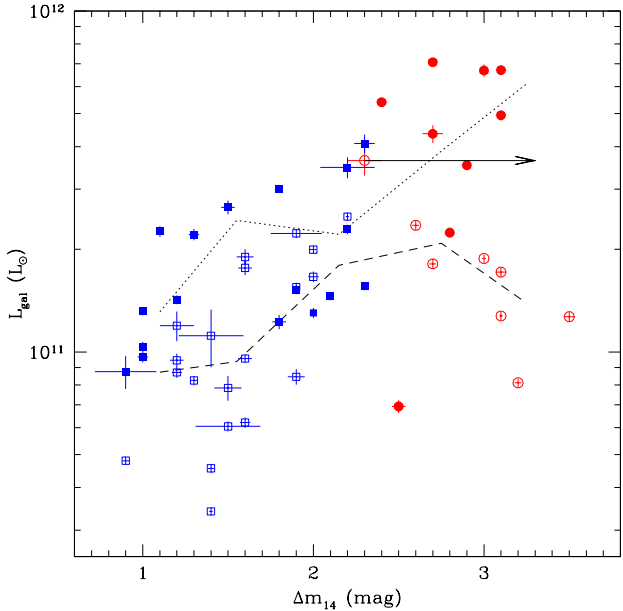}
\caption{Dominant galaxy luminosity as a function of $\Delta
  m_{14}$. These plots show that luminous dominant galaxies are likely
  to be found in systems with large magnitude gaps. The red circles
  are the FGs and the blue squares are the XCS BCGs. The dashed line
  and the dotted line are the medians of the low-mass and high-mass
  systems. The data point with the arrow is J124425.9 (see the text
  for details).}
\label{opt_gap}
\end{figure}

\begin{figure}
\includegraphics[width=0.48\textwidth]{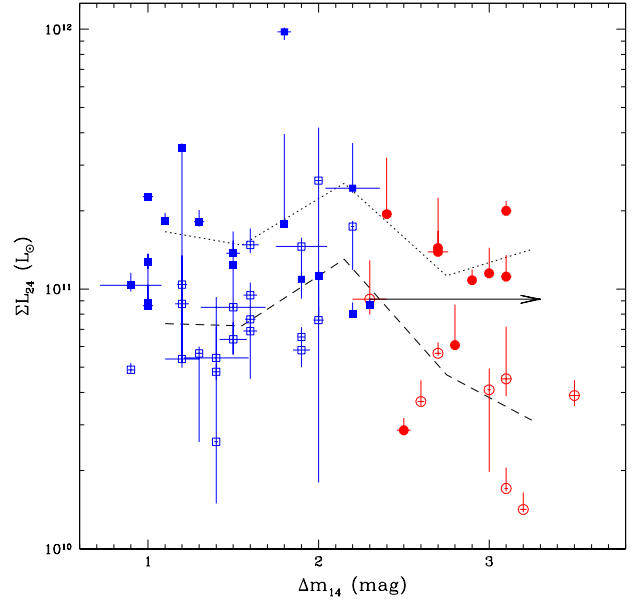}
\caption{$\Sigma L_{24}$ as a function of $\Delta m_{14}$. The light
  contained within $\Sigma L_{24}$ remains constant or decreases as
  the magnitude gap increases. The red circles are the FGs and the
  blue squares are the XCS BCGs. The dashed line and the dotted line
  are the medians of the low-mass and high-mass systems. The data
  point with the arrow is J124425.9 (see the text for details).}
\label{m24_gap}
\end{figure}

In Figure \ref{opt_gap}, we show how the optical luminosity of the
dominant galaxy changes with magnitude gap and in Figure \ref{m24_gap}
we show how the total optical luminosity of the second-to-fourth
brightest galaxies ($\Sigma L_{24}$) changes with magnitude gap. The
symbols and line styles are the same as those used in Figure
\ref{frac_opt_gap}. From Figure \ref{opt_gap} we see that the overall
trend is for the luminosity of the dominant galaxy to increase as the
magnitude gap grows. These two quantities are significantly correlated
($r_S=0.52$, $p<0.00005$). Similar to Figure \ref{frac_opt_gap} we see
the suggestion of a bi-modality at high magnitude gaps. In Figure
\ref{m24_gap}, $\Sigma L_{24}$ is either constant or slightly
decreasing suggesting that very little merging (and certainly no major
merging) is occurring among these galaxies. As in Figure
\ref{frac_opt_gap}, the high-mass outlier is J030659.8.

\section{Discussion}\label{discussion}

The debate about what FSs truly represent is far from over. Whether
they are the end-point of mergers in groups/clusters or not is a
contentious issue. If FSs do not represent a distinct class of objects
then the cause of their magnitude gap, which is their distinguishing
feature, needs to be addressed. Two of the leading possibilities are
an early formation epoch coupled with a favorable configuration for
the rapid and efficient accretion of companions \citep{jones03,
khosroshahi04, d'onghia05, khosroshahi07, dariush07,
vonbendabeckmann08} and the failed group scenario \citep{mulchaey99}.

The results of this study provide some answers to this question. The
observed facts that: (1) our FSs exist at all system masses; (2) our
FSs are not deficient in $L^*$ galaxies (see the CMDs in Appendix
\ref{notes}); and (3) the FGs have the largest stellar mass of any
cluster BCG (as a function of the system mass), all rule out the
failed group scenario. These are normal groups and clusters with
populated red sequences but with an extremely massive BCG.

The correlations with magnitude gap (Figures \ref{gap_mass},
\ref{frac_opt_gap}, \ref{opt_gap}, and \ref{m24_gap}) support the idea
that the gap is an indicator of the evolutionary phase of an FS, with
a large gap indicative of an early-forming system that contains an FG
that has undergone many small mergers. This observational result is
entirely consistent with the simulations of
\citet{d'onghia05}. Therefore, any model of FS formation and evolution
must do the following as the magnitude gap grows.

\begin{itemize}
\item The luminosity of the FG must increase.
\item The fraction of the total optical luminosity of the system contained
  in the FG must increase.
\item The total fraction of light contained in the second-to-fourth
brightest galaxies must not increase.
\end{itemize}

If repeated mergers within the system are responsible for the above
points then there are two likely scenarios and the fact that the
luminosity contained within the second-to-fourth brightest galaxies
stays constant with magnitude gap is important. First, the FG may grow
by accreting many smaller galaxies leaving the second-to-fourth
brightest galaxies untouched. This scenario de-emphasizes dynamical
friction and favors fast and efficient mergers with galaxies with low
impact parameters. It is supported by the fact that in all our FSs
only two (J030659.8 and J123338.5) have fourth brightest galaxies that
are less luminous than $L^*$. In all other systems the FG is so
luminous that it is more than 2.5 mag brighter than $L^*$. Second, the
FG could undergo fewer mergers with smaller galaxies and then develop
the magnitude gap in a single equal-mass merger. This requires that
the second-to-fifth brightest galaxies be of similar luminosity, so
that when one is accreted the luminosity contained within the
second-to-fourth brightest galaxies remains roughly constant. Both of
these scenarios imply that the merging is more efficient in FSs than
non-FSs, i.e., a higher fraction of the stellar mass is retained in
the FG.

For the cluster-sized FSs there is an additional formation
scenario. The FG could form in a group where the relative velocities
are conducive to merging, before accreting (or being accreted by)
another system. This second system, however, must be lacking in bright
galaxies otherwise the magnitude gap would vanish and the system would
no longer qualify as an FS, making this mode of formation unlikely,
yet not impossible. \citet{schirmer10} found J0454-0309 to consist of
two systems, a sparse cluster and an infalling FS, the latter they
believe will seed the FG. They also find, outside a radius of 1.5 Mpc,
two filaments that extend over 4 Mpc. In this study, we find two
systems (J141627.7 and J141657.5) that are possibly interacting and
represent a possible example of this scenario.

If only one process leads to the development of an FS then it must be
possible for this process to occur in a system of any mass, i.e., we
have FSs ranging in mass from groups to clusters. Looking at Figures
\ref{frac_opt_gap}, \ref{opt_gap}, and \ref{m24_gap} the case could be
made for two different formation mechanisms: one in the high-mass
systems and another in the low-mass systems. Although our data are not
definitive we will nonetheless speculate about this possibility.

It is possible that the high-mass FSs represent those that evolved to
have a magnitude gap via mergers while the low-mass FSs are those that
were formed with a magnitude gap. The systems that evolved to be FSs
would contain FGs that have undergone numerous mergers and so would be
expected to have large masses. They would also exist in rich systems
and so the chances of them developing large magnitude gaps would be
small. The systems that were formed as fossils, on the other hand,
would contain FGs that had not undergone many (if any) mergers and so
would be relatively less massive. Being poor systems they are
statistically more likely to have larger magnitude gaps.

The results from Figure \ref{tx_mass} suggest that $z\sim 1$ BCGs
cannot be the progenitors of low-redshift FSs without a major merger
or an abnormal number of minor mergers at $z<1$. According to the fits
in this figure, at 3 keV, an average FG has $M_*=8 \times
10^{11}\,M_\odot$ and an average BCG has $M_*=5 \times
10^{11}\,M_\odot$. The FG is therefore 60\% more massive than the
BCG. If a BCG was to undergo a single major merger to form an FG then
this would require a merger ratio of less than 1:1.6 (i.e., a nearly
equal-mass merger). However, a $z\sim 1$ galaxy undergoes a single
major merger (merger ratios 1:1--1:4) in $\sim 8$ Gyr and a single
minor merger (merger ratios 1:4--1:10) in $\sim 3$ Gyr
\citep{lotz11}. The look-back time from $z=0.15$ (the mean redshift of
the FG sample) to $z\sim 1$ is $\sim 6$ Gyr.

Essentially the FG must be formed from the merger of two BCGs, a rare
event at low redshift. In the SDSS 90\% of all major mergers occur
between galaxies that are fainter than $L^*$ \citep{patton08}. To have
two BCGs merging requires the merger of two clusters---another rare
event at low redshifts. 50\% of massive halos at $z=0$ had their last
major merger at $z\ge 1$ \citep{fakhouri10}. The above considerations
suggest that FSs are not a phase of normal group/cluster evolution
\citep{vonbendabeckmann08}.

Closely linked to the formation of the magnitude gap is the formation
of the FG. Why are they among the most massive galaxies for a given
system mass? Are FGs formed later than BCGs in non-FSs through
gas-poor mergers \citep[dry mergers;][]{diaz-gimenez08}? Are they
formed via gas-rich mergers \citep[wet mergers;][]{khosroshahi06b}? Or
is it a combination of the two wet mergers early with most of their
mass being assembled later through dry mergers
\citep{mendezabreu12}. The results of our stellar population analysis,
which found no differences between FGs and BCGs, rule out late-time
wet mergers. However, it is possible that the stellar population
models we are using are unable to resolve such small differences.

The secret to determining the formation method of FGs and FSs may lie
in the diffuse stellar component (DSC). The DSC is an important
component of a cluster's overall luminosity and is composed of
material that has been stripped from cluster galaxies during dynamical
interactions \citep{feldmeier02, rudick06, rudick09}. Anywhere from
10\% to 40\% of a system's total luminosity can be found in the DSC,
with the quantity increasing with time due to mergers. It is likely
that there is no universal DSC fraction, but that different
groups/clusters will have different DSC levels, depending on their
specific evolution and history \citep{murante04, rudick06,
conroy07}. Also, differing dynamical interactions are found to create
distinct structures within the DSC. For example, tidal streams are
associated with fast, close encounters between a galaxy and the
dominant galaxy. The DSC can provide a wealth of information on the
dynamical history of both the dominant galaxy and the cluster itself
\citep{rudick11}.

The detection of structure in the DSC would indicate recent mergers
and that the FG is still rapidly evolving at $z=0.2$
\citep[e.g.,][]{brough11}. The lack of structure (but the existence of
DSC) would indicate that the FG has not evolved through major mergers
since $z>1$ \citep[e.g.,][]{stott08}. On the other hand, a lack of DSC
altogether would indicate that the FG formed its mass without major
mergers \citep[e.g.,][]{mulchaey99}).

\section{Conclusions}\label{conclusions}

Using X-ray data from the XCS combined with optical data from the SDSS
we have defined a sample of 17 FSs at $z\leq 0.25$, of which 14 have
not been classified as such previously. This catalog represents not
only an increase in the number of known FSs but also an increase in
the quality of the X-ray data used to study such systems. Using the
data from XCS we examine the X-ray scaling relations of FSs. For the
FGs, we estimate stellar masses, ages, and metallicities using {\sc
starlight} and obtain star-formation rates from MPA-JHU. Using these
data compare the stellar mass assembly and the stellar populations of
FGs to two samples of BCGs, one optically selected and one X-ray
selected. The main results from this paper are as follows.

\begin{enumerate}
\item FSs, i.e., systems with a large magnitude gap, have masses that
range from those of galaxy groups to those of galaxy clusters.
\item At fixed halo mass, the stellar mass of the dominant galaxy in
FSs is larger than those in non-FSs.
\item The fraction of light in the dominant galaxy, as well as the
luminosity of the dominant galaxy, increases with magnitude gap for
all galaxy groups and clusters.
\end{enumerate}

A scenario whereby FSs form at high redshift and FGs grow to high
masses through fast and efficient mergers could explain most of the
results in this paper. A study of the intracluster light in both FSs
and non-FSs could help to decide whether the above explanation is
plausible or not.

\section*{Acknowledgments}

The authors thank the anonymous referee who helped to improve the
paper through a careful reading of the manuscript.

This work is based on observations obtained with \textit{XMM}, an ESA
science mission with instruments and contributions directly funded by
ESA Member States and NASA. It has made use of data or software
provided by the US National Virtual Observatory, which is sponsored by
the National Science Foundation; the NASA/IPAC Extragalactic Database
(NED) which is operated by the Jet Propulsion Laboratory, California
Institute of Technology, under contract with the National Aeronautics
and Space Administration; and the SDSS. Funding for the SDSS and
SDSS-II has been provided by the Alfred P. Sloan Foundation, the
Participating Institutions, the National Science Foundation, the
U.S. Department of Energy, the National Aeronautics and Space
Administration, the Japanese Monbukagakusho, the Max Planck Society,
and the Higher Education Funding Council for England. The SDSS Web
Site is http://www.sdss.org/. The SDSS is managed by the Astrophysical
Research Consortium for the Participating Institutions. The
Participating Institutions are the American Museum of Natural History,
Astrophysical Institute Potsdam, University of Basel, University of
Cambridge, Case Western Reserve University, University of Chicago,
Drexel University, Fermilab, the Institute for Advanced Study, the
Japan Participation Group, Johns Hopkins University, the Joint
Institute for Nuclear Astrophysics, the Kavli Institute for Particle
Astrophysics and Cosmology, the Korean Scientist Group, the Chinese
Academy of Sciences (LAMOST), Los Alamos National Laboratory, the
Max-Planck-Institute for Astronomy (MPIA), the Max-Planck-Institute
for Astrophysics (MPA), New Mexico State University, Ohio State
University, University of Pittsburgh, University of Portsmouth,
Princeton University, the United States Naval Observatory, and the
University of Washington.

Financial support for this project includes: The Science and
Technology Facilities Council (STFC) through grants ST/F002858/1
and/or ST/I000976/1 (for E.L.-D., A.K.R., N.M., A.R.L., and M.S.),
ST/H002391/1 and PP/E001149/1 (for C.A.C. and J.P.S.), and through a
studentship (for N.M.), The University of KwaZulu-Natal and the
Leverhulme Trust (for M.H.), FP7-PEOPLE-2007-4-3-IRG n 20218 (for
B.H.), Funda\c{c}\~{a}o para a Ci\^{e}ncia e a Tecnologia through the
project PTDC/CTE-AST/64711/2006 (for P.T.P.V.), The Swedish Research
Council (V.R.) through the Oskar Klein Centre for Cosmoparticle
Physics (for M.S.), The U.S.\ Department of Energy, National Nuclear
Security Administration by the University of California, Lawrence
Livermore National Laboratory under contract no.\ W-7405-Eng-48 (for
S.A.S.), and The National Science Foundation Cyber-Enabled Discovery
and Innovation (CDI) grant 0941742 (J.W.R.).

\clearpage

\setcounter{table}{0}

\begin{landscape}
\begin{table*}
\scriptsize
\caption{The details of the fossil systems in our sample. The ID
  matches a system to a fossil galaxy in Table
  \ref{fg_details}. $\Delta m_{12}$ and $\Delta m_{14}$ are the
  \citet{jones03} and \citet{dariush10} magnitude gaps (quoted errors
  are from the photometry, while those in parentheses are those that
  arise due to the errors on $R_{200}$); $R_{200}$ is in Mpc; $T_X$ is
  in keV; $L_X$ is the X-ray luminosity measured inside $R_{200}$ in
  $10^{44}\,h_{70}^{-2}\,\mathrm{erg\,s}^{-1}$; $L_\mathrm{tot}$ is the total
  $r$-band luminosity inside $0.5R_{200}$ in $10^{11}\,L_\odot$; and
  $\Sigma L_{24}$ is the total $r$-band luminosity of the second to
  fourth brightest galaxies in $10^{11}\,L_\odot$. FS references:
  (1) \citet{diaz-gimenez08}; (2) \citet{jones03};
  (3) \citet{cypriano06}; (4) \citet{khosroshahi06a};
  (5) \citet{khosroshahi06b}; (6) \citet{voevodkin10}; (7) \citet{proctor11}; 
  (8) \citet{santos07}. $^a$ The system was an \textit{XMM}
  target. $^b$ The magnitude of the FG has not been corrected so the
  magnitude gap of this system is a lower limit; see Appendix
  \ref{J124425} for details.}\label{fs_details}
\begin{center}
\begin{tabular}{lr@{}lccccccccc}
\hline
ID & \multicolumn{2}{c}{XCS Name} & Literature Name & $\Delta m_{12}$
& $\Delta m_{14}$ & $R_{200}$ & $T_X$ & $L_X$ & $L_\mathrm{tot}$ &
$\Sigma L_{24}$ & FS refs.\\
\hline
1  &  XMMXCS J015315.0+010214.2 &      & WHL J015315.2+010220          & 1.7$\pm 0.02$ & 2.7$\pm 0.02$         & $0.664_{-0.008}^{+0.007}$ & $1.08_{-0.02}^{+0.02}$ & $0.05_{-0.03}^{+0.11}$    &   $3.06_{-0.06}^{+0.06}$ & $0.57_{-0.01}^{+0.06}$    & ---        \\
2  &  XMMXCS J030659.8+000824.9 &      & SDSS CE J046.719402+00.163919 & 1.3$\pm 0.06$ & 2.5$\pm 0.04$         & $1.01_{-0.18}^{+0.27}$    & $2.3_{-0.7}^{+1.3}$    & $0.014_{-0.010}^{+0.020}$ &   $1.05_{-0.03}^{+0.03}$ & $0.29_{-0.01}^{+0.03}$    & ---        \\ 
3  &  XMMXCS J073422.2+265143.9 & $^a$ & [DMM2008] IV                  & 2.4$\pm 0.30$ & 3.0$\pm 0.01$($+0.5$) & $0.67_{-0.05}^{+0.10}$    & $1.1_{-0.1}^{+0.2}$    & $0.21_{-0.19}^{+0.79}$    &   $2.86_{-0.74}^{+0.74}$ & $0.41_{-0.21}^{+0.09}$    & 1          \\ 
4  &  XMMXCS J083454.8+553420.9 &      & WHL J083454.9+553421          & 2.4$\pm 0.03$ & 3.0$\pm 0.03$         & $1.17_{-0.07}^{+0.06}$    & $3.9_{-0.4}^{+0.4}$    & $9.66_{-2.74}^{+2.75}$    &  $12.47_{-0.29}^{+0.43}$ & $1.15_{-0.03}^{+0.29}$    & ---        \\
5  &  XMMXCS J092540.0+362711.1 &      & NSC J092521+362758            & 1.9$\pm 0.02$ & 2.8$\pm 0.01$($-0.3$) & $1.14_{-0.09}^{+0.12}$    & $3.0_{-0.4}^{+0.6}$    & $1.03_{-0.87}^{+0.87}$    &   $4.53_{-0.23}^{+0.23}$ & $0.61_{-0.01}^{+0.26}$    & ---        \\ 
6  &  XMMXCS J101703.6+390250.7 & $^a$ & A0963                    & 2.2$\pm 0.02$ & 2.7$\pm 0.02$         & $1.63_{-0.02}^{+0.02}$    & $6.6_{-0.1}^{+0.1}$    & $15.80_{-0.25}^{+0.29}$   &  $25.50_{-0.81}^{+0.81}$ & $1.44_{-0.03}^{+0.81}$    & ---        \\
7  &  XMMXCS J104044.4+395710.4 &      & A1068                    & 2.3$\pm 0.02$ & 3.1$\pm 0.03$         & $1.217_{-0.006}^{+0.006}$ & $3.54_{-0.03}^{+0.03}$ & $8.39_{-0.16}^{+0.17}$    &  $11.44_{-0.20}^{+0.20}$ & $2.00_{-0.03}^{+0.18}$    & ---        \\
8  &  XMMXCS J123024.3+111127.8 &      & BLOX J1230.6+1113.3 ID        & 2.1$\pm 0.18$ & 3.5$\pm 0.03$         & $0.54_{-0.01}^{+0.01}$    & $0.80_{-0.03}^{+0.03}$ & $0.018_{-0.002}^{+0.002}$ &   $1.61_{-0.06}^{+0.06}$ & $0.39_{-0.03}^{+0.06}$    & ---        \\ 
9  &  XMMXCS J123338.5+374114.9 &      & ---                           & 2.6$\pm 0.01$ & 3.2$\pm 0.02$         & $0.58_{-0.04}^{+0.03}$    & $0.9_{-0.1}^{+0.1}$    & $0.03_{-0.01}^{+0.01}$    &   $0.95_{-0.02}^{+0.02}$ & $0.142_{-0.003}^{+0.023}$ & ---        \\
10 &  XMMXCS J124425.9+164758.0 & $^b$ & WHL J124425.4+164756          & 0.5$\pm 0.20$ & 2.3$\pm 0.10$         & $0.63_{-0.06}^{+0.08}$    & $1.3_{-0.2}^{+0.3}$    & $0.06_{-0.03}^{+0.05}$    &   $5.85_{-0.38}^{+0.38}$ & $0.91_{-0.11}^{+0.38}$    & ---        \\
11 &  XMMXCS J130749.6+292549.2 &      & ZwCl 1305.4+2941              & 2.6$\pm 0.18$ & 3.1$\pm 0.03$         & $1.04_{-0.03}^{+0.03}$    & $3.2_{-0.2}^{+0.2}$    & $1.94_{-0.11}^{+0.10}$    &  $12.83_{-0.40}^{+0.40}$ & $1.12_{-0.03}^{+0.23}$    & ---        \\
12 &  XMMXCS J131145.1+220206.0 &      & MaxBCG J197.94248+22.02702    & 2.1$\pm 0.06$ & 2.7$\pm 0.06$         & $1.16_{-0.06}^{+0.06}$    & $3.4_{-0.2}^{+0.2}$    & $1.57_{-0.39}^{+0.69}$    &  $10.97_{-0.30}^{+0.29}$ & $1.39_{-0.03}^{+0.28}$    & ---        \\
13 &  XMMXCS J134825.6+580015.8 &      & ---                           & 2.0$\pm 0.02$ & 2.6$\pm 0.02$         & $0.78_{-0.05}^{+0.07}$    & $1.6_{-0.2}^{+0.3}$    & $0.08_{-0.04}^{+0.03}$    &   $3.55_{-0.08}^{+0.08}$ & $0.37_{-0.01}^{+0.08}$    & ---        \\ 
14 &  XMMXCS J141627.7+231525.9 & $^a$ & ZwCl 1413.9+2330              & 1.8$\pm 0.02$ & 2.9$\pm 0.02$         & $1.25_{-0.06}^{+0.06}$    & $3.7_{-0.2}^{+0.2}$    & $1.42_{-0.51}^{+1.10}$    &   $7.99_{-0.14}^{+0.14}$ & $1.08_{-0.02}^{+0.11}$    & 2,3,4,5,6,7\\
15 &  XMMXCS J141657.5+231239.2 &      & ---                           & 2.8$\pm 0.02$ & 3.1$\pm 0.01$         & $0.56_{-0.03}^{+0.04}$    & $0.9_{-0.1}^{+0.1}$    & $0.017_{-0.004}^{+0.005}$ &   $1.51_{-0.04}^{+0.04}$ & $0.171_{-0.004}^{+0.035}$ & ---        \\ 
16 &  XMMXCS J160129.8+083856.3 &      & ---                           & 2.4$\pm 0.03$ & 3.1$\pm 0.03$($-0.3$) & $0.77_{-0.04}^{+0.05}$    & $1.7_{-0.2}^{+0.2}$    & $0.84_{-0.35}^{+0.35}$    &   $2.77_{-0.09}^{+0.09}$ & $0.45_{-0.07}^{+0.27}$    & ---        \\ 
17 &  XMMXCS J172010.0+263724.7 & $^a$ & SDSS-C4 3072                  & 1.9$\pm 0.03$ & 2.4$\pm 0.03$         & $1.54_{-0.01}^{+0.01}$    & $5.53_{-0.04}^{+0.04}$ & $21.58_{-7.85}^{+23.45}$  &  $18.64_{-1.26}^{+1.26}$ & $1.94_{-0.04}^{+1.26}$    & 8          \\
\hline
\end{tabular}
\end{center}
\end{table*}
\clearpage
\end{landscape}

\appendix

\section{The robustness of the fossil system definition}\label{robust}

The most commonly used definition of an FS was first set out in
\citet[][the exact definition is given in the
introduction]{jones03}. For a system to be classified as an FS in this
study it must have $L_{X,\mathrm{bol}}\gtrsim 5\times
10^{41}\,h_{70}^{-2}$ erg s$^{-1}$ and a magnitude gap of 2.5 in the
$r$ band between the brightest and the fourth brightest galaxies
located within half the virial radius, which we denote as $\Delta
m_{14}$. As mentioned in the introduction, one of the main reasons why
there are so few confirmed FSs in the literature is the lack of high
quality X-ray data. Low S/N X-ray data make it extremely difficult to
not only detect extended sources but to also estimate their
luminosity. It is for this reason that many samples are defined as
`optical fossil' samples, i.e., samples that are only known to satisfy
the optical criterion.

The source detection algorithm used in the XCS, the XCS Automated
Pipeline Algorithm ({\sc Xapa}), has made this project possible. Many
difficulties can arise when trying to detect and measure X-ray
sources, which usually have low counts: a point-spread function and
sensitivity that varies over the instrument's field-of-view;
deblending of point and extended sources; and background determination
to name but a few (LD11, M11). However, once in possession of accurate
$L_X$s determining whether or not a system satisfies the X-ray
criterion is a straight forward matter.

Determining whether or not the optical criterion is satisfied,
however, is a more complicated matter. To be able to determine the
magnitude gap of a system requires an estimate of the virial radius
and redshift information for each of the galaxies, which is required
to determine whether a galaxy is a member of the system or not. Many
FS studies start with the \citeauthor{jones03} definition and then
modify it in various ways depending on circumstances. For example, if
no estimate of the virial radius is available then a fixed aperture
may be used.

Studies have been performed on the effect the adopted definition of
the magnitude gap has on the properties of FS
\citep[e.g.,][]{dariush10} but to date no study has been done on the
robustness of the FS definition. In this appendix we test the
robustness our FS classifications by varying some of the parameters
used during the search for FSs, e.g., the size of the redshift
cuts. We also investigate how the results of this study change when we
(1) use different radii to define the magnitude gap of our systems and
(2) use the \citeauthor{jones03} magnitude gap instead of the
\citeauthor{dariush10}.

In this study we relied on the redshift data provided by the SDSS to
determine membership. Specifically, any galaxy with a spectroscopic
redshift less than $\Delta cz=2000$ km s$^{-1}$ or a photometric
redshift less than $\Delta z_\mathrm{phot}=0.1$ away from the
potential FG was considered a system member. Given that the average
photometric redshift error for our sample was 0.04, this cut is quite
a generous and errs on the side of caution. The magnitude gap was then
calculated based on the galaxies that populated the red sequence of
the systems CMD, within $\pm 0.2$ mag in color of the potential FG.

To test the robustness of our classifications we varied some of these
values to see how that would affect the classifications. We varied the
spectroscopic-redshift cut in steps of 250 km s$^{-1}$ up to 2000 km
s$^{-1}$ and the search radius in steps of $0.25R_{200}$ from
$0.25R_{200}$ to $R_{200}$. We also considered two values for the
color cut, 0.15 and 0.2 mag, and the photometric redshift cut, $\Delta
z=0.1$ and 0.2.   

We find that the FS definition that we use here is robust to both the
spectroscopic-redshift and photometric-redshift cuts, and the
color-cut. The only quantity that it depends on is the size of the
search radius. In all cases, reducing the search radius from $R_{200}$
to $0.25R_{200}$ increases the magnitude gap, in some cases doubling
it. Our sample is inherently robust to variations in the radius, since
for many (more than half) of our systems we could have increased the
search radius well beyond $0.5R_{200}$ and the system would have still
been classified as an FS. We also examined the effects on our measured
magnitude gaps (and therefore FS classification) of the errors on
$R_{200}$. Systems whose gaps change when varying $R_{200}$ within the
errors are noted in the individual notes. These changes in $R_{200}$
would not reduce our measured $L_X$s below that required to define an
FS.

\begin{figure}
\includegraphics[width=1.0\textwidth]{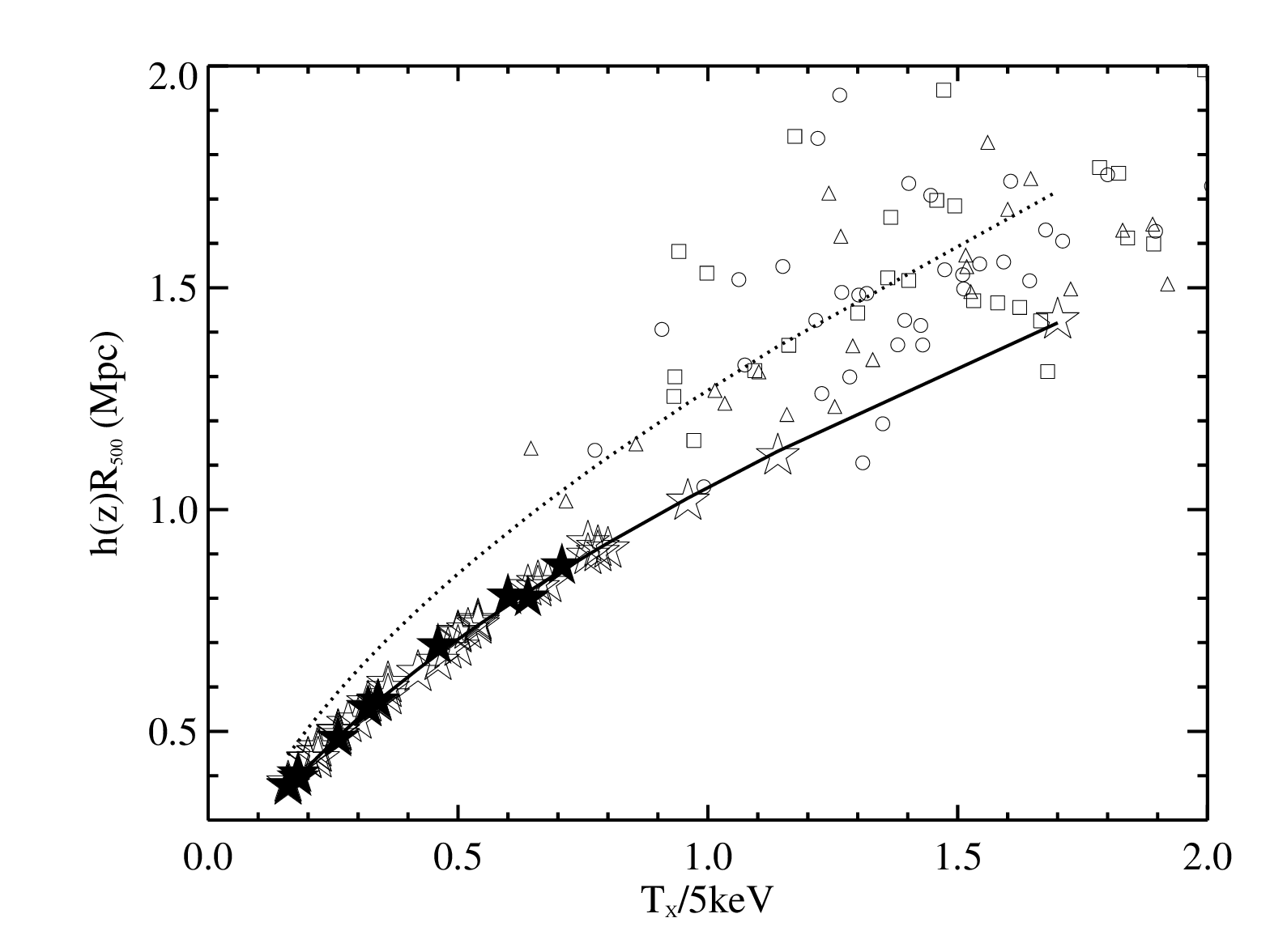}
\caption{Comparison of the $R_{500}$--$T_X$ relation from
  \citet[][solid line]{arnaud05} to the relation determined from the
  $T_X$ and $R_{500}$ data from \citet[][dotted line]{mantz10}. The
  stars are the data used in this paper (open for the XCS clusters and
  closed for the FS). Open circles, triangles, and squares are the
  BCS, REFLEX, MACS clusters respectively, where we show the core
  excised temperatures and radii from \citet{mantz10}.}
\label{mantz}
\end{figure}

As mentioned in Section \ref{fs sample}, our $R_{200}$s are estimated
from the $T_X$--$R_{200}$ relation found in \citet{arnaud05}, which,
although not as good as actually measuring $R_{200}$, is perfectly
reasonable. There are, however, other methods of estimating $R_{200}$
when a direct measurement is not possible. Here, we investigate the
effects on our results that the choice of $R_{200}$ has by repeating
our analysis using four different estimates of $R_{200}$: a fixed 0.35
Mpc aperture, a fixed 0.5 Mpc aperture, an $R_{200}$ estimated from
the $N_{200}$--$R_{200}$ relation of \citet{johnston07}, and an
$R_{200}$ estimated from the $T_X$--$R_{200}$ derived from the data in
\citet{mantz10}. In Figure \ref{mantz}, we compare the $R_{500}$--$T_X$
relation from \citet[][solid line]{arnaud05} to the relation
determined from the $T_X$ and $R_{500}$ data from \citet[][dotted
line]{mantz10}. The stars are the data used in this paper (open for
the XCS clusters and closed for the FS). Open circles, triangles, and
squares are the BCS, REFLEX, MACS clusters, respectively, where we show
the core excised temperatures and radii from \citet{mantz10}. The fit
to the \citeauthor{mantz10} data produces a slightly larger $R_{500}$
than the \citeauthor{arnaud05} fit ($\sim 20\%$ at $T_X=1.0$ keV). We
obtain the \citeauthor{mantz10} $R_{200}$ estimates by scaling the
$R_{200}$ estimates from \citeauthor{arnaud05} by the ratio of the
\citeauthor{mantz10} and \citeauthor{arnaud05} $R_{500}$ estimates. We
find that this small shift in the $R_{200}$--$T_X$ scaling relations
only weakly affects our sample definition and induces only a small
amount of scatter into relevant figures. Similar effects are found
when using the fixed 0.35 Mpc apertures.

The other two estimators, the fixed 0.5 Mpc aperture and the $N_{200}$
estimate, produce larger effects. Compared to our sample of 17 FSs,
the former finds only 13 and the latter finds 21, i.e., the number of
FSs found at these extremes is $\pm 25\%$ of that found in this
study. The reason for the decrease in the case of the fixed 0.5 Mpc
aperture is due to the fact that almost half of our sample have
$0.5R_{200}$ less than 0.5 Mpc. The search radius is therefore larger
than necessary and includes more galaxies, which results in a smaller
magnitude gap. The reason for the increase in the case of the
$R_{200}$ estimates from the $N_{200}$--$R_{200}$ relation is most
likely due to the fact that true FSs, if they are the end-products of
galaxy merging in groups/clusters, should have an $R_{200}$ that is
too large for their richness compared to that of normal
groups/clusters. Therefore if such a relation, based on normal
groups/clusters, is used then the $R_{200}$ estimates (and the search
radii) will be too small resulting in larger magnitude gaps. Such a
shift in sample size will have a significant impact on the results of
an FS number density studies.

\begin{figure}
\plottwo{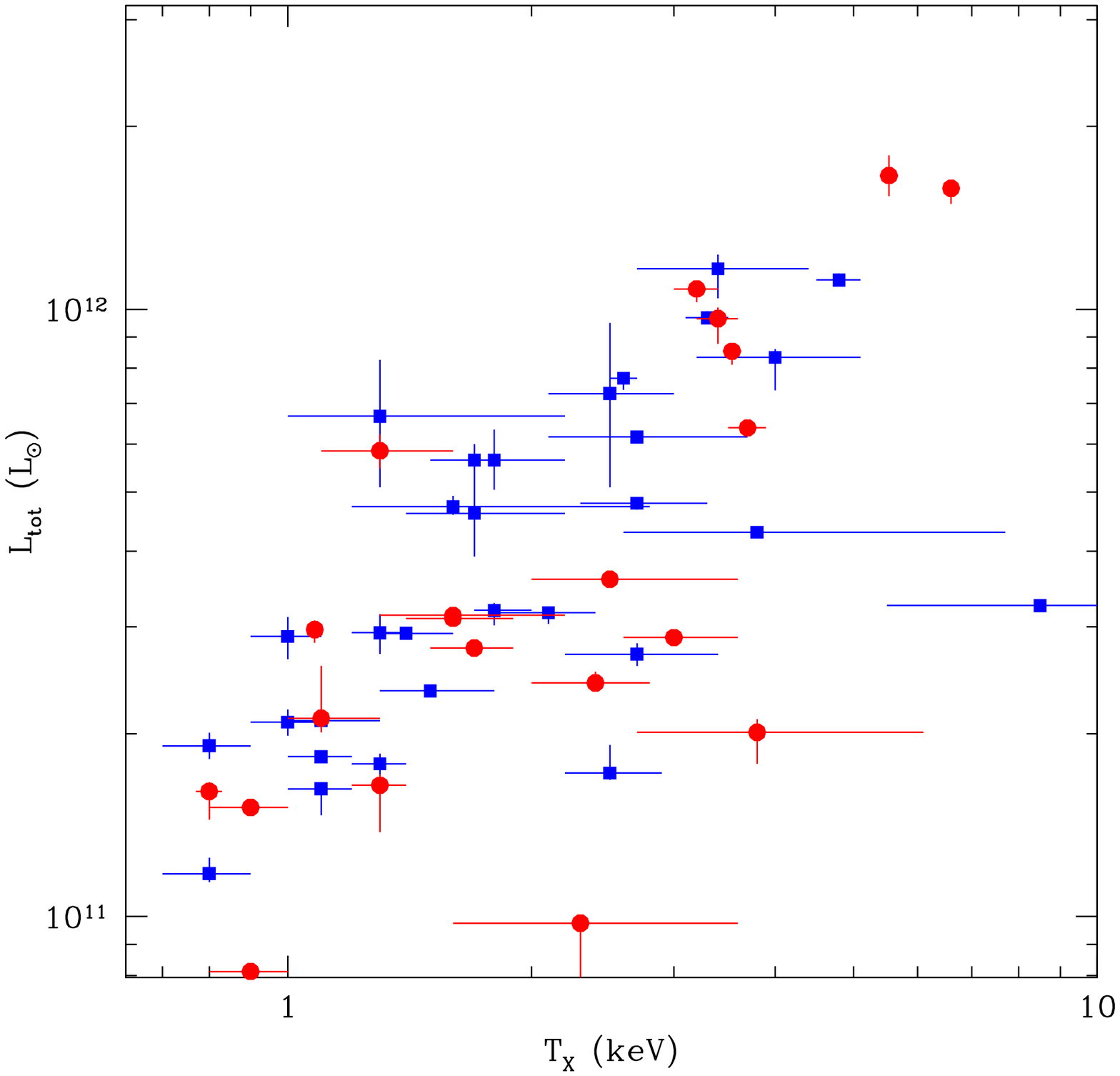}{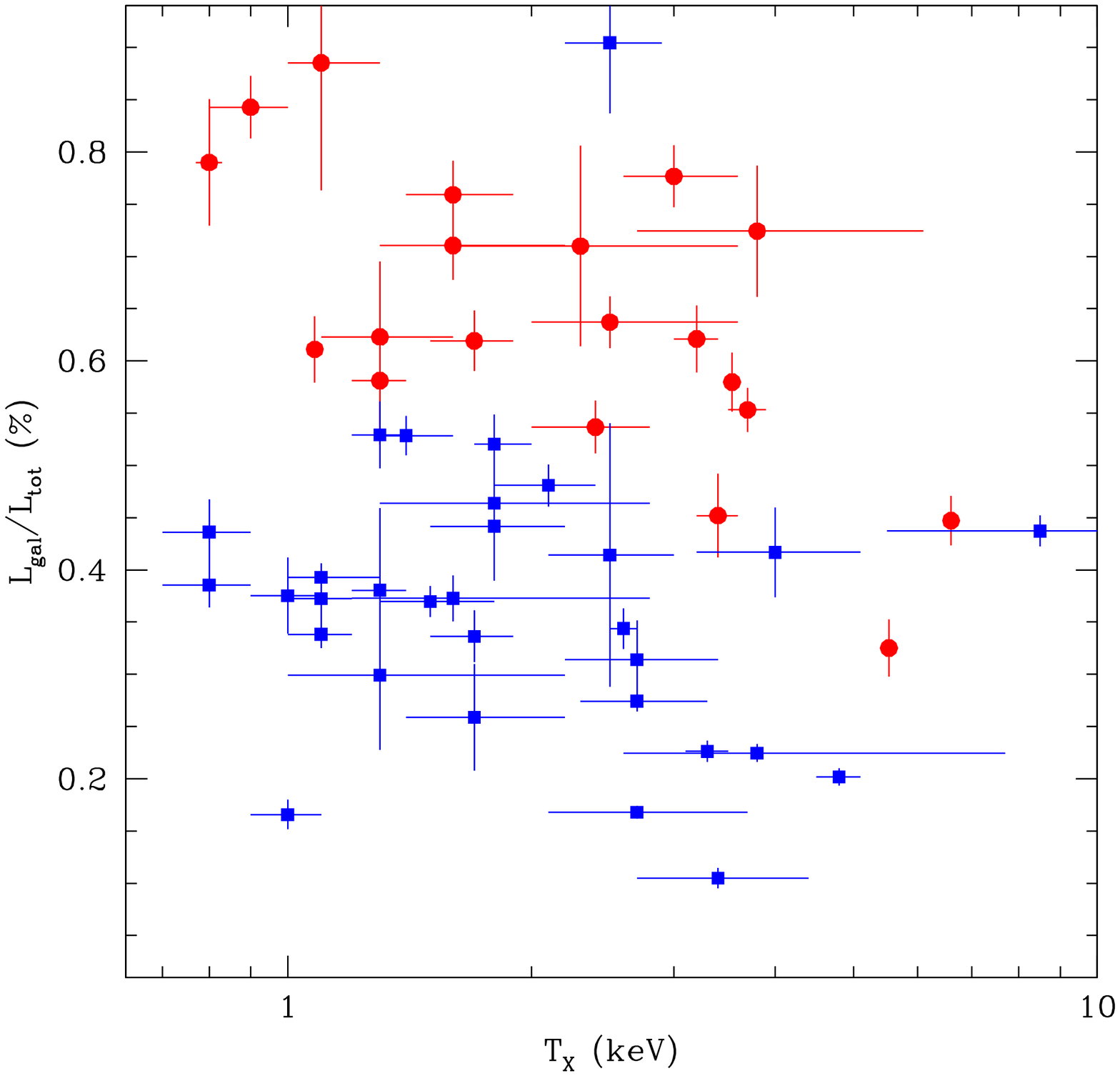}
\caption{Optical luminosity of the galaxy system within $0.5R_{200}$
  as a function of $T_X$ (left) and the fraction of the optical
  luminosity in the system contained within the dominant galaxy as a
  function of $T_X$ (right) these plots are the same as Figures
  \ref{tx_ltot} and \ref{frac_opt_tx} but the $R_{200}$ used to define
  the fossil systems and calculate $L_\mathrm{tot}$ is estimated from
  the $N_{200}$--$R_{200}$ relation of \citet{johnston07} and they
  exhibit larger scatter.}
\label{n200}
\end{figure}

These two estimators also introduce scatter into many of the tight
relations that FSs were found to follow. This can be seen by comparing
the left panel of Figure \ref{n200} to Figure \ref{tx_ltot} and the
right panel to Figure \ref{frac_opt_tx}. Figure \ref{n200} shows the
optical luminosity of the galaxy system within $0.5R_{200}$ as a
function of $T_X$ (left panel) and the fraction of that light
contained within the dominant galaxy as a function of $T_X$
(right). Note that using $N_{200}$ to estimate $R_{200}$, as we do
here, changes the radius within which we calculate $L_\mathrm{tot}$
(and therefore changes $L_\mathrm{gal}$/$L_\mathrm{tot}$), but should
not affect $T_X$ because at these radii the $T_X$ profile should be
flat. In the case of the left plot the scatter in this relation (for
both FSs and XCS clusters) increases from 0.2 dex to 0.25 dex when
using $N_{200}$ to estimate $R_{200}$. Note that by using $N_{200}$,
we are also entirely removing any induced correlation between
$L_\mathrm{tot}$ and $T_{X}$ in Figure \ref{tx_ltot}, since the radius
we used to calculate $L_\mathrm{tot}$ is independent of $T_{X}$ (see
Section \ref{assem}). So the small increase in scatter in Figure A2
compared to Figure 6 could be from removing this correlation. In the
right plot, apart from the increased scatter, we also see a boosting
of the fraction at high $T_X$ because these systems are richer and
therefore a small decrease in radius can reduce the number of galaxies
used to calculate the magnitude gap more than for a low-$T_X$
system. This results in a smaller $L_\mathrm{tot}$ and a
correspondingly larger fraction of the system light being contained in
these dominant galaxies.

If we use the \citet{jones03} magnitude gap rather than the
\citet{dariush10} magnitude gap we find that less systems are
classified as an FS ($\sim 30\%$ less) and we lose some statistical
power. There are no changes to the overall results of this study but
we find that the tight trends we find for FSs (such as in Figure
\ref{frac_opt_tx} ) are now more sparsely populated as some of our FSs
have been re-classified as non-FSs. Interestingly, in the study of the
magnitude gap made by \citet{dariush10} it was found that the
\citeauthor{jones03} magnitude gap was better at finding high-mass
halos. We expected to lose mostly low-mass FSs when making this change
but actually found that it was the intermediate mass FSs that fell out
of the sample.

In summary, we find that our definition of an FS is robust to changes
to the cuts used in redshift and color space but that it is sensitive
to changes in $R_{200}$. We, therefore, tested various estimates of
$R_{200}$ and found that the number of systems classified as an FS can
vary by $\pm 25\%$. All our results hold for the samples defined using
various other estimates of $R_{200}$ but there is increased scatter in
the correlations. Our results are also insensitive to our choice of
magnitude gap definition. It is, therefore, important to have accurate
estimates of $R_{200}$ when defining FS samples, especially if
estimating their number density.

\section{Notes on individual systems}\label{notes}

In this appendix we provide notes on each of the 17 FSs, along with
SDSS images with XCS contours overlaid and CMDs. In the CMDs, green
triangles are galaxies that, based on their SDSS spectroscopic
redshift, are considered system members (i.e., $\Delta cz \le 2000$ km
s$^{-1}$) while red triangles are those that are not. Black circles
are galaxies that, based on their SDSS photometric redshift, are
considered system members (i.e., $\Delta z_\mathrm{phot} \le
0.1$). The blue diamonds mark the galaxies used to calculate $\Delta
m_{12}$ and $\Delta m_{14}$. The dotted lines are the color cuts
employed when calculating the magnitude gaps. Where possible velocity
dispersions have been calculated using the bi-weight estimator of
\citet{beers90} and for those systems for which it was not possible to
estimate a velocity dispersion values were obtained from the
literature (if available).

\pagebreak

\subsection{XMMXCS J015315.0+010214.2}

\begin{figure}[h]
\plottwo{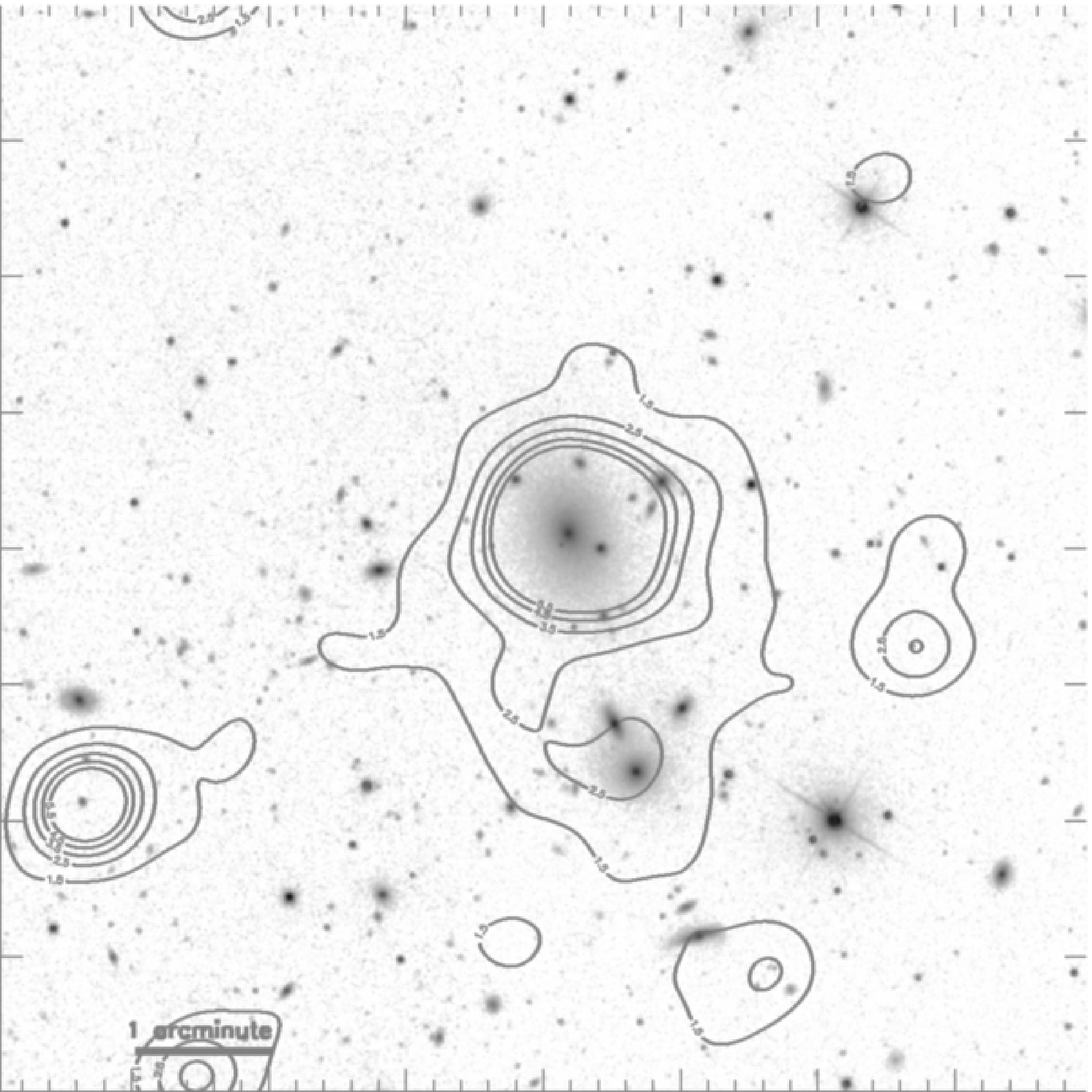}{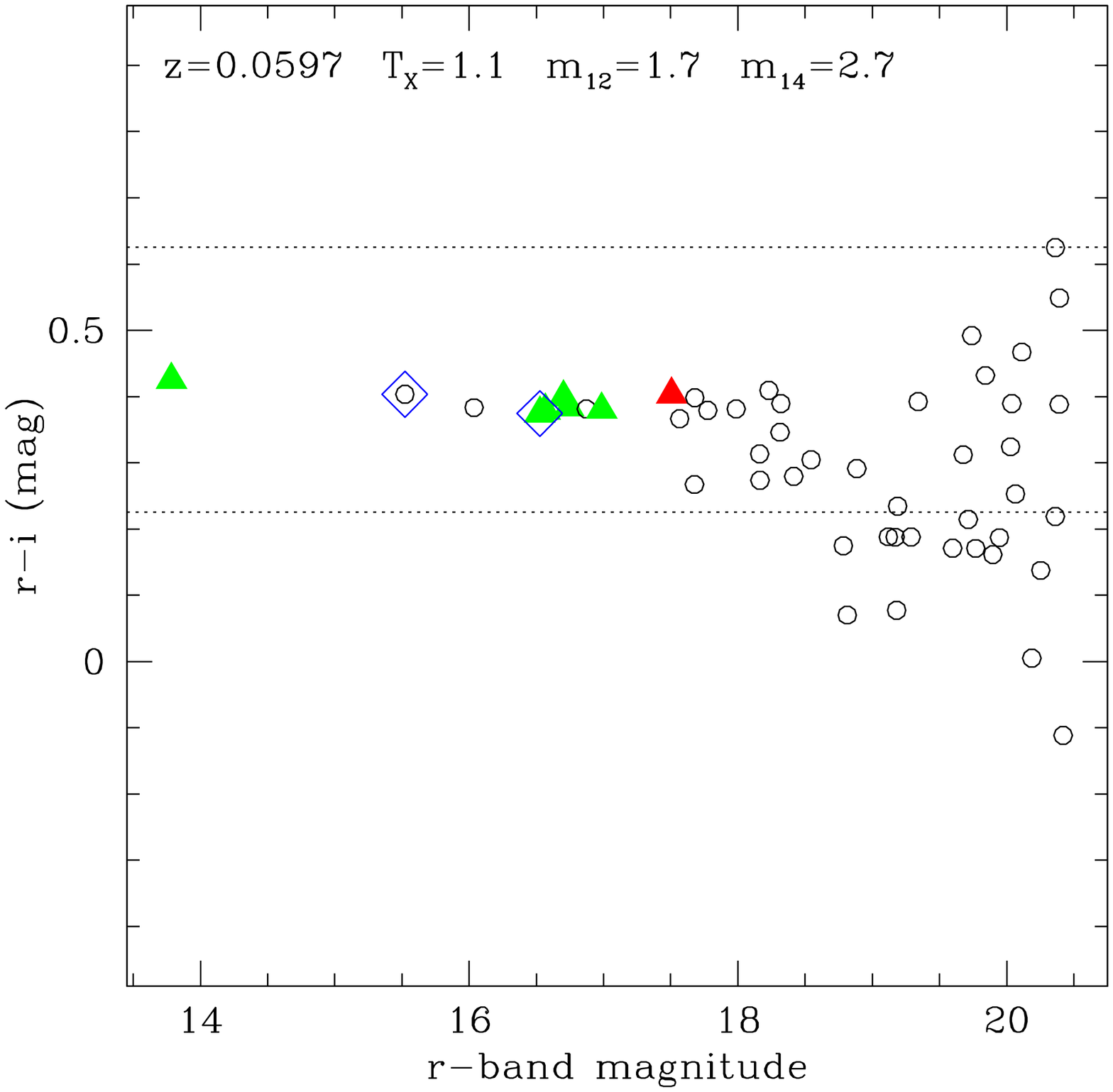}
\caption{$z$=0.0597 system located at 01:53:15.0 +01:02:14.2.}
\label{j015315}
\end{figure}

This $z=0.0597$ system is located at 01:53:15.0 +01:02:14.2 (Figure
\ref{j015315}). The CMD shows an obvious RS, which all spectroscopic
members lie on, and the magnitude gap is 2.7 based on SDSS photometric
data, i.e., not all of the four brightest galaxies have a
spectroscopic redshift. However, the two galaxies that only have
photometric redshifts both lie on the RS as well. The X-ray
temperature of the system is $T_X=1.1$ keV and the X-ray emission peak
lies $\sim 8.6$ kpc from the FG. The system is located near the edge
of the SDSS footprint but at a distance of $\sim 1.4R_{200}$ both
$\Delta m_{14}$ and $L_\mathrm{tot}$ should be unaffected. The X-ray
source is extended with $R_{200}=0.66$ Mpc or $\sim 30R_{90}$
($R_{90}$ is the radius within which 90\% of the galaxy's light is
contained, in this case the FG). The system has a velocity dispersion
of 266 km s$^{-1}$ based on 18 galaxies.

\subsection{XMMXCS J030659.8+000824.9}\label{J0306}

\begin{figure}[h]
\plottwo{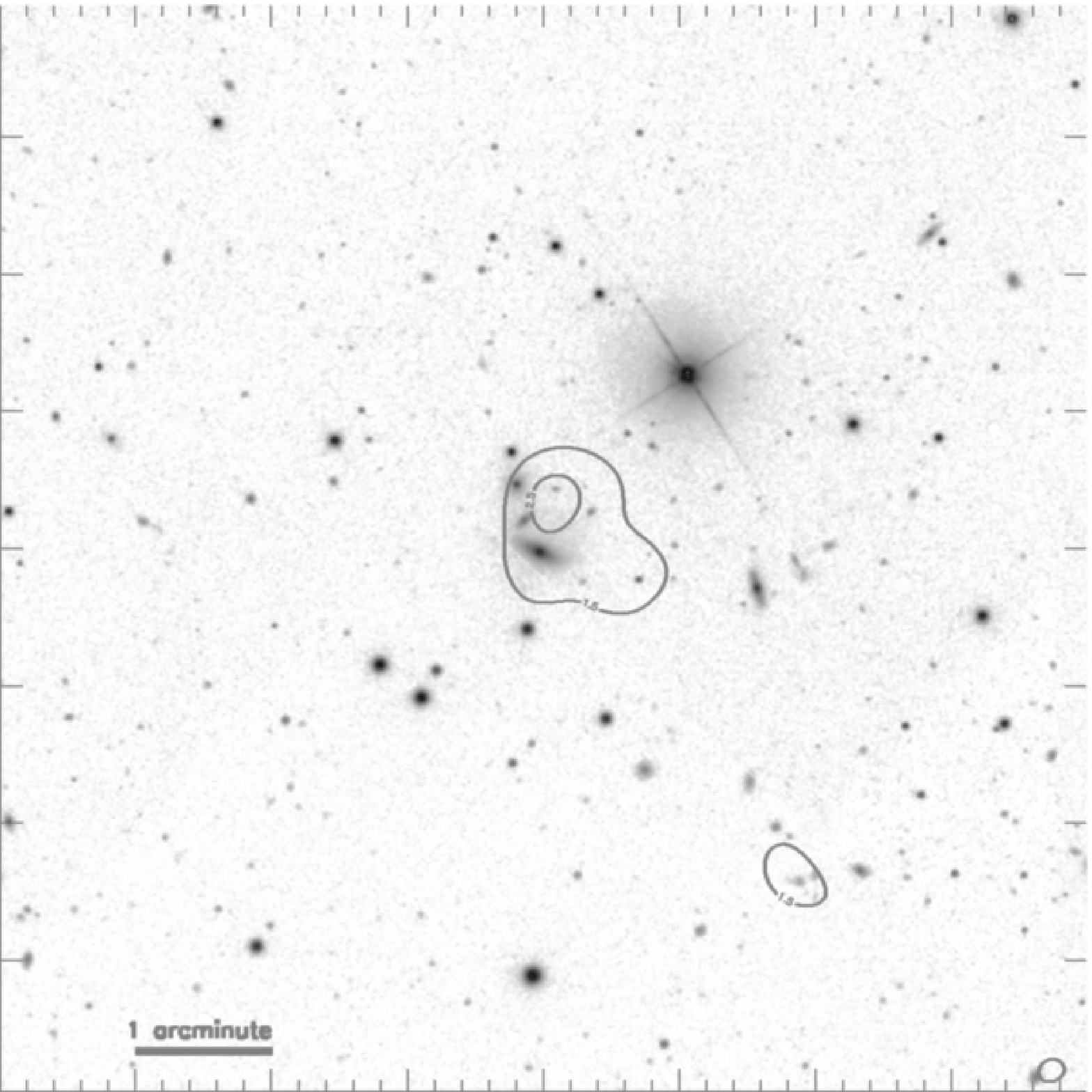}{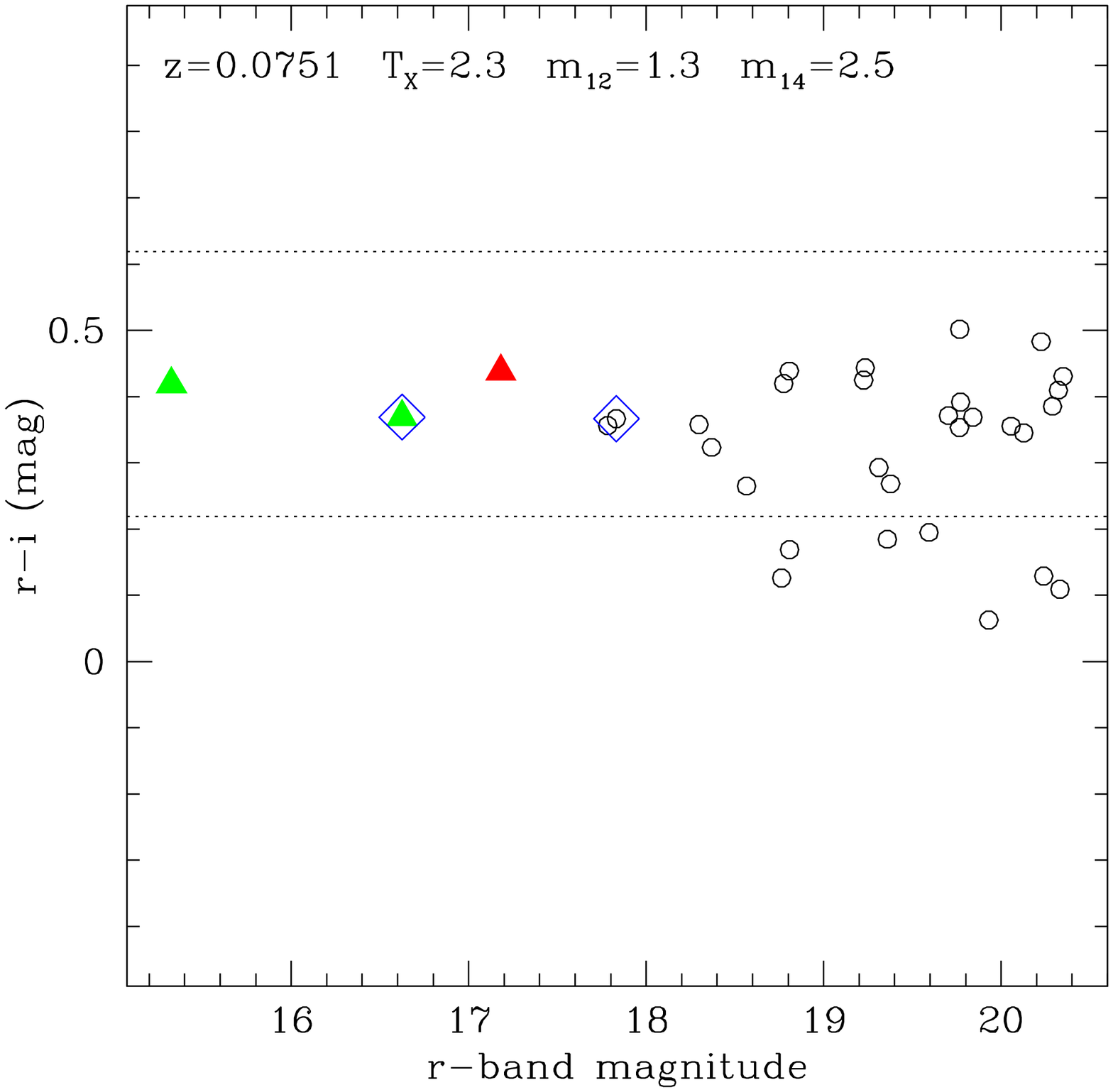}
\caption{$z=0.0751$ system located at 03:06:59.8 +00:08:24.9.}
\label{j030659}
\end{figure}

This $z=0.0751$ system is located at 03:06:59.8 +00:08:24.9 (Figure
\ref{j030659}). The CMD shows an RS and both spectroscopic members
lay on it. The magnitude gap is 2.5 based on SDSS photometric data and
the galaxy that was excluded has a spectroscopic redshift that is
$\sim 38,000$ km s$^{-1}$ away from the FG. Of the four galaxies used
to calculate $\Delta m_{14}$, the two that only have photometric
redshifts both lie on the RS as well. The X-ray emission has $T_X=2.3$
keV and the peak lies $\sim 26.2$ kpc from the FG. The system is
located near the edge of the SDSS footprint but at a distance of $\sim
7.0R_{200}$ both $\Delta m_{14}$ and $L_\mathrm{tot}$ are
unaffected. The X-ray source is extended with $R_{200}=1.01$ Mpc or
$\sim 90R_{90}$ (the largest of any system). We note that this system
is an outlier in many of the plots and despite satisfying all the
criteria necessary to be classified as an FS we acknowledge that this
classification is uncertain. The system has a velocity dispersion of
1082 km s$^{-1}$ based on 13 galaxies.

\subsection{XMMXCS J073422.2+265143.9}

\begin{figure}[h]
\plottwo{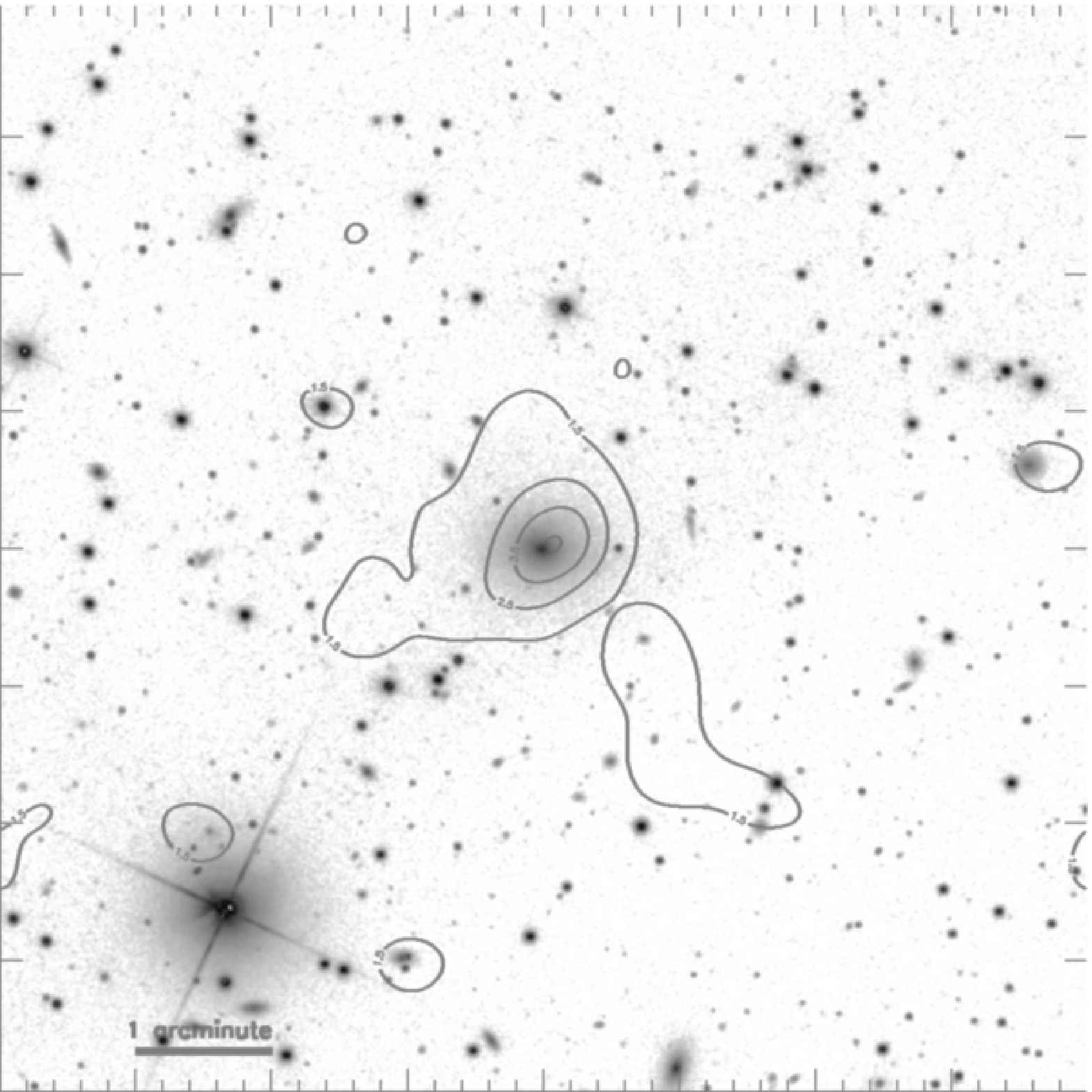}{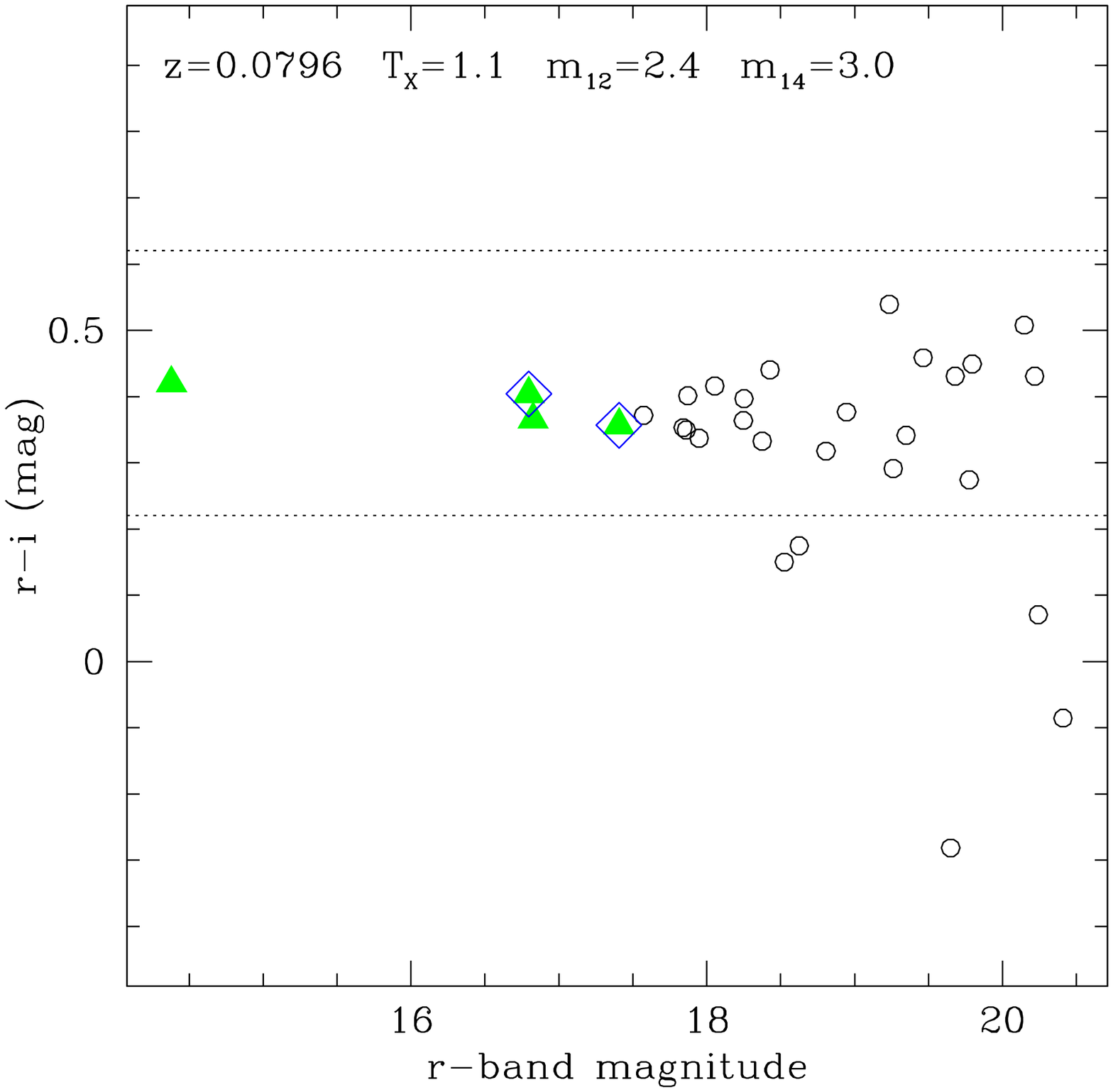}
\caption{$z=0.0796$ system located at 07:34:22.2 +26:51:43.9.}
\label{j073442}
\end{figure}

This $z=0.0796$ system is located at 07:34:22.2 +26:51:43.9 (Figure
\ref{j073442}. The CMD shows an RS and all spectroscopic members lay
on it. The magnitude gap is 3.0 based on SDSS spectroscopic data,
i.e., all of the four brightest galaxies have a spectroscopic
redshift. The X-ray emission has $T_X=1.1$ keV and the peak lies $\sim
1.5$ kpc from the FG. The system is located near the edge of the SDSS
footprint but at a distance of $\sim 10.0R_{200}$ both $\Delta m_{14}$
and $L_\mathrm{tot}$ are unaffected. The X-ray source is extended with
$R_{200}=0.67$ Mpc or $\sim 30R_{90}$. Reducing $R_{200}$ by its error
would increase the magnitude gap by 0.5. The system has a velocity
dispersion of 411 km s$^{-1}$ based on 18 galaxies. This system was
classified as a fossil in \citet{diaz-gimenez08} and was an
\textit{XMM} target in a program to study
FSs. \citeauthor{diaz-gimenez08} quote a virial radius of 1.6 Mpc
(twice as large as that found here) and a velocity dispersion of 551
km s$^{-1}$.

\pagebreak

\subsection{XMMXCS J083454.8+553420.9}

\begin{figure}[h]
\plottwo{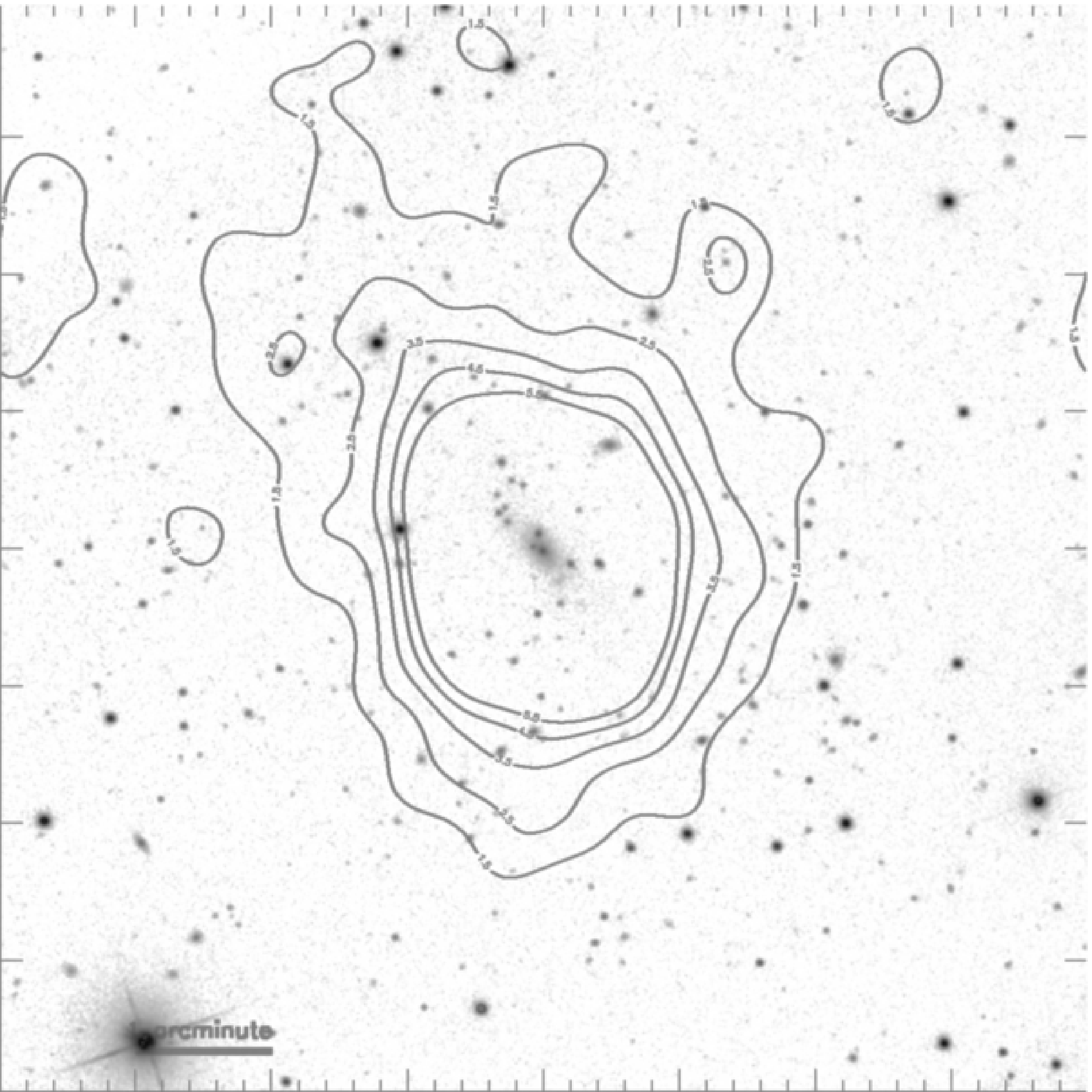}{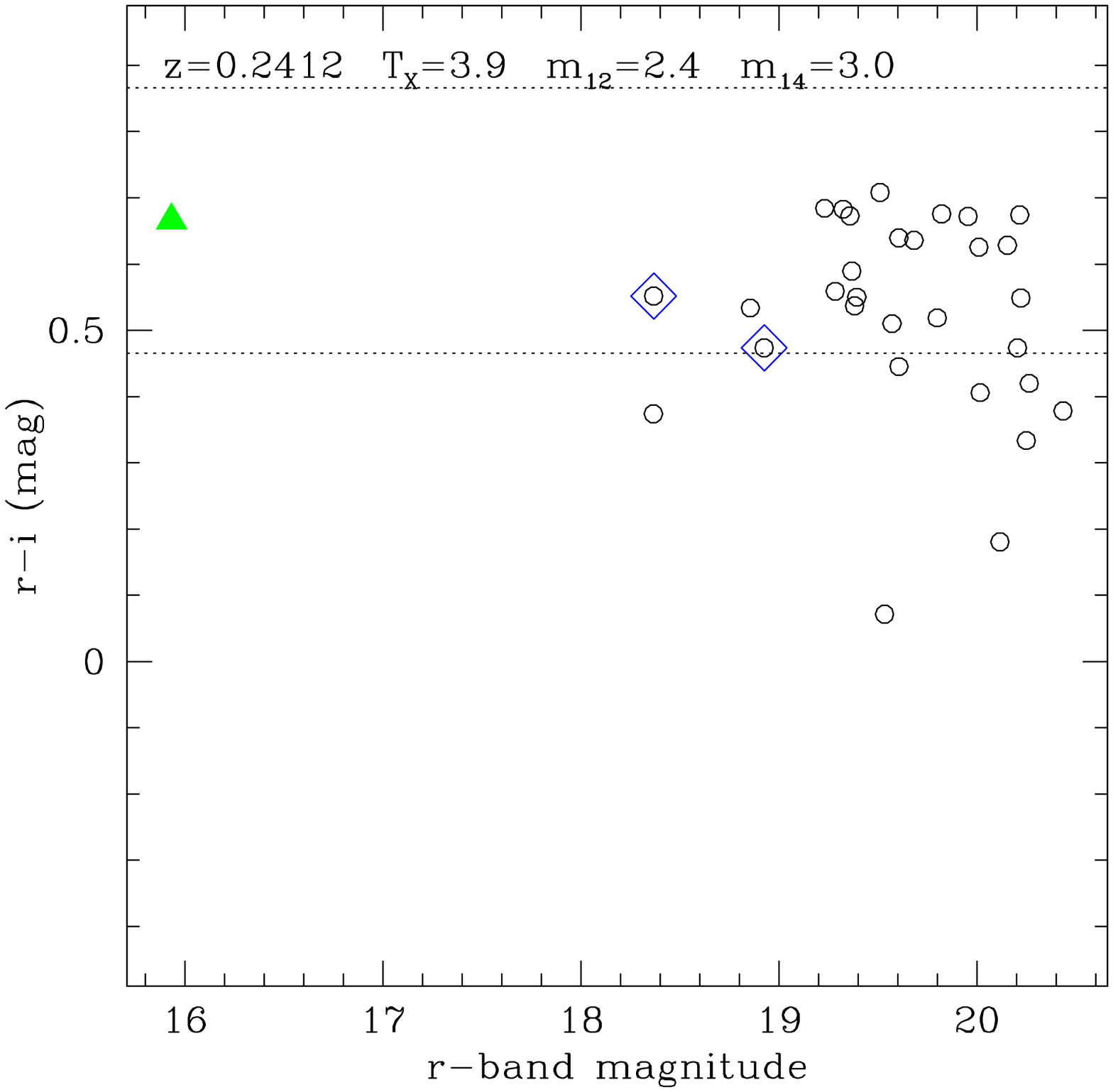}
\caption{$z=0.2412$ system located at 08:34:54.8 +55:34:20.9.}
\label{j083454}
\end{figure}

This $z=0.2412$ system is located at 08:34:54.8 +55:34:20.9 (Figure
\ref{j083454}). There is only the slightest hint of an RS in the
CMD. The magnitude gap is 3.0 based on SDSS photometric data. The
X-ray emission has $T_X=3.9$ keV and the peak lies $\sim 3.3$ kpc from
the FG. The X-ray source is extended with $R_{200}=1.17$ Mpc or $\sim
30R_{90}$.

\subsection{XMMXCS J092540.0+362711.1}

\begin{figure}[h]
\plottwo{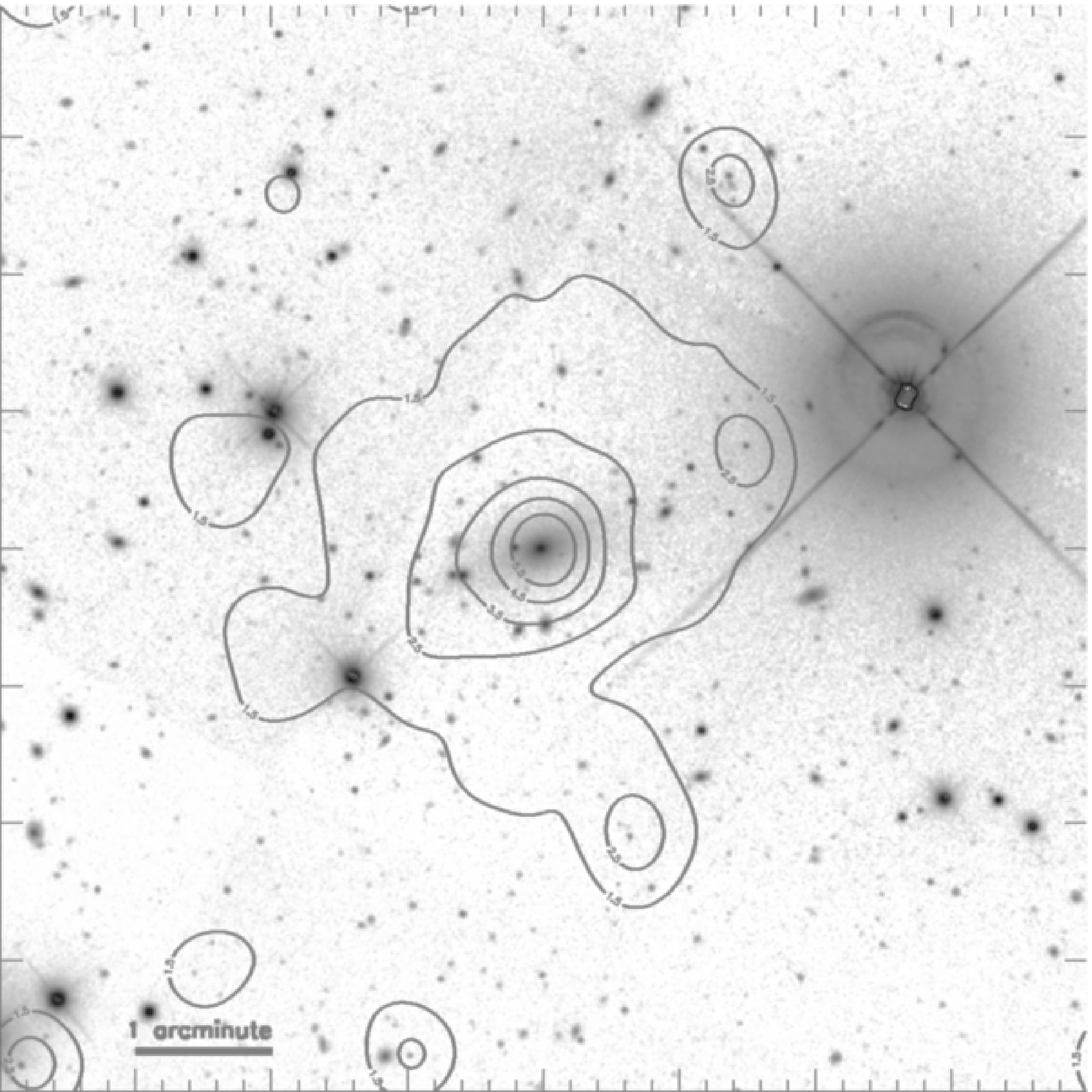}{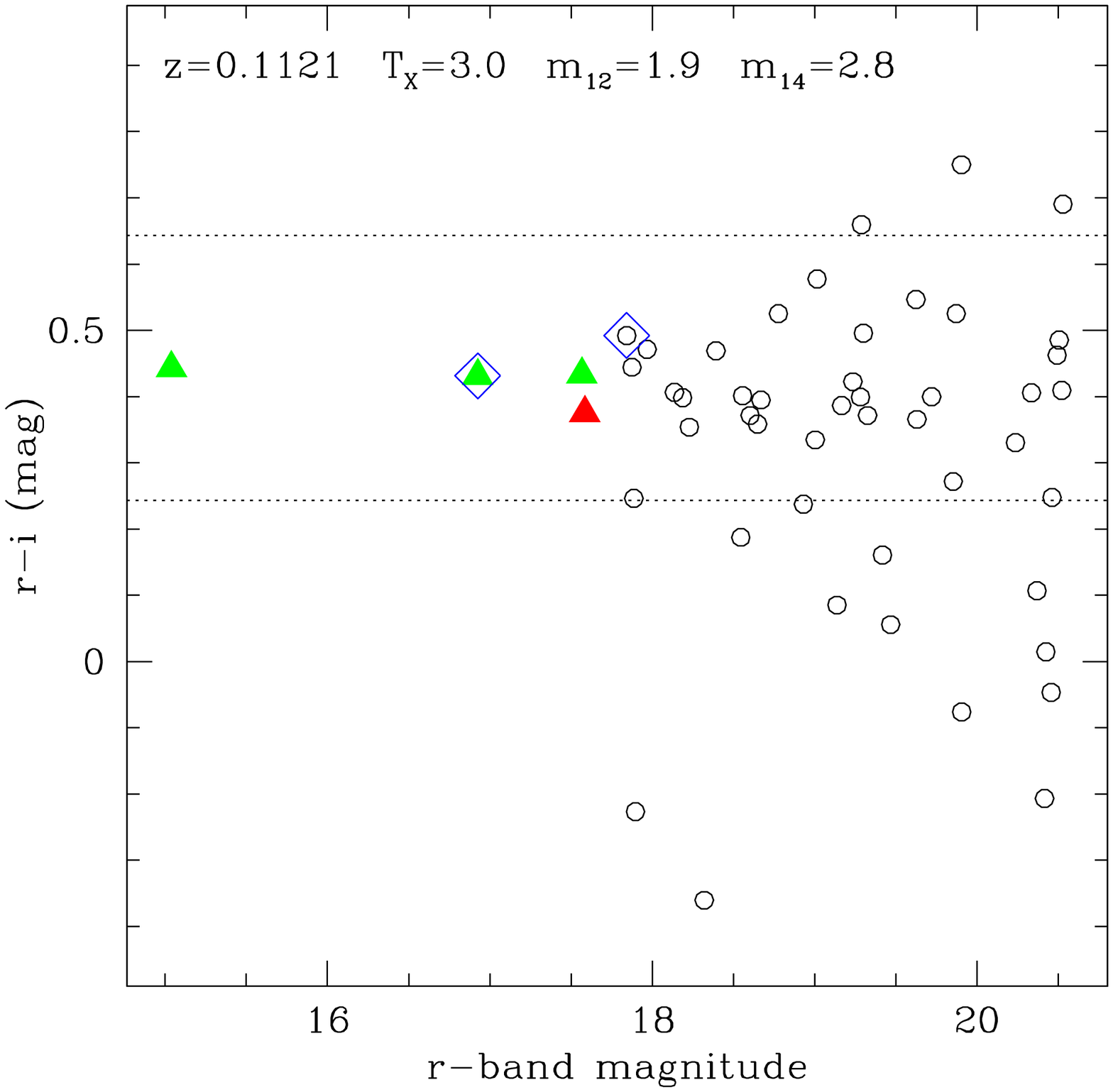}
\caption{$z=0.1121$ system located at 09:25:40.0 +36:27:11.1.}
\label{j092540}
\end{figure}

This $z=0.1121$ system is located at 09:25:40.0 +36:27:11.1 (Figure
\ref{j092540}). The CMD shows an RS and all spectroscopic members lay
on it. The magnitude gap is 2.8 based on SDSS photometric data. The
rejected galaxy lies $\sim7800$ km s$^{-1}$ away from the FG. The
X-ray emission has $T_X=3.0$ keV and the peak lies $\sim 26.1$ kpc
from the FG. The X-ray source is extended with $R_{200}=1.14$ Mpc or
$\sim 40R_{90}$. Increasing $R_{200}$ by its error would decrease the
magnitude gap by 0.3 but the system would still be classified as a
fossil. We note that there is a bright galaxy just outside
$0.5R_{200}$ that would change $\Delta m_{14}$ if it were
included. The system has a velocity dispersion of 435 km s$^{-1}$
based on 22 galaxies.

\pagebreak

\subsection{XMMXCS J101703.6+390250.7}

\begin{figure}[h]
\plottwo{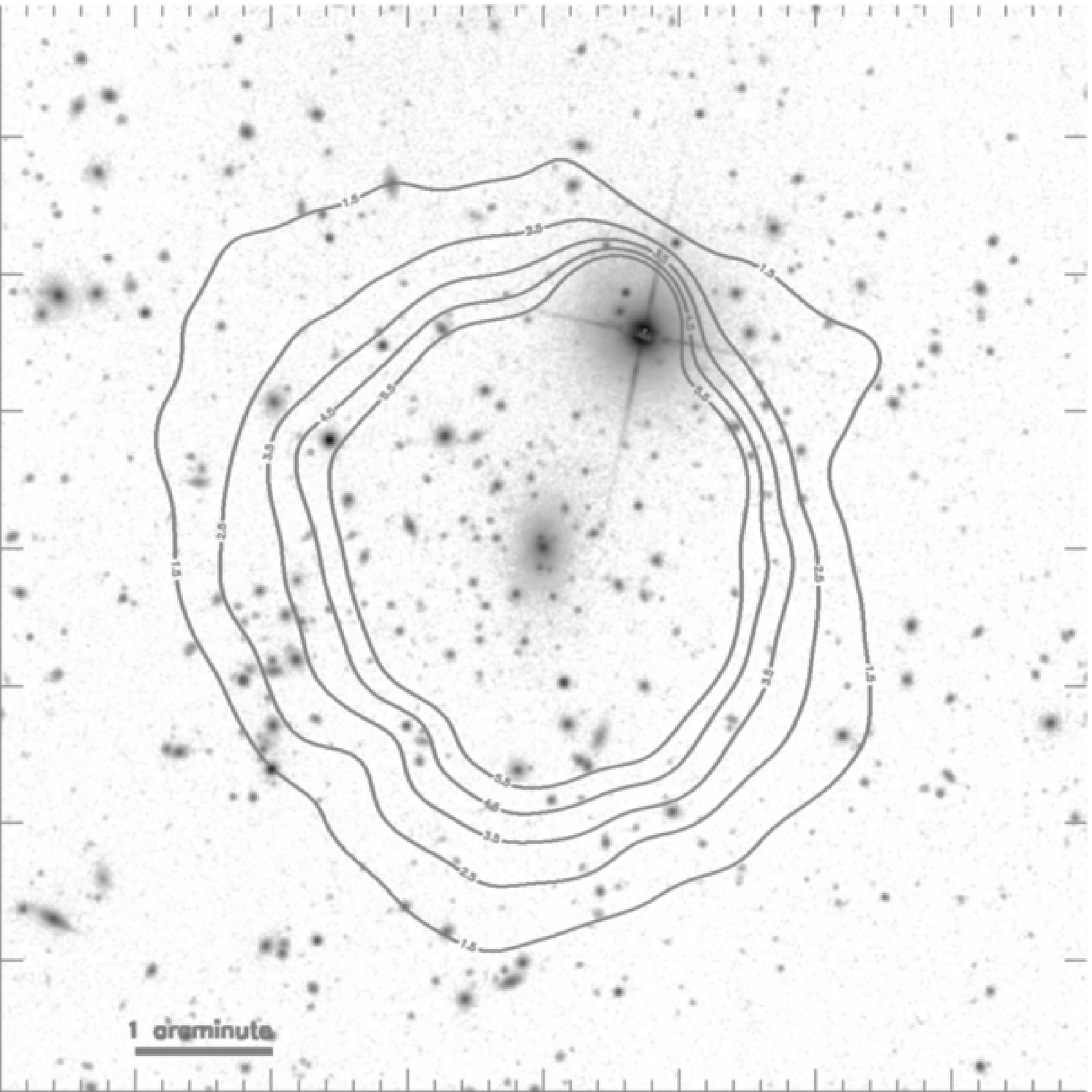}{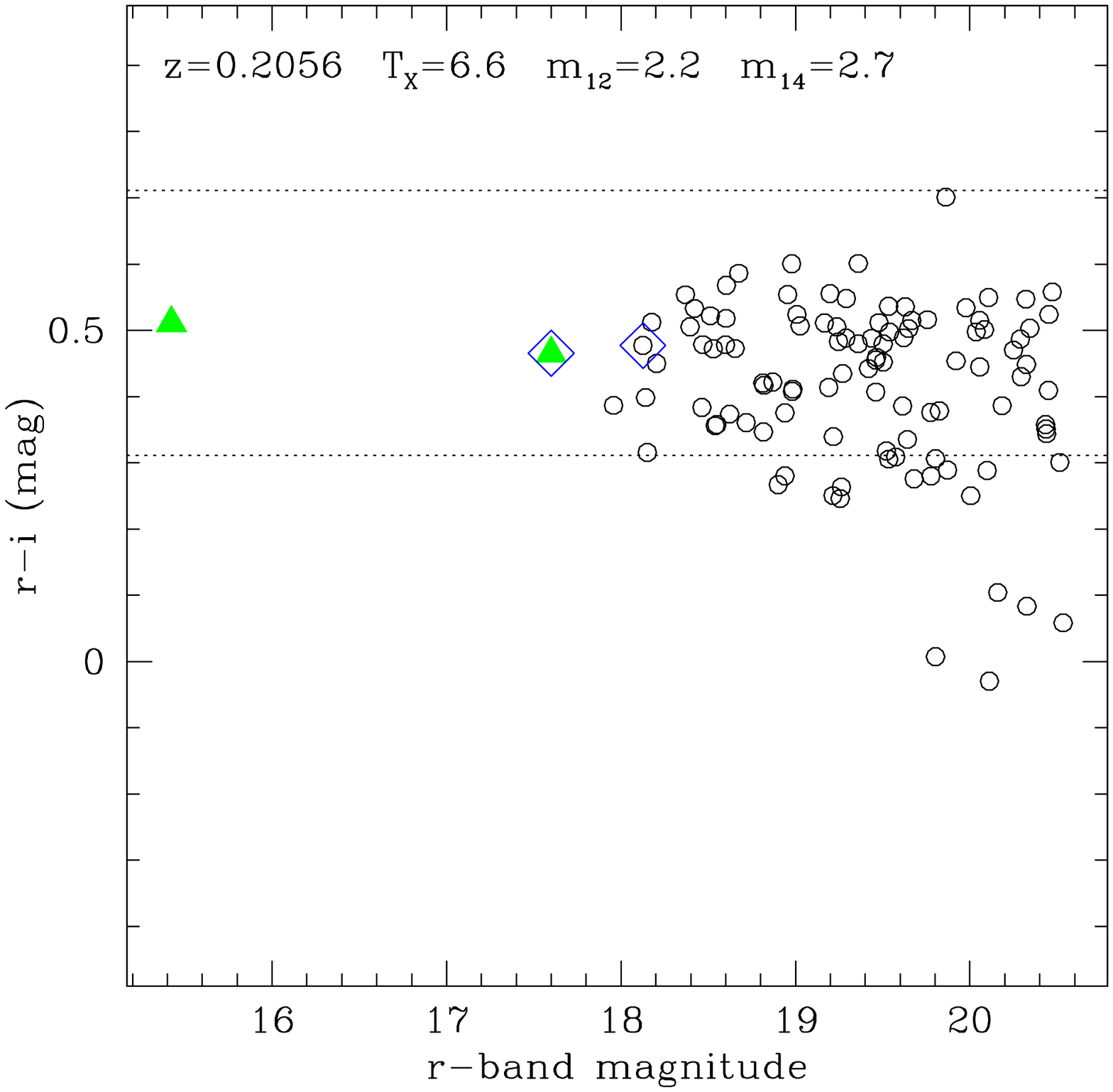}
\caption{$z=0.2056$ system located at 10:17:03.6 +39:02:50.7.}
\label{j101703}
\end{figure}

This $z=0.2056$ system is located at 10:17:03.6 +39:02:50.7 (Figure
\ref{j101703}). The CMD shows the hint of an RS and both spectroscopic
members lay on it. The magnitude gap is 2.7 based on SDSS photometric
data. The X-ray emission has $T_X=6.6$ keV (the highest of any system)
and the peak lies $\sim 4.6$ kpc from the FG. The X-ray source is
extended with $R_{200}=1.63$ Mpc (the largest of any system) or $\sim
30R_{90}$. We note that there is a bright galaxy just outside
$0.5R_{200}$ that would change $\Delta m_{14}$ if it were
included. The system has a velocity dispersion of 1013 km s$^{-1}$
based on 10 galaxies. This system is also known as A0963 and was
an \textit{XMM} target. From the literature, $L_X=6.1\times 10^{44}$
erg s$^{-1}$ \citep[][half of that found here]{soltan83} and
$\sigma=1350\pm 200$ km s$^{-1}$ \citep{lavery98}. It is an X-ray
lensing cluster that is unusually relaxed with $<5\%$ substructure
\citep{smith05}.

\subsection{XMMXCS J104044.4+395710.4}

\begin{figure}[h]
\plottwo{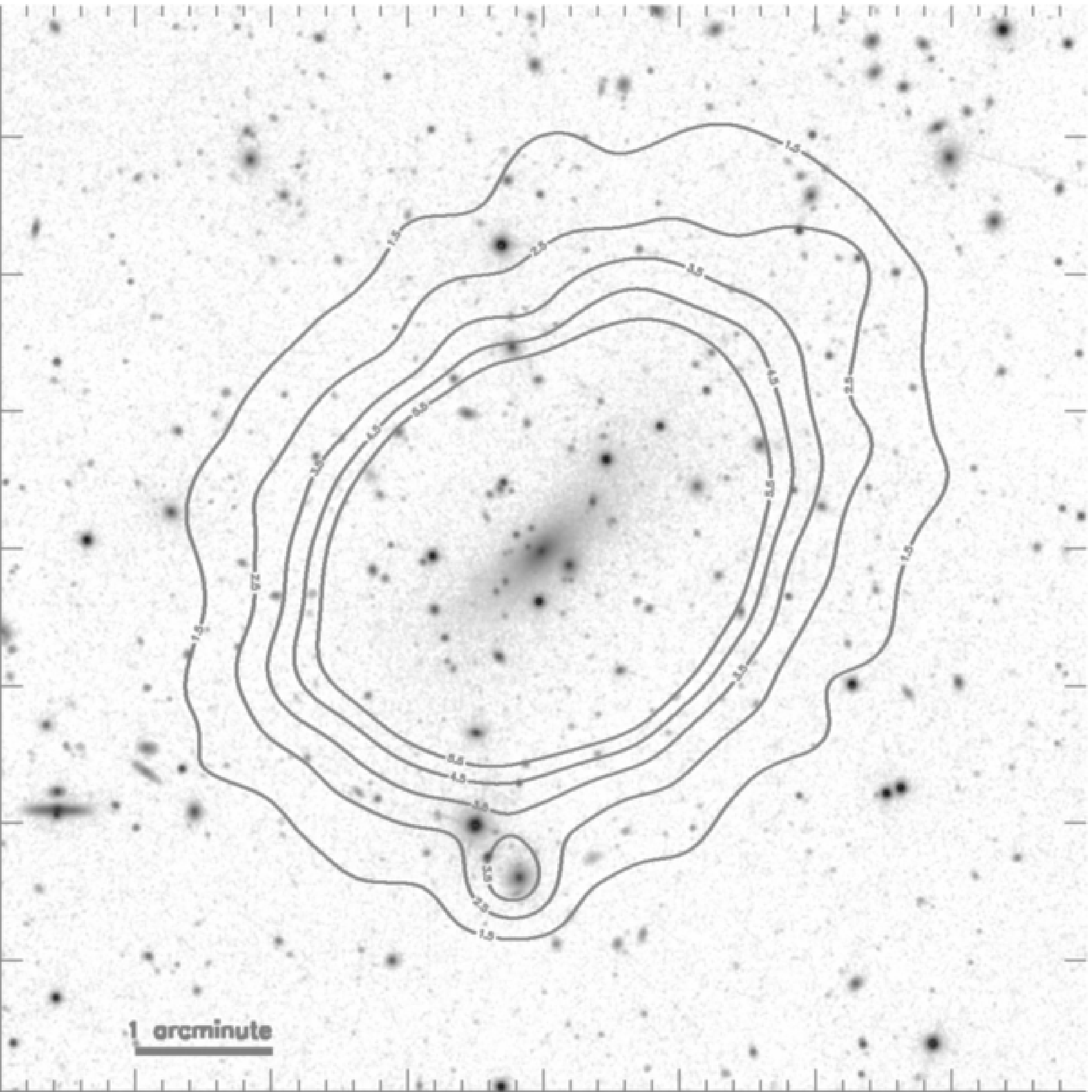}{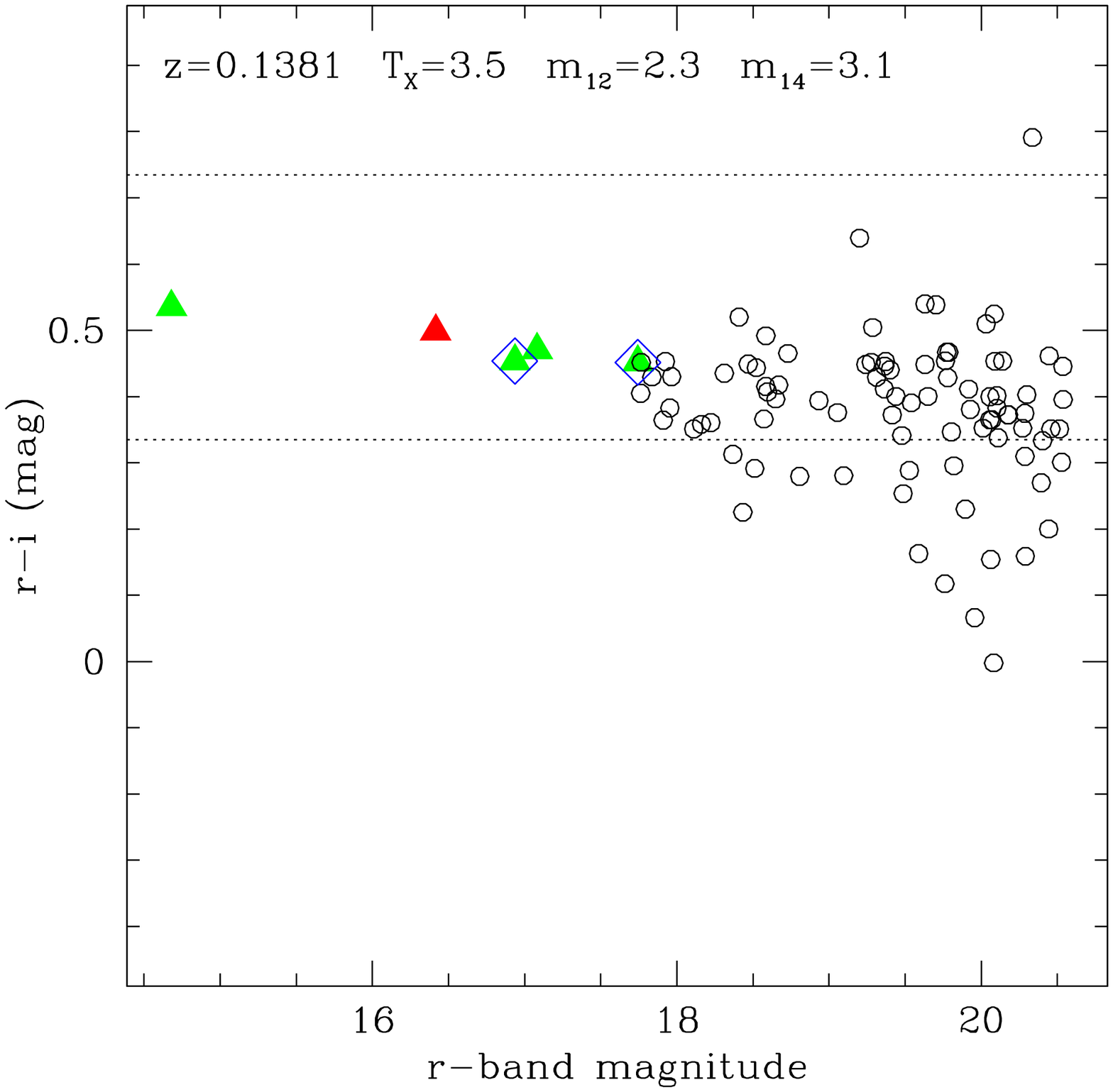}
\caption{$z=0.1381$ system located at 01:40:44.4 +39:57:10.4.}
\label{j104044}
\end{figure}

This $z=0.1381$ system is located at 01:40:44.4 +39:57:10.4 (Figure
\ref{j104044}). The CMD shows an obvious RS, which all spectroscopic
members lie on, and the magnitude gap is 3.1 based on SDSS
spectroscopic data. The X-ray temperature of the system is $T_X=3.5$
keV and the X-ray emission peak lies $\sim 3.2$ kpc from the FG. The
X-ray source is extended with $R_{200}=1.22$ Mpc or $\sim
25R_{90}$. The system has a velocity dispersion of 1248 km s$^{-1}$
based on 18 galaxies. This system is also known as A1068 and was an
\textit{XMM} target. It is also a cooling-flow cluster with
$L_X=5\times 10^{44}$ erg s$^{-1}$ \citep{quillen08}.

\subsection{XMMXCS J123024.3+111127.8}

\begin{figure}[h]
\plottwo{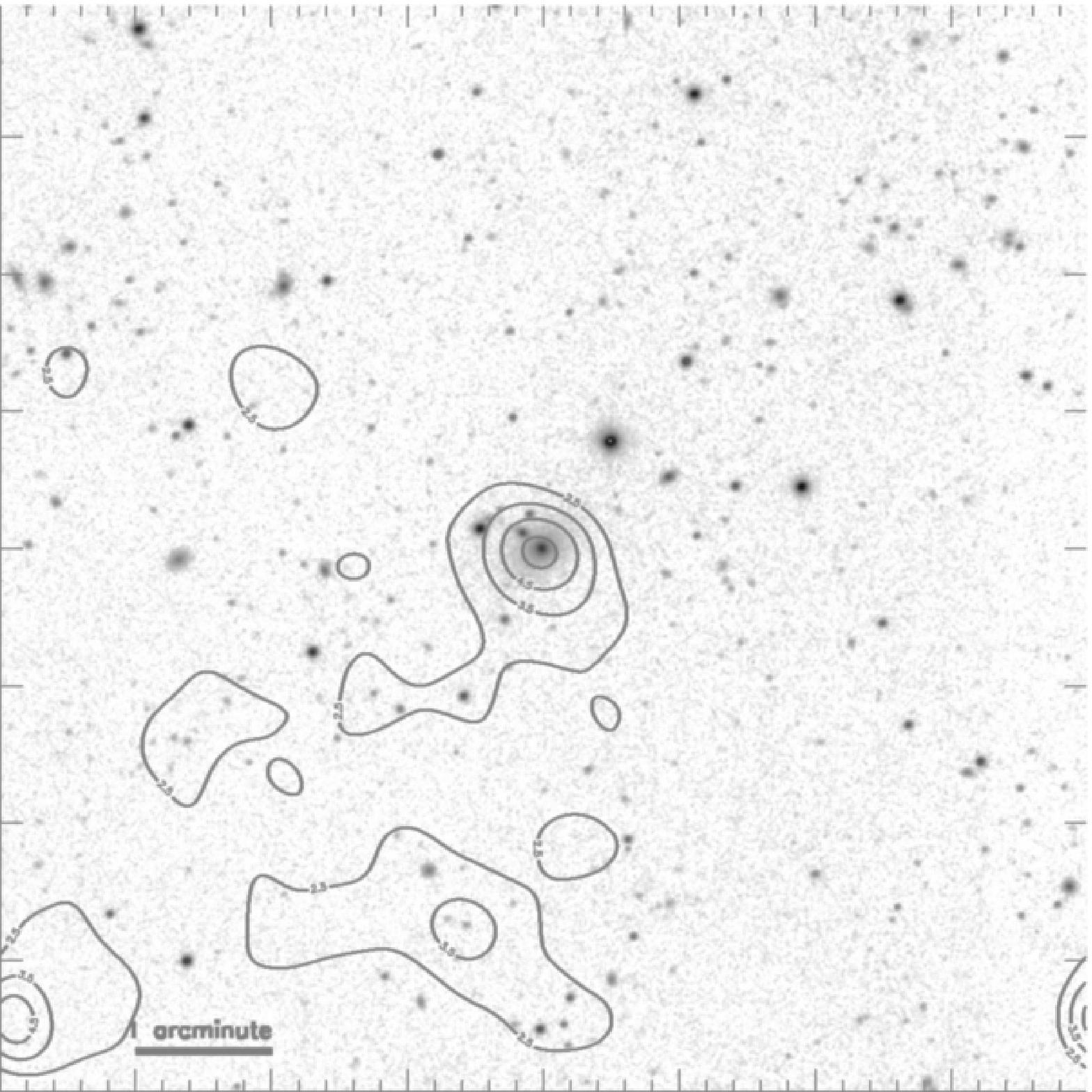}{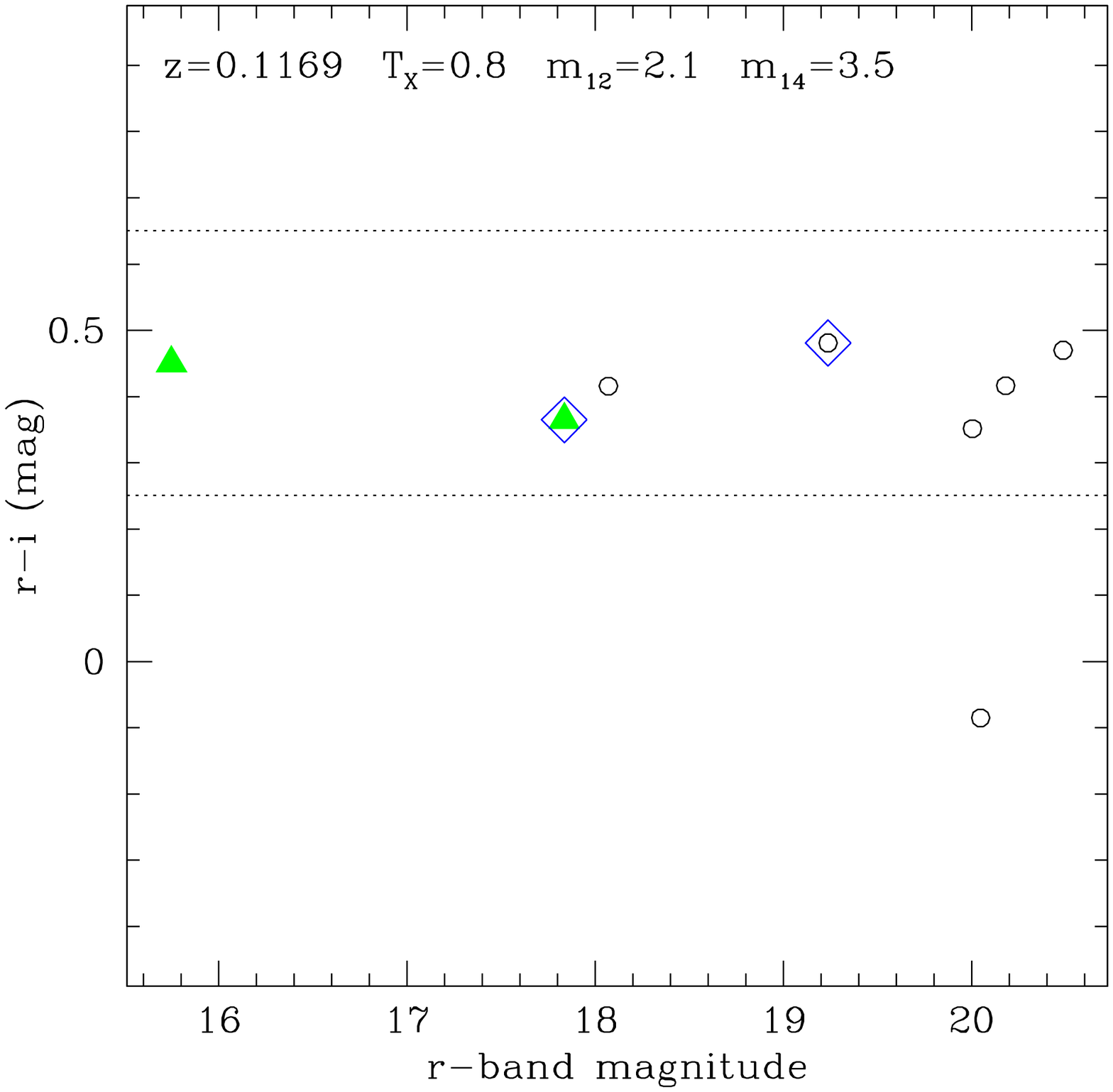}
\caption{$z=0.1169$ system located at 12:30:24.3 +11:11:27.8.}
\label{j123024}
\end{figure}

This $z=0.1169$ system is located at 12:30:24.3 +11:11:27.8 (Figure
\ref{j123024}). The few galaxies that are in the vicinity of the FG
all have redshifts (one spectroscopic and two photometric) similar to
that of the FG. The CMD shows the hint of an RS, which both
spectroscopic members lie on, and the magnitude gap is 3.5 (the
largest of any system) based on SDSS photometric data. The X-ray
temperature of the system is $T_X=0.8$ keV and the X-ray emission peak
lies $\sim 15.6$ kpc from the FG. The X-ray source is extended with
$R_{200}=0.54$ Mpc or $\sim 20R_{90}$, but is located close to the
edge of an \textit{XMM} field. We note that there is a bright galaxy
just outside $0.5R_{200}$ that would change $\Delta m_{14}$ if it were
included. No velocity dispersion could be measured for this system.

\pagebreak

\subsection{XMMXCS J123338.5+374114.9}

\begin{figure}[h]
\plottwo{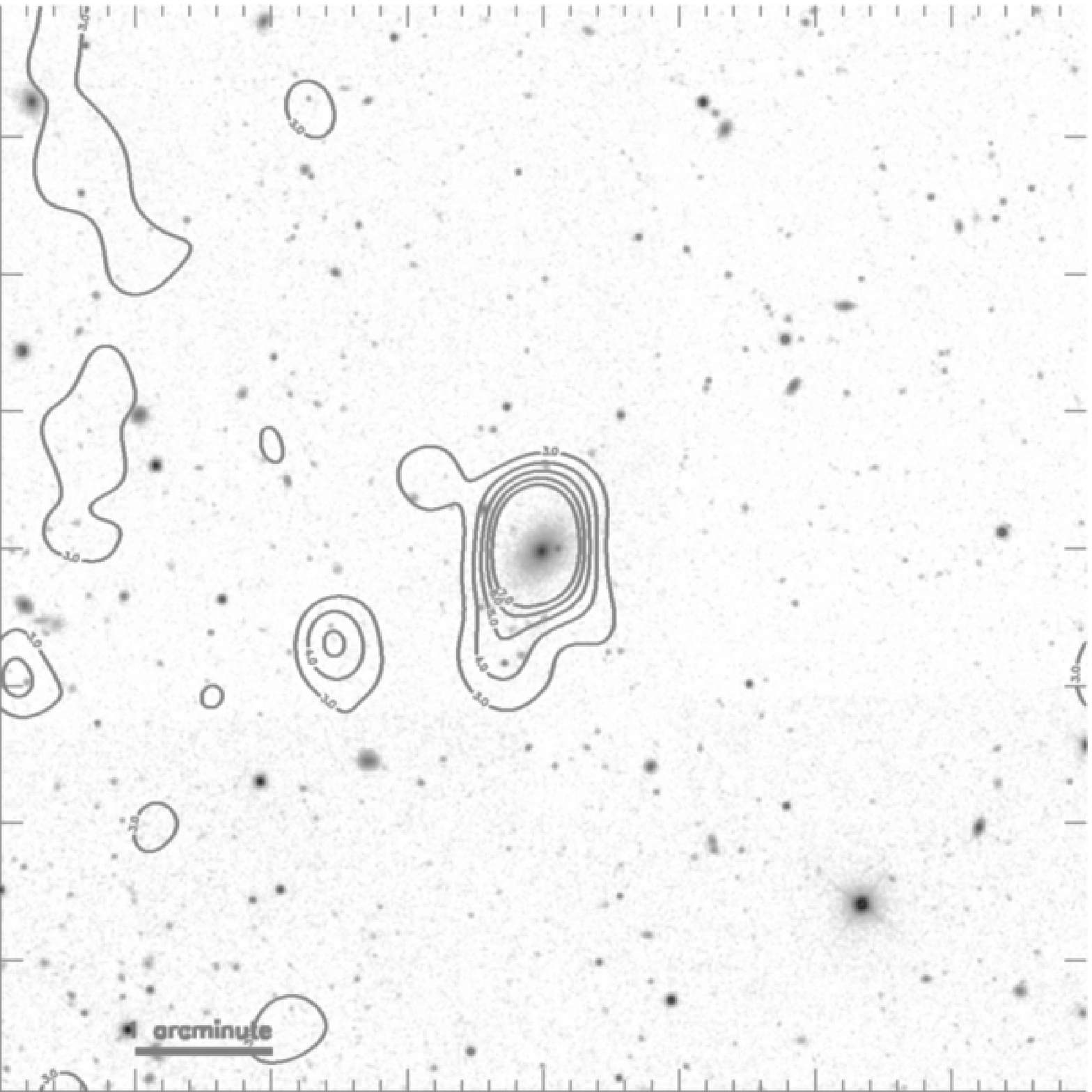}{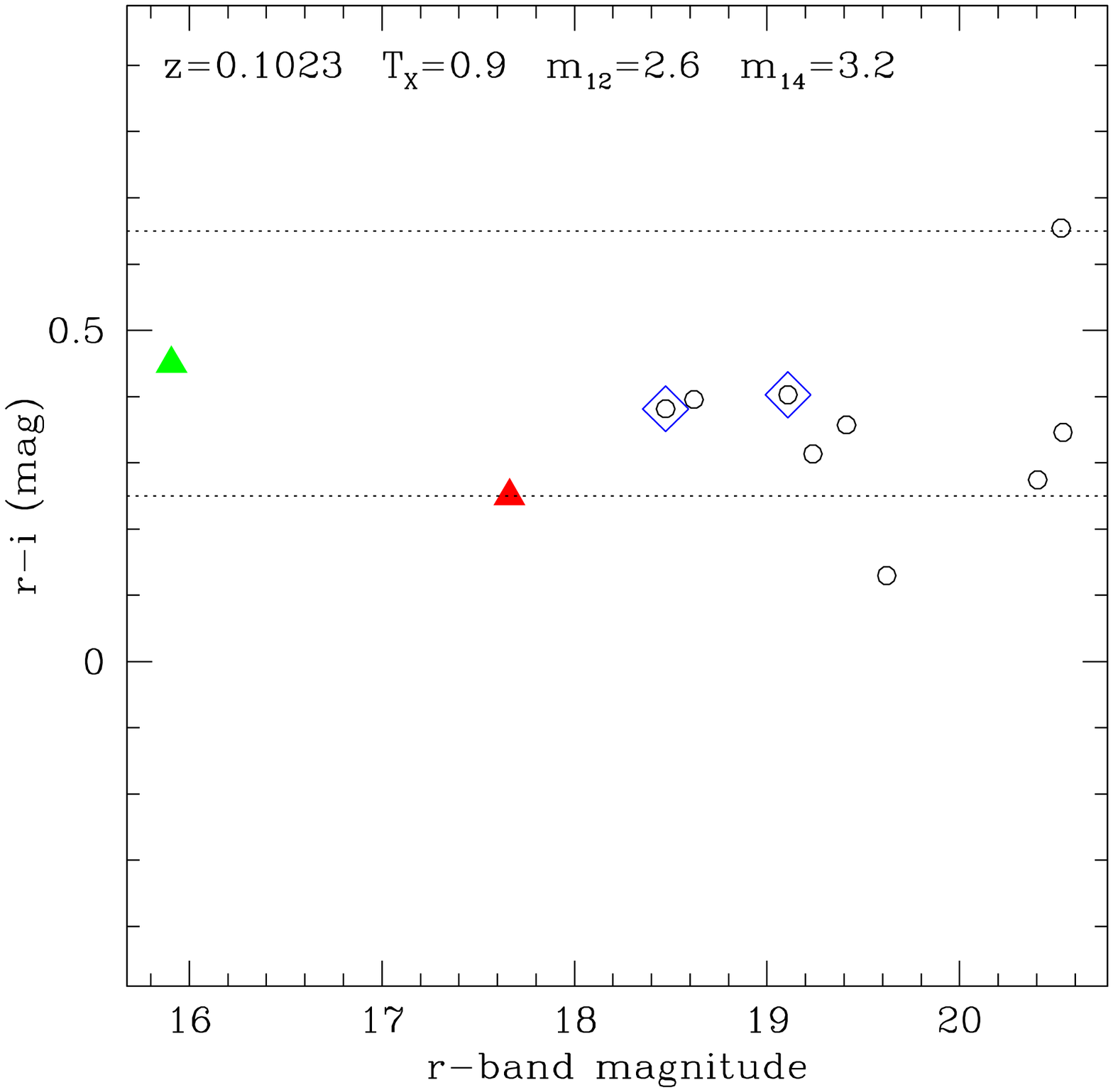}
\caption{$z=0.1023$ system located at 12:33:38.5 +37:41:14.9.}
\label{j123338}
\end{figure}

This $z=0.1023$ system is located at 12:33:38.5 +37:41:14.9 (Figure
\ref{j123338}). Again, there are very few galaxies in the vicinity of
the FG, however, the CMD shows the hint of an RS. The only
spectroscopic system member (the FG) lies on the RS and the magnitude
gap is 3.2 based on SDSS photometric data. The rejected galaxy is
$\sim 11,100$ km s$^{-1}$ away from the FG. The X-ray temperature of
the system is $T_X=0.9$ keV and the X-ray emission peak lies $\sim
21.6$ kpc from the FG. The X-ray source is extended with
$R_{200}=0.58$ Mpc or $\sim 30R_{90}$. No velocity dispersion could be
measured for this system.

\subsection{XMMXCS J124425.9+164758.0}\label{J124425}

\begin{figure}[h]
\plottwo{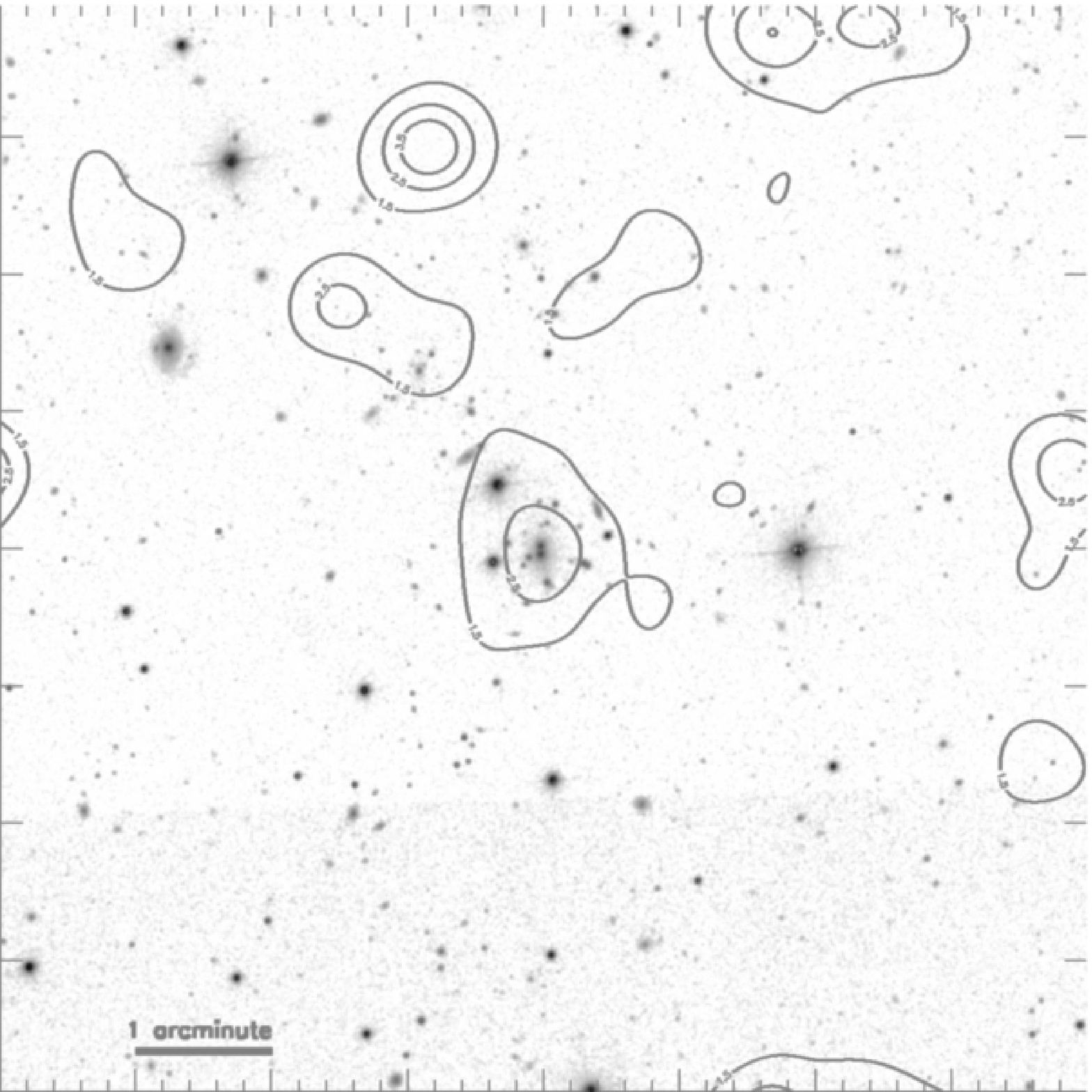}{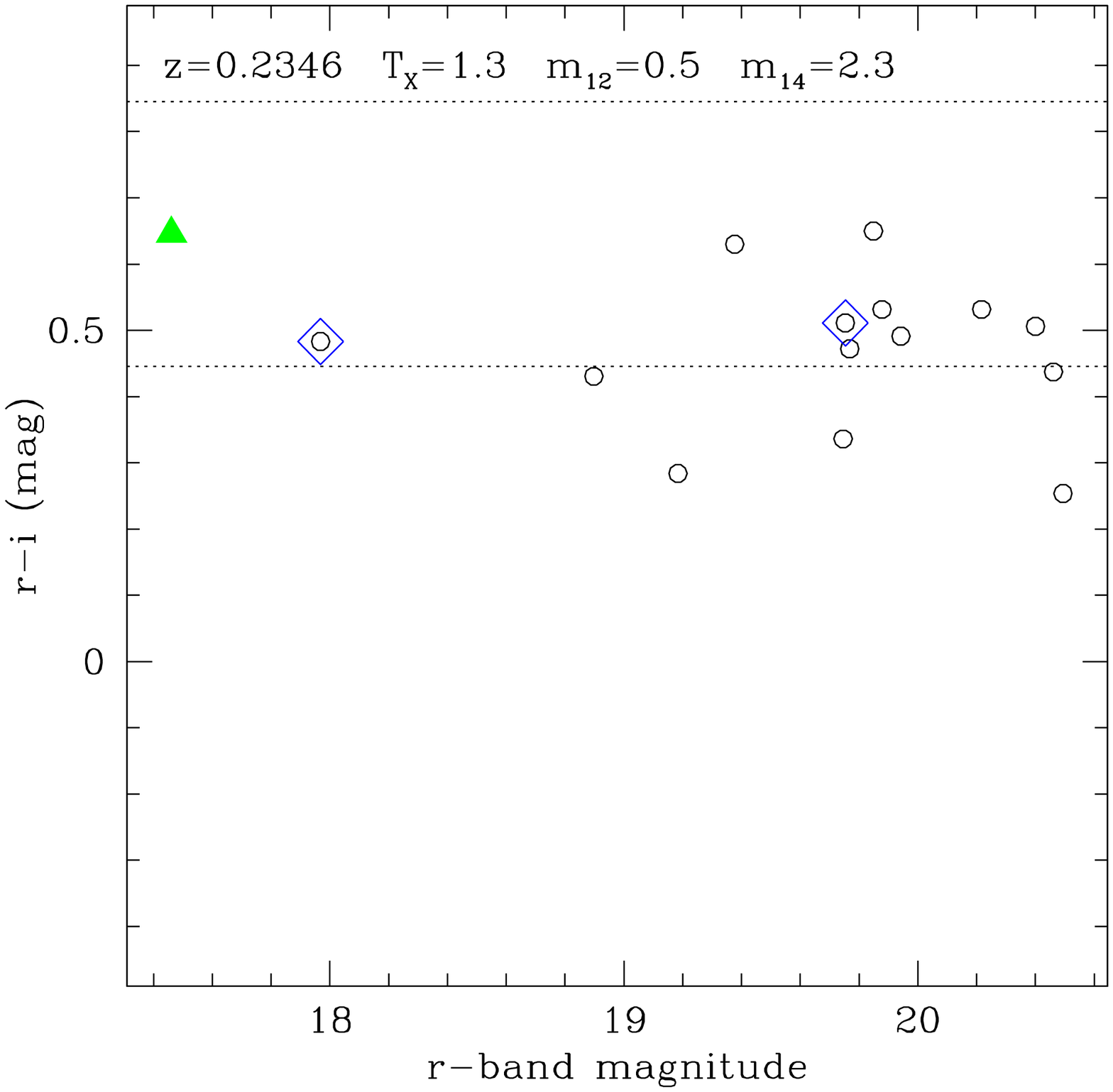}
\caption{$z=0.2346$ system located at 12:44:25.9 +16:47:58.0.}
\label{j124425}
\end{figure}

This $z=0.2346$ system is located at 12:44:25.9 +16:47:58.0 (Figure
\ref{j124425}). The FG in this system has a double core, therefore we
cannot trust the sky-subtraction correction and so have not applied
it. This means that the magnitude gap of 2.3 is a lower limit. If we
applied the estimated correction the gap would be 3.3 and if we only
applied the average correction from the other 16 FSs the gap would be
2.6, therefore we accept this as an FS. The X-ray temperature of the
system is $T_X=1.3$ keV and the X-ray emission peak lies $\sim 25.8$
kpc from the FG. The X-ray source is extended with $R_{200}=0.63$ Mpc
or $\sim 20R_{90}$. No velocity dispersion could be measured for this
system.

\pagebreak

\subsection{XMMXCS J130749.6+292549.2}

\begin{figure}[h]
\plottwo{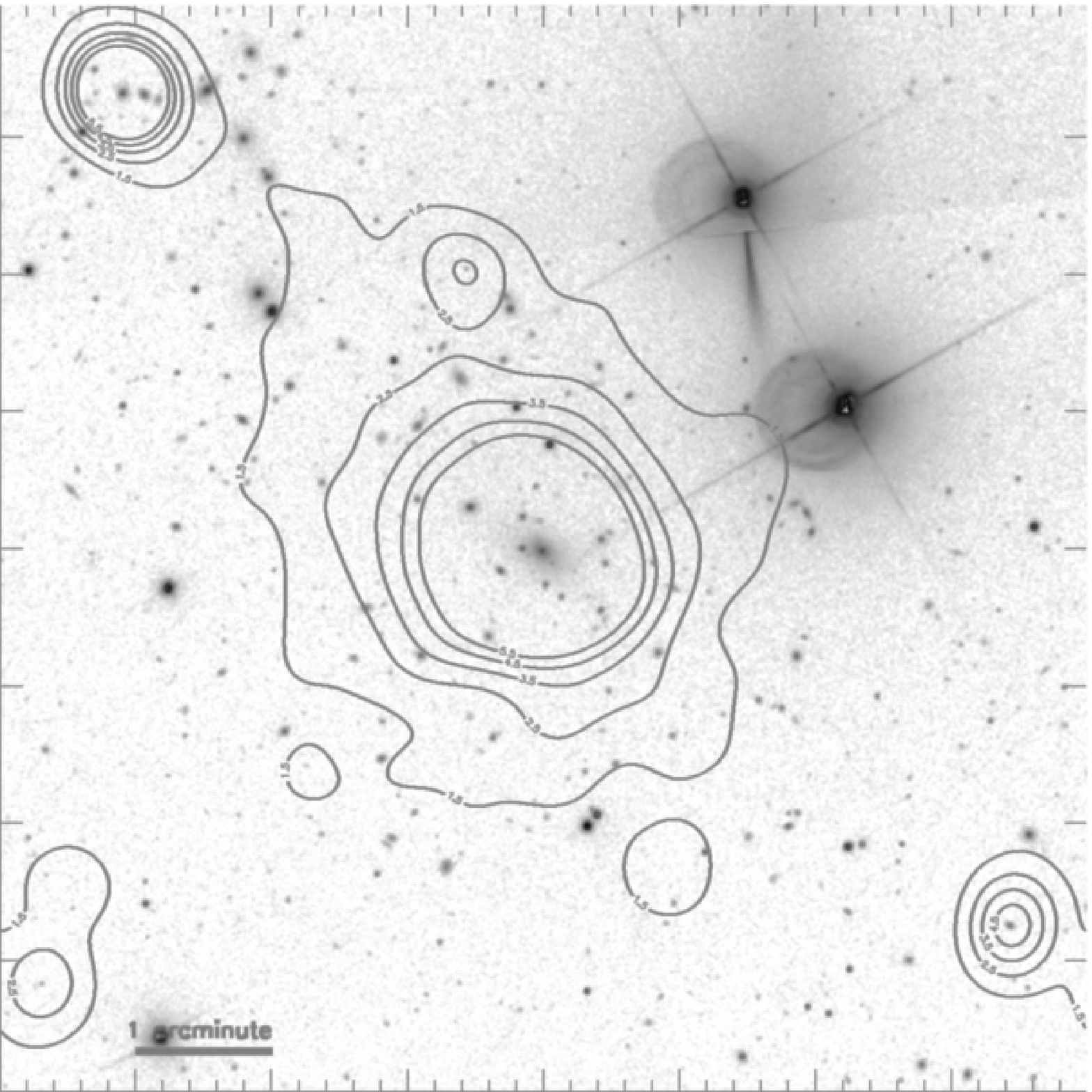}{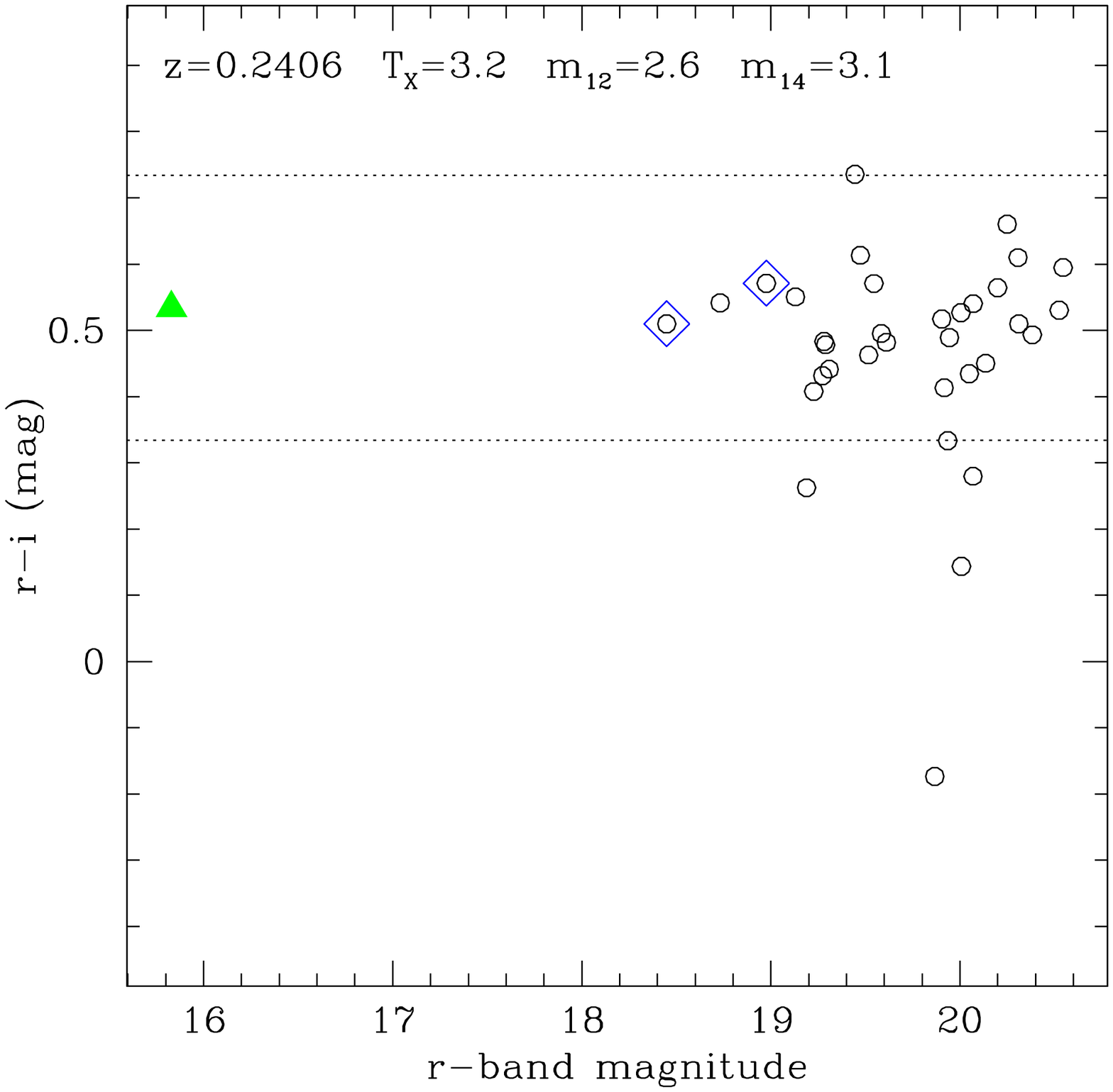}
\caption{$z=0.2406$ system located at 13:07:49.6 +29:25:49.2.}
\label{j130749}
\end{figure}

This $z=0.2406$ system is located at 13:07:49.6 +29:25:49.2 (Figure
\ref{j130749}). There is the hint of an RS in the CMD, which the FG
lies on, and the magnitude gap is 3.1 based on SDSS photometric
data. There are no other spectroscopic objects in the vicinity of the
FG, however, those objects that have photometric redshifts are at the
same redshift. The X-ray temperature of the system is $T_X=3.2$ keV
and the X-ray emission peak lies $\sim 18.9$ kpc from the FG. The
X-ray source is extended with $R_{200}=1.04$ Mpc or $\sim
25R_{90}$. No velocity dispersion could be measured for this
system. This system is also known as ZwCl
1305.4+2941. \citet{gastaldello08} find $T_X=3.17\pm 0.19$ KeV (fully
consistent with our measurement) and $L_{X,500}=(1.25\pm 0.16)\times
10^{44}\,h_{70}^{-2}$ erg s$^{-1}$.

\subsection{XMMXCS J131145.1+220206.0}

\begin{figure}[h]
\plottwo{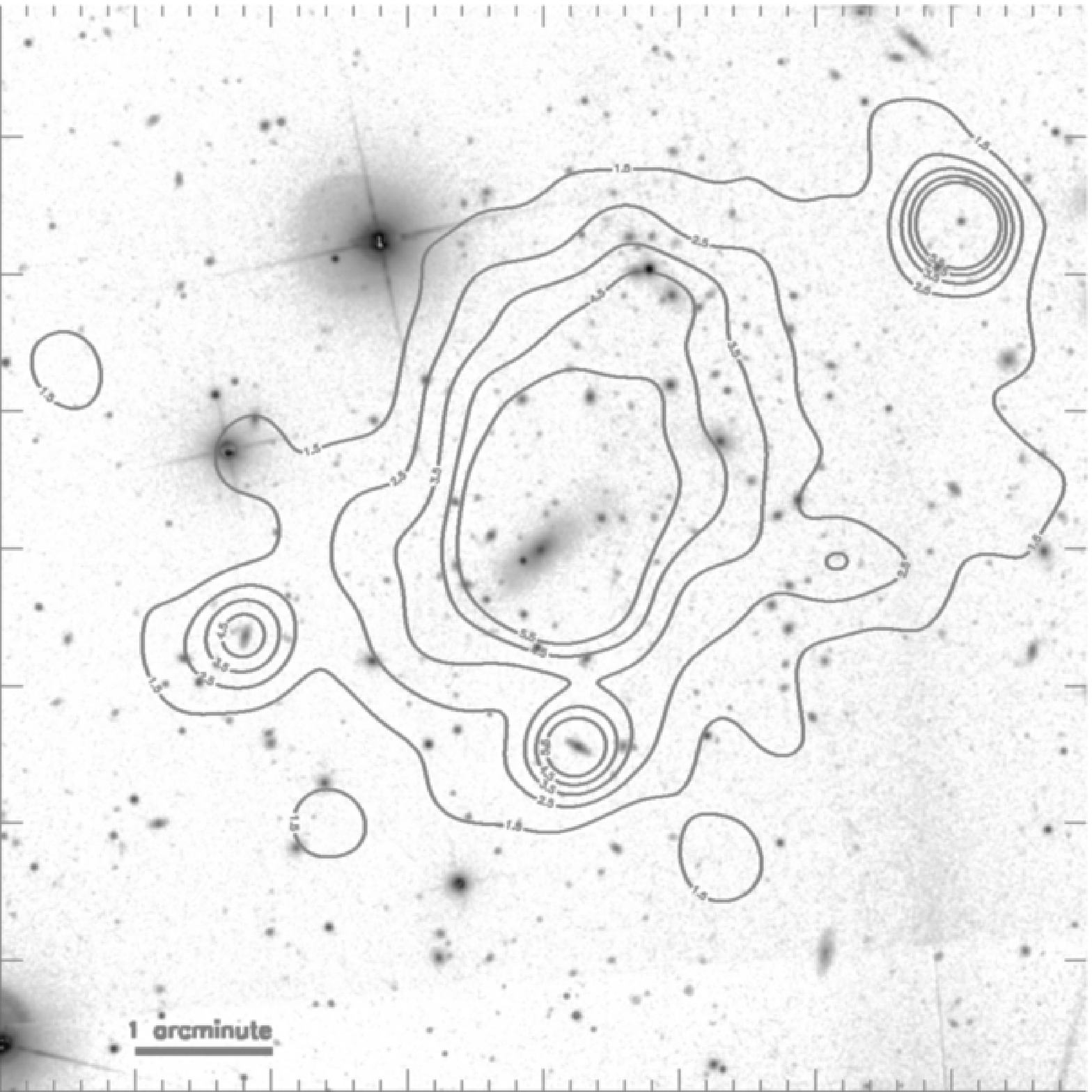}{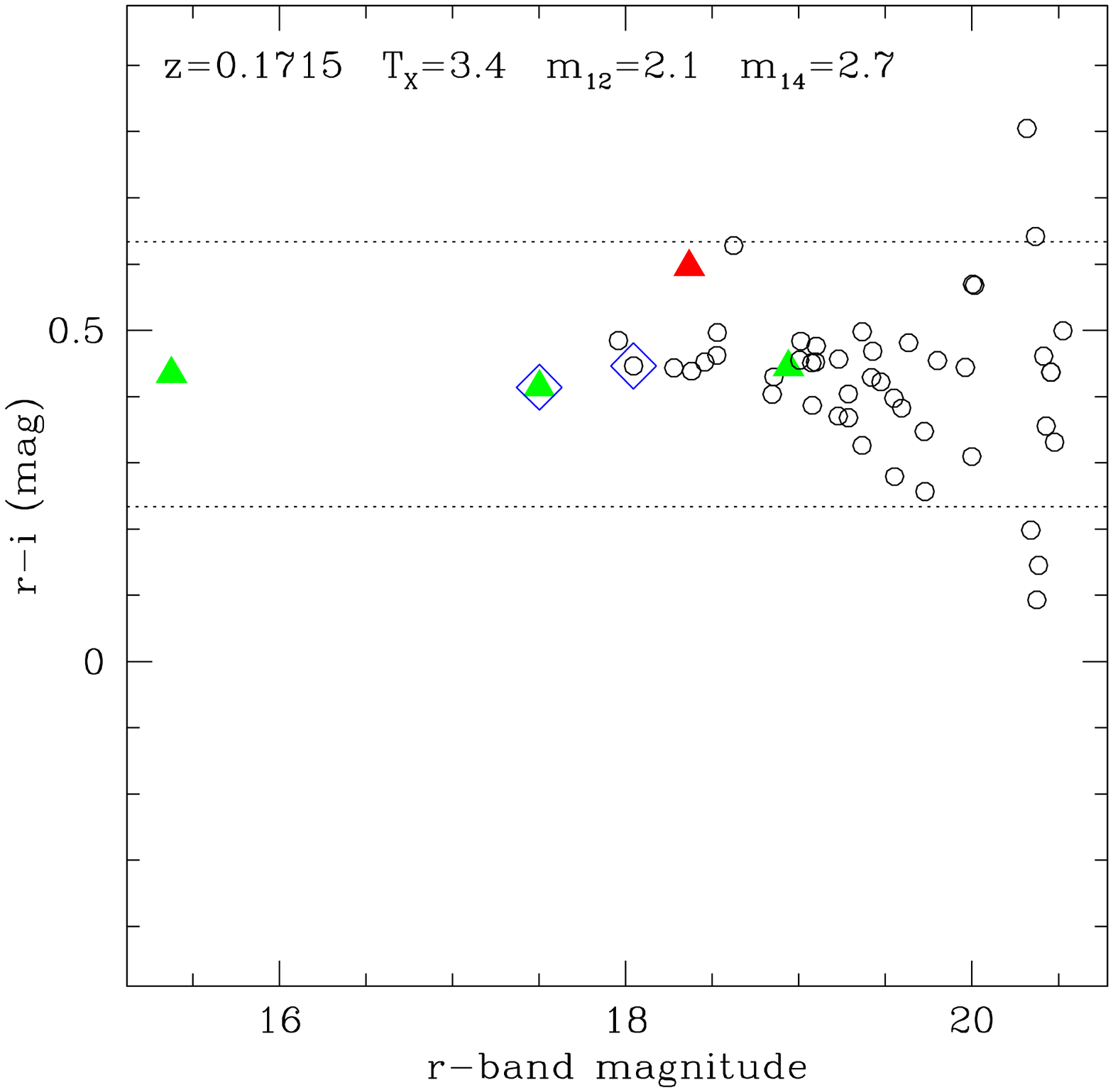}
\caption{$z=0.1715$ system located at 13:11:45.1 +22:02:06.0.}
\label{j131145}
\end{figure}

This $z=0.1715$ system is located at 13:11:45.1 +22:02:06.0 (Figure
\ref{j131145}). The CMD shows an obvious RS, which all spectroscopic
members lie on, and the magnitude gap is 2.7 based on SDSS photometric
data. The rejected galaxy is $\sim 50,000$ km s$^{-1}$ away from the
FG. The X-ray temperature of the system is $T_X=3.4$ keV and the X-ray
emission peak lies $\sim 95.3$ kpc from the FG. The X-ray source is
extended with $R_{200}=1.16$ Mpc or $\sim 25R_{90}$. The system has a
velocity dispersion of 362 km s$^{-1}$ based on 10 galaxies. This
system is also known as MaxBCG J197.94248+22.02702.

\pagebreak

\subsection{XMMXCS J134825.6+580015.8}

\begin{figure}[h]
\plottwo{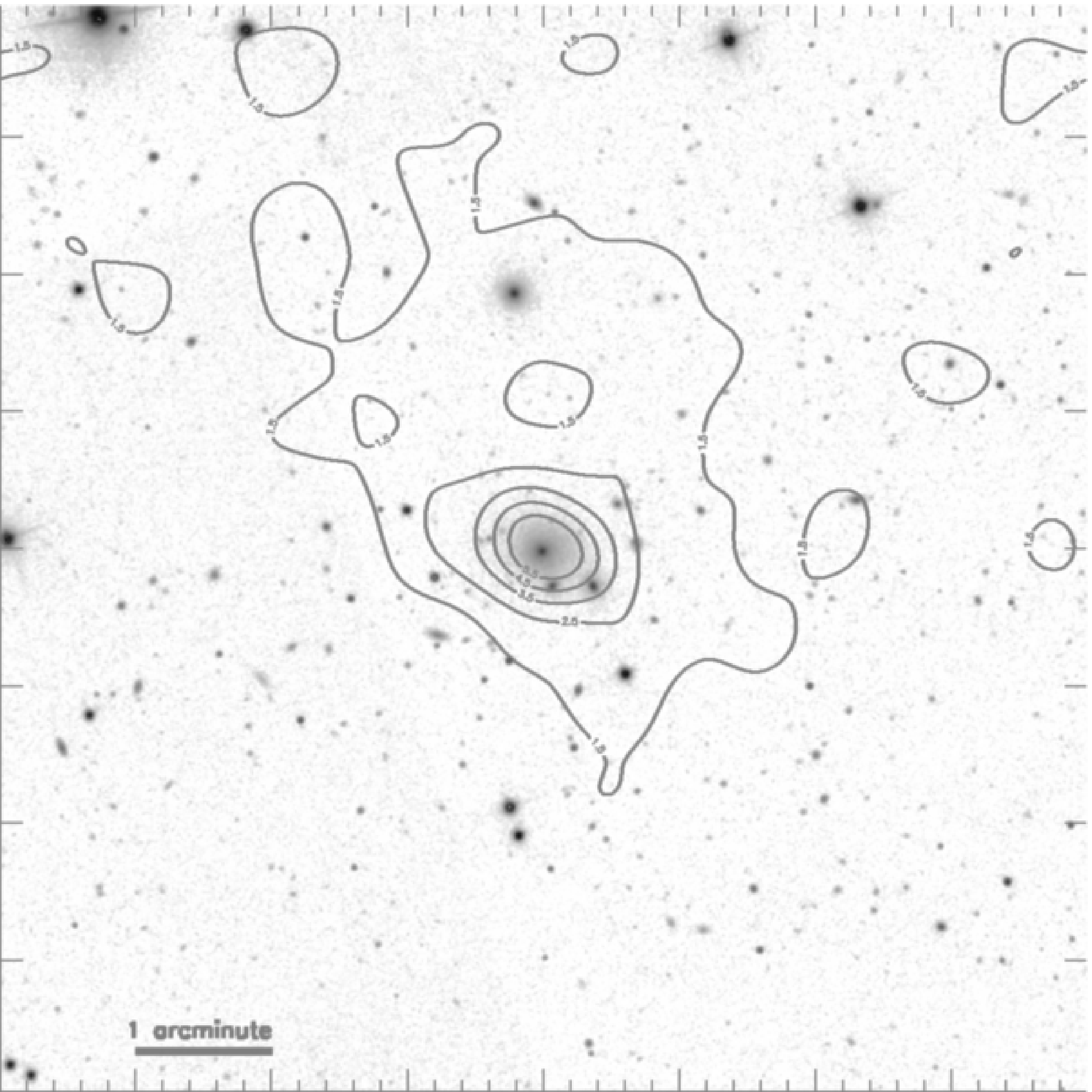}{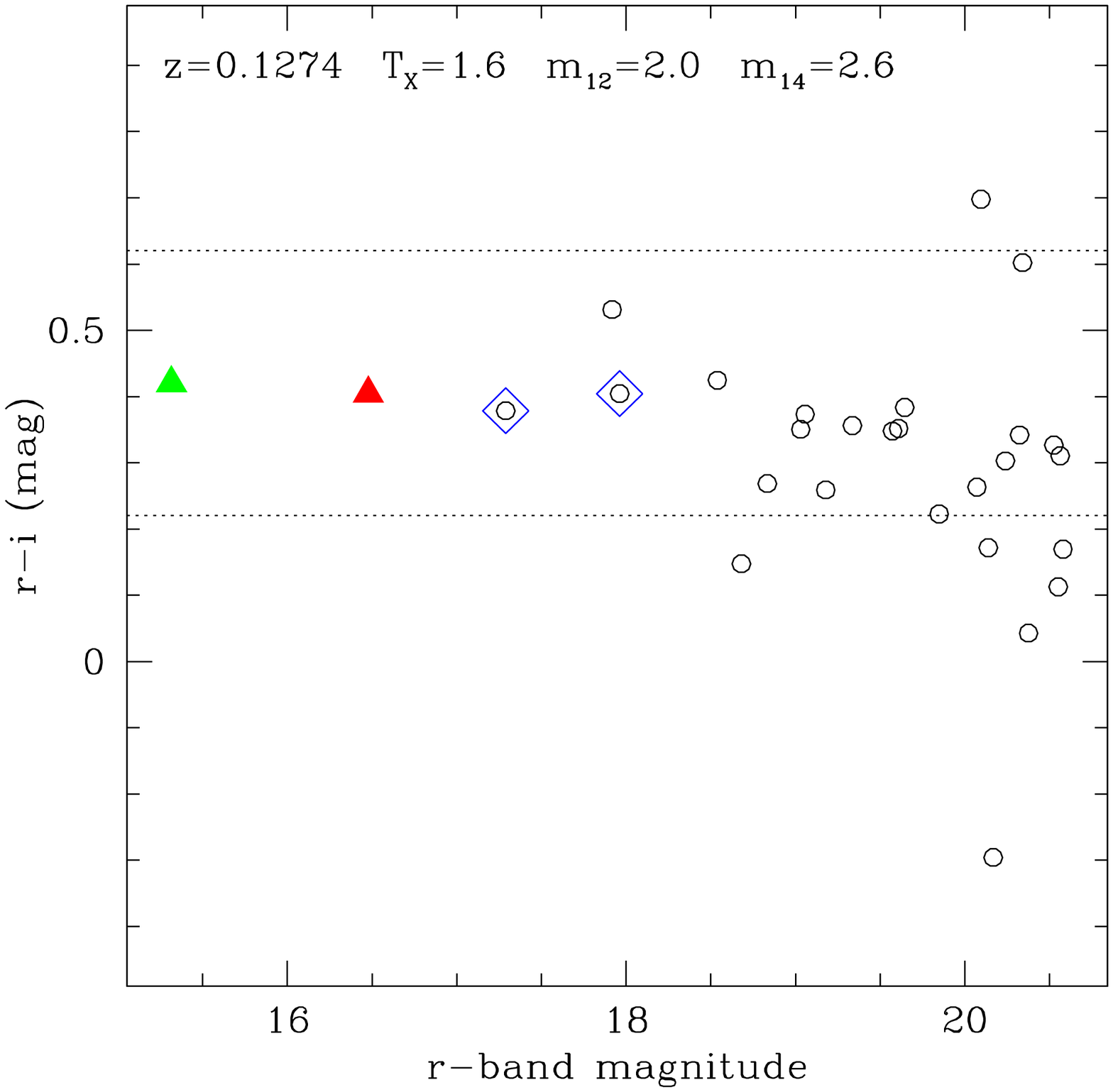}
\caption{$z=0.1274$ system located at 13:48:25.6 +58:00:15.8.}
\label{j134825}
\end{figure}

This $z=0.1274$ system is located at 13:48:25.6 +58:00:15.8 (Figure
\ref{j134825}). The CMD shows an obvious RS, which the only
spectroscopic members (the FG) lies on, and the magnitude gap is 2.6
based on SDSS photometric data. The rejected galaxy is $\sim 10000$ km
s$^{-1}$ away from the FG. The X-ray temperature of the system is
$T_X=1.6$ keV and the X-ray emission peak lies $\sim 7.4$ kpc from the
FG. The X-ray source is extended with $R_{200}=0.78$ Mpc or $\sim
20R_{90}$, but it is located near the edge of the \textit{XMM}
chip. The system has a velocity dispersion of 526 km s$^{-1}$ based on
five galaxies.

\subsection{XMMXCS J141627.7+231525.9}

\begin{figure}[h]
\plottwo{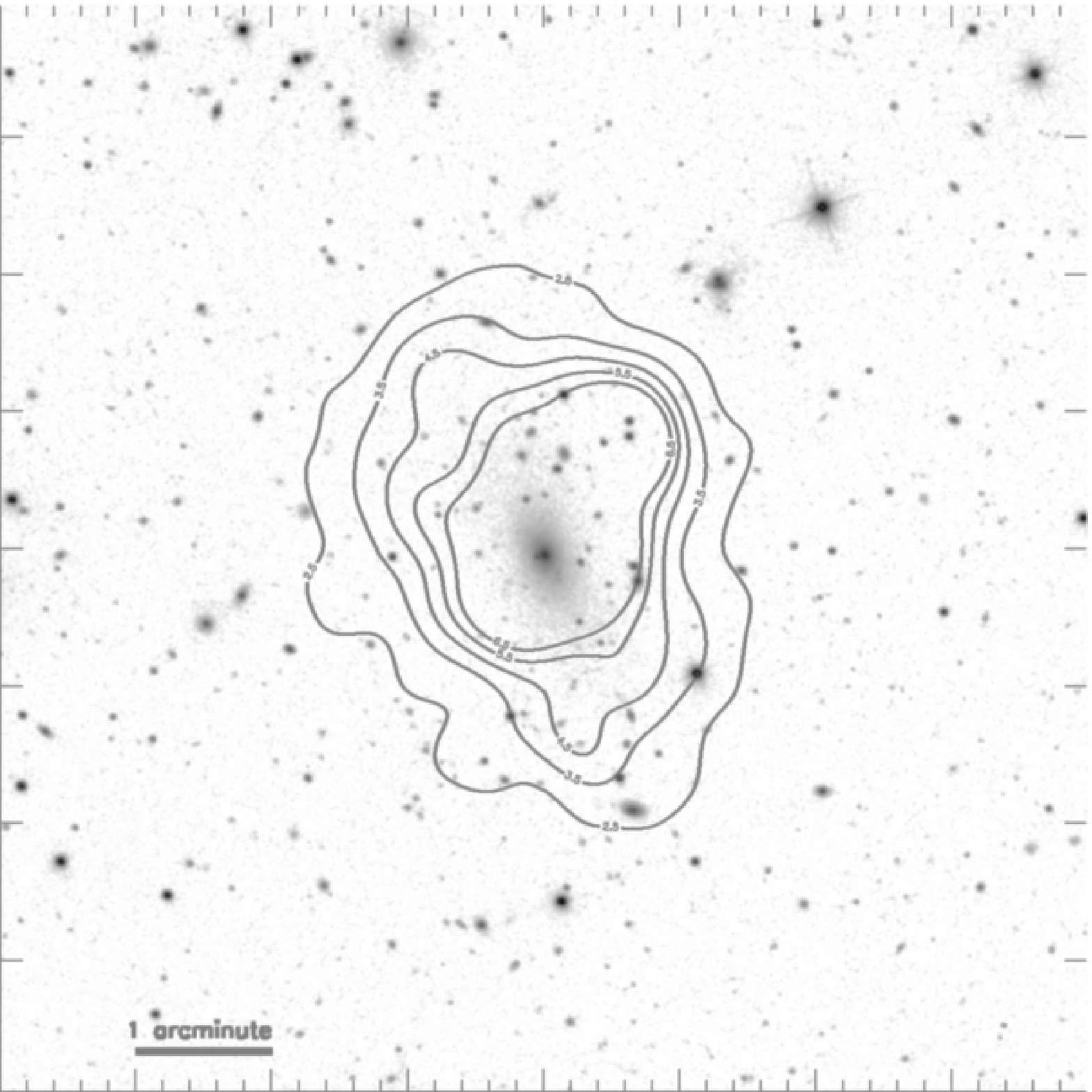}{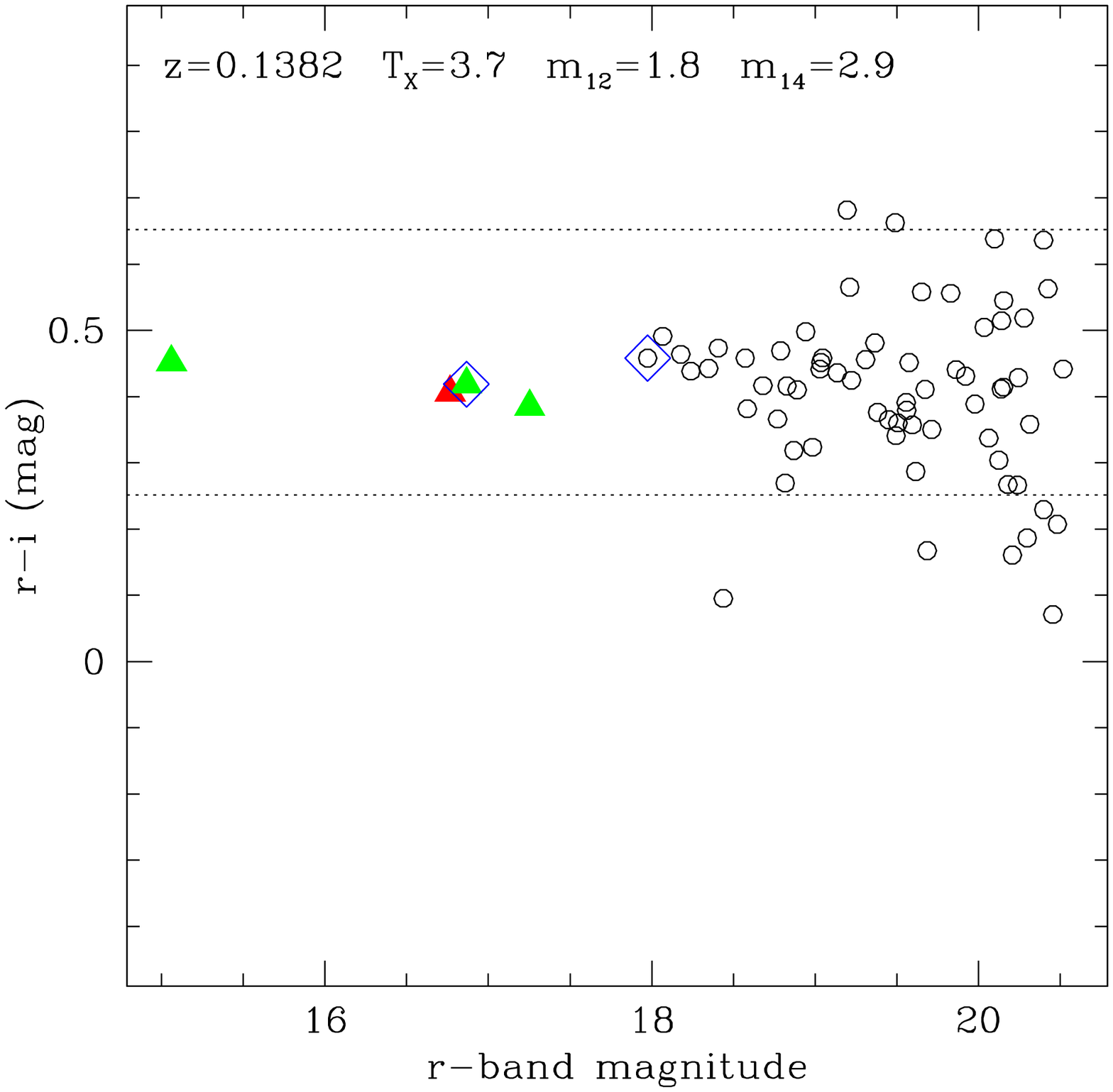}
\caption{$z=0.1382$ system located at 14:16:27.7 +23:15:25.9.}
\label{j141627}
\end{figure}

This $z=0.1382$ system is located at 14:16:27.7 +23:15:25.9 (Figure
\ref{j141627}). The CMD shows an obvious RS, which all spectroscopic
members lie on, and the magnitude gap is 2.9 based on SDSS photometric
data.  The rejected galaxy is $\sim 10500$ km s$^{-1}$ away from the
FG. The X-ray temperature of the system is $T_X=3.7$ keV and the X-ray
emission peak lies $\sim 13.9$ kpc from the FG. The X-ray source is
extended with $R_{200}=1.25$ Mpc or $\sim 25R_{90}$. The system has a
velocity dispersion of 646 km s$^{-1}$ based on 21 galaxies. This
system was an \textit{XMM} target and was classified as an FS in
\citet{jones03} and subsequently studied in \citet{voevodkin10},
\citet{cypriano06}, and \citet{khosroshahi06a, khosroshahi06b}. It is
also known as ZwCl 1413.9+2330. From the literature: $\Delta
m_{12}=2.4$, $L_X=2.2\times 10^{44}\,h_{50}^{-2}$ erg s$^{-1}$, and
$T_X=1.53\pm 0.35$ keV \citep{jones03}; $L_X=1.11\times
10^{44}\,h^{-2}_{70}$ erg s$^{-1}$ \citep{cypriano06}; $T_X\sim 4$
keV, $\sigma \sim 700$ km s$^{-1}$ \citep{khosroshahi06a}; and $\Delta
m_{12}=1.7$ $L_X=6.09\times 10^{43}$ erg s$^{-1}$; $R_{500}=0.89$ Mpc
and $\sigma=652$ km s$^{-1}$ \citep{voevodkin10}. This system appears
to be associated with XMMXCS J141657.5+231239.2.

\subsection{XMMXCS J141657.5+231239.2}

\begin{figure}[h]
\plottwo{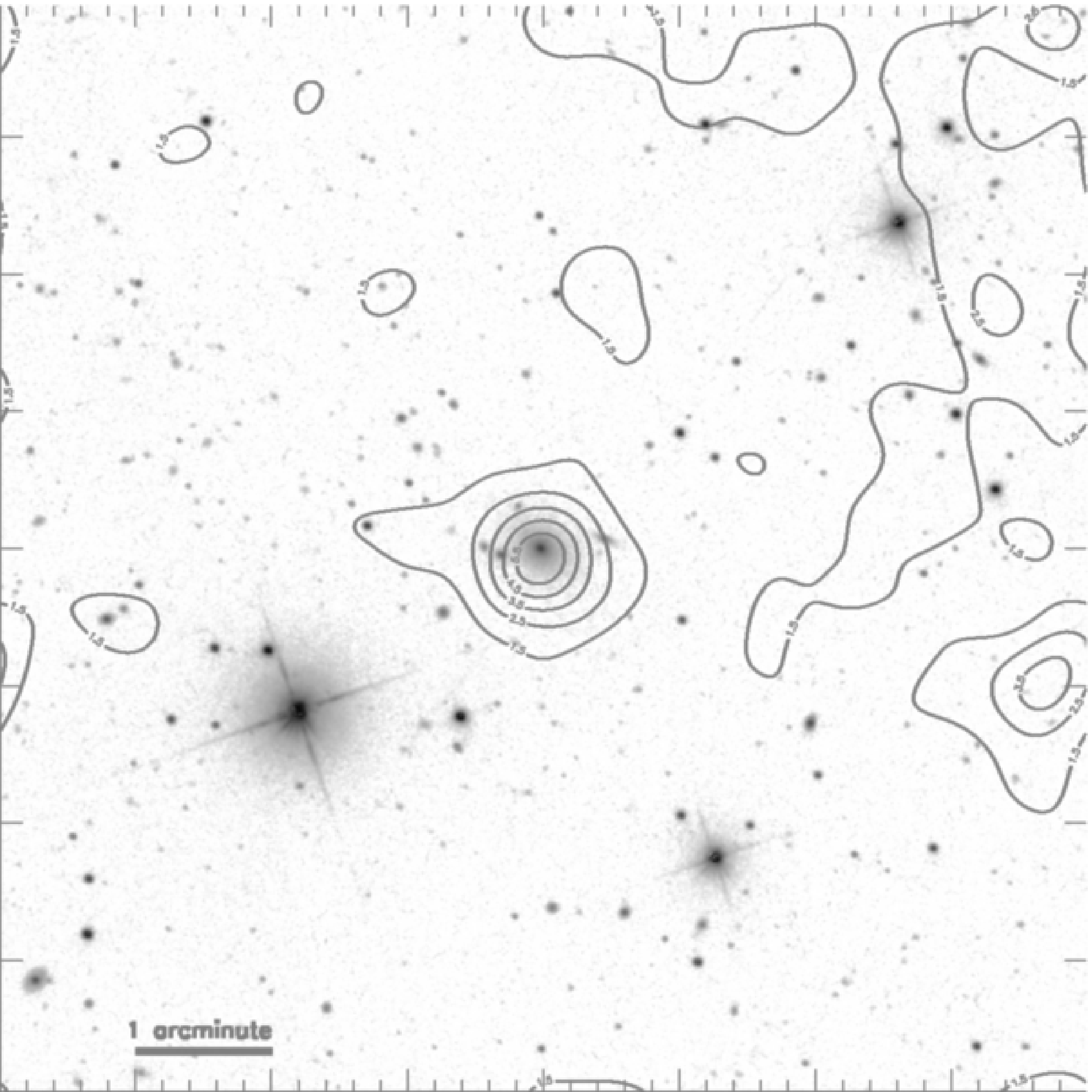}{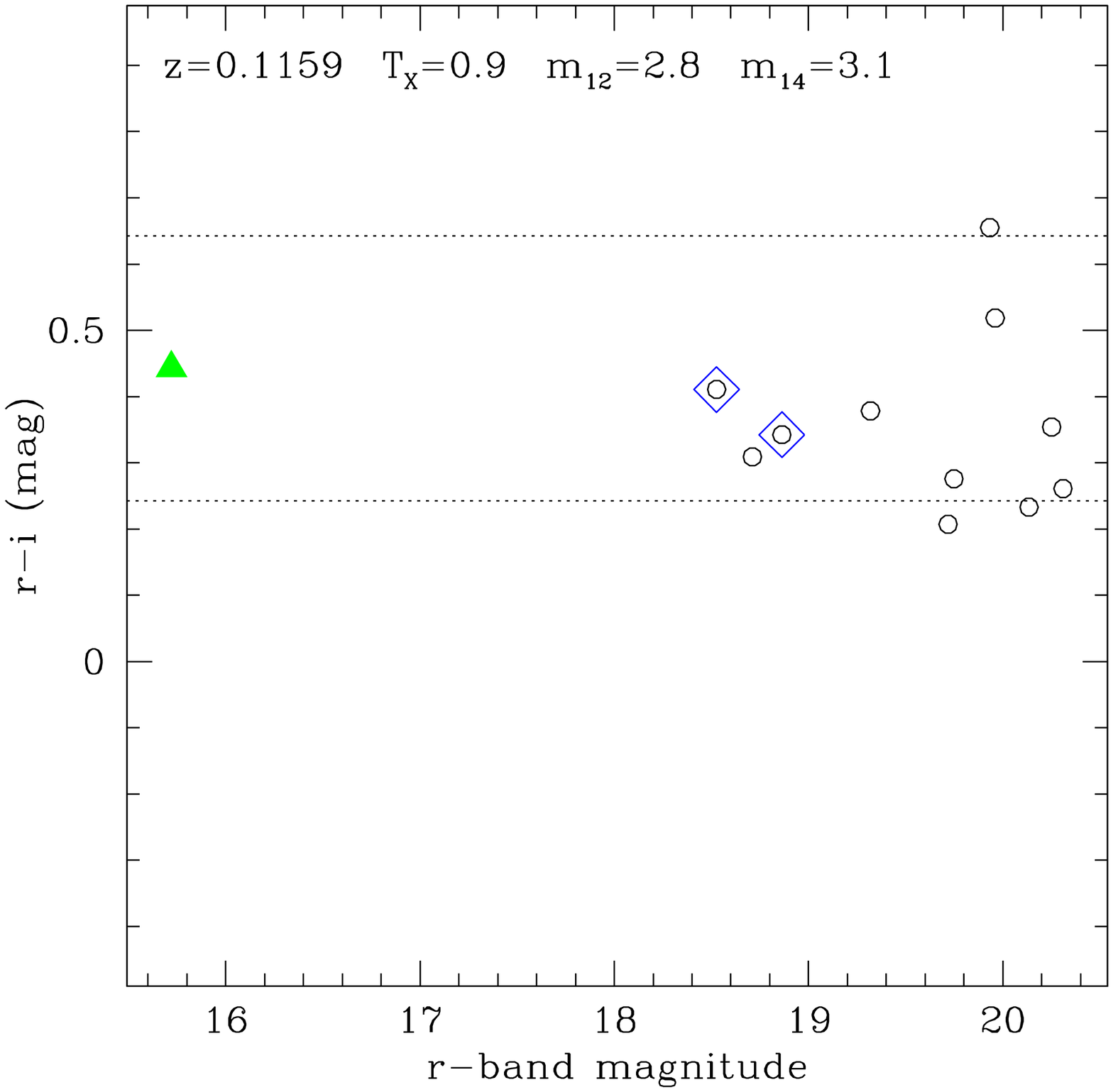}
\caption{$z=0.1159$ system located at 14:16:57.5 +23:12:39.2.}
\label{j141657}
\end{figure}

This $z=0.1159$ system is located at 14:16:57.5 +23:12:39.2 (Figure
\ref{j141657}). The CMD shows an RS and the magnitude gap is 3.1 based
on SDSS photometric data. We note that there is a bright galaxy just
outside $0.5R_{200}$ that would change $\Delta m_{14}$ if it were
included. The X-ray temperature of the system is $T_X=0.9$ keV and the
X-ray emission peak lies $\sim 7.0$ kpc from the FG. The X-ray source
is extended with $R_{200}=0.56$ Mpc or $\sim 20R_{90}$. No velocity
dispersion could be measured for this system. This system appears to
be associated with XMMXCS J141627.7+231525.9.

\subsection{XMMXCS J160129.8+083856.3}

\begin{figure}[h]
\plottwo{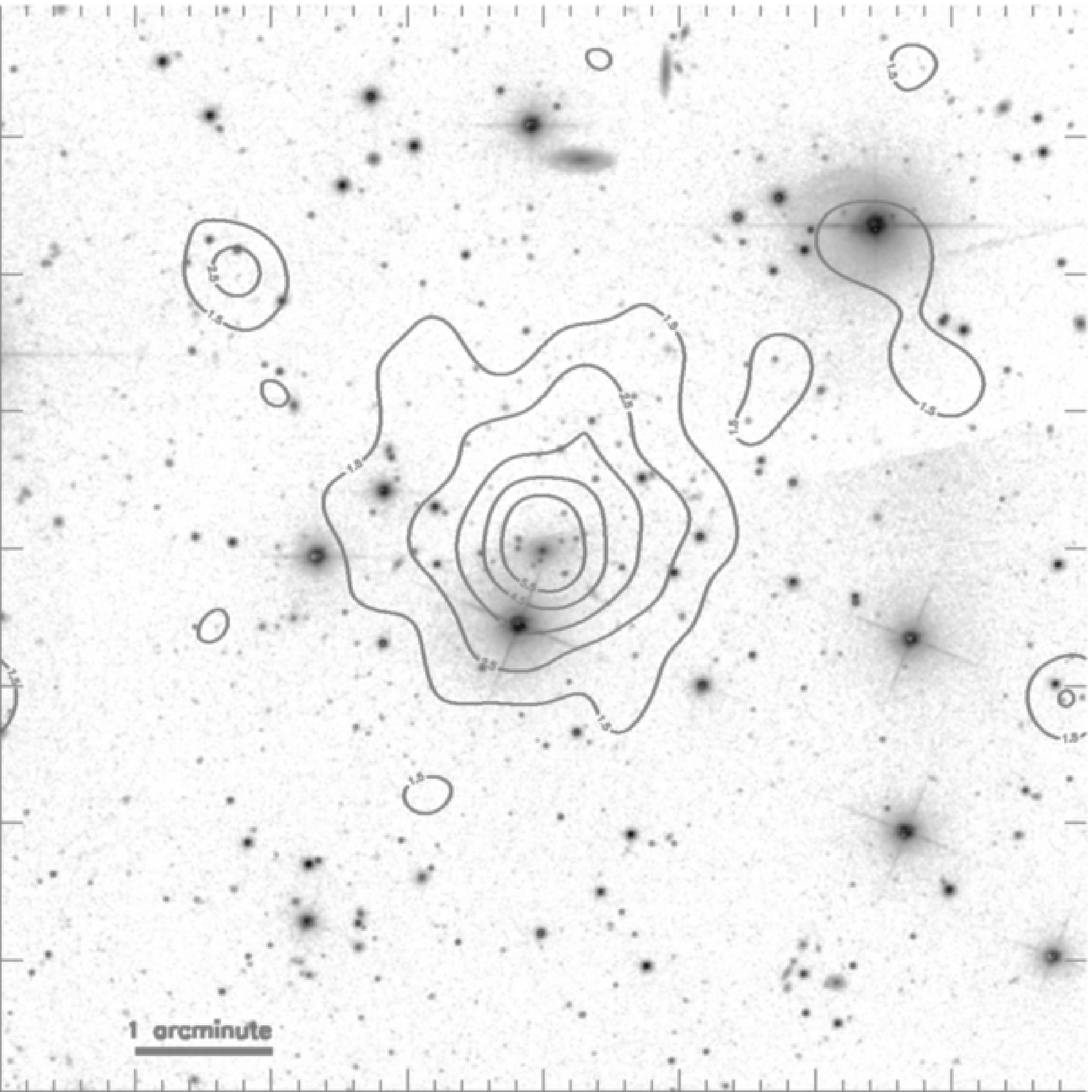}{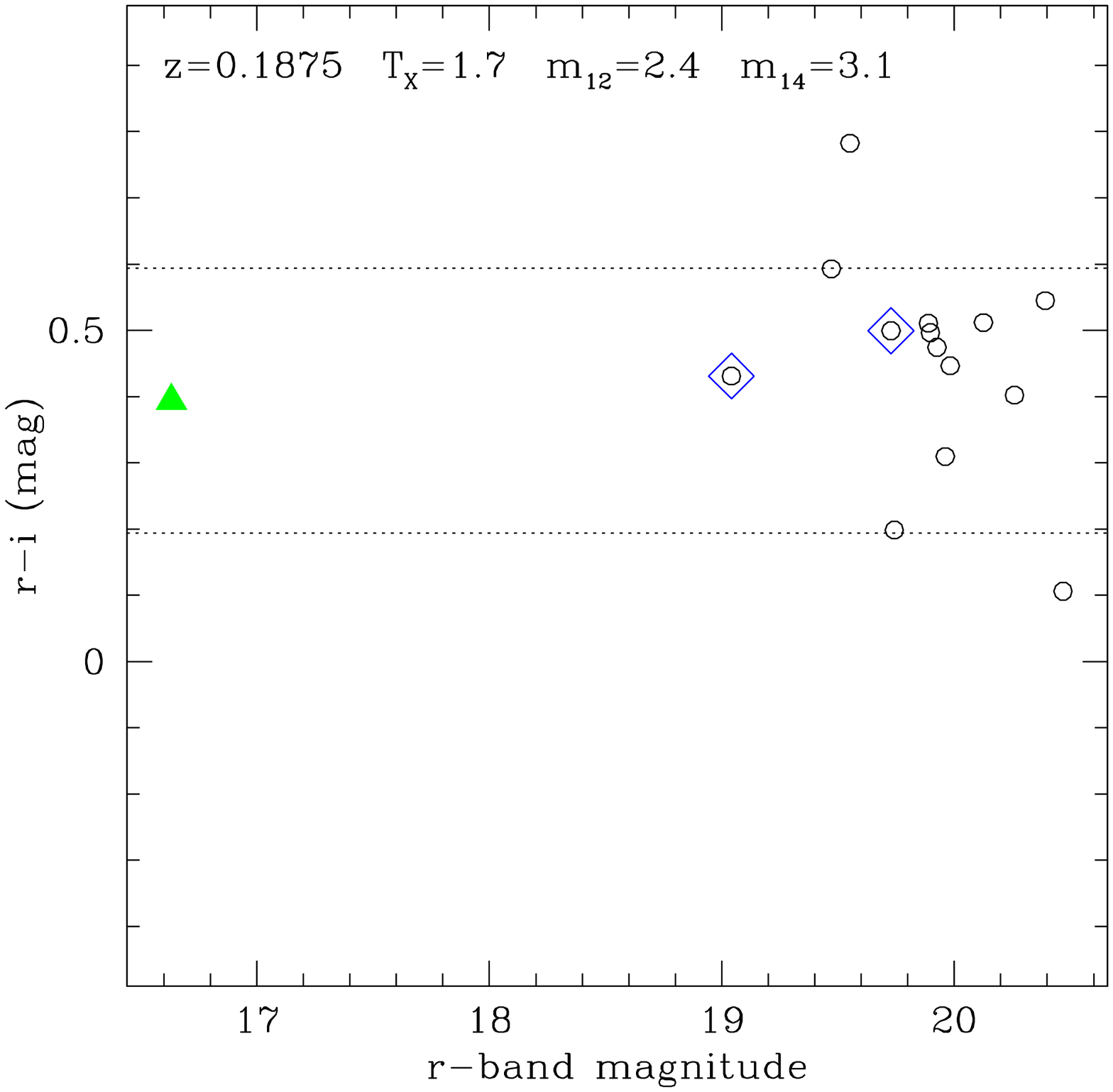}
\caption{$z=0.1875$ system located at 16:01:29.8 +08:38:56.3.}
\label{j160129}
\end{figure}

This $z=0.1875$ system is located at 16:01:29.8 +08:38:56.3 (Figure
\ref{j160129}). The CMD shows no sign of an RS and the magnitude gap
is 3.1 based on SDSS photometric data. Increasing $R_{200}$ by its
error would decrease the magnitude gap by 0.3 but the system would
still be classified as an FS. We note that there is a bright galaxy
just outside $0.5R_{200}$ that would change $\Delta m_{14}$ if it were
included. The X-ray temperature of the system is $T_X=1.7$ keV and the
X-ray emission peak lies $\sim 18.1$ kpc from the FG. The X-ray source
is extended with $R_{200}=0.77$ Mpc or $\sim 35R_{90}$. No velocity
dispersion could be measured for this system.

\subsection{XMMXCS J172010.0+263724.7}

\begin{figure}[h]
\plottwo{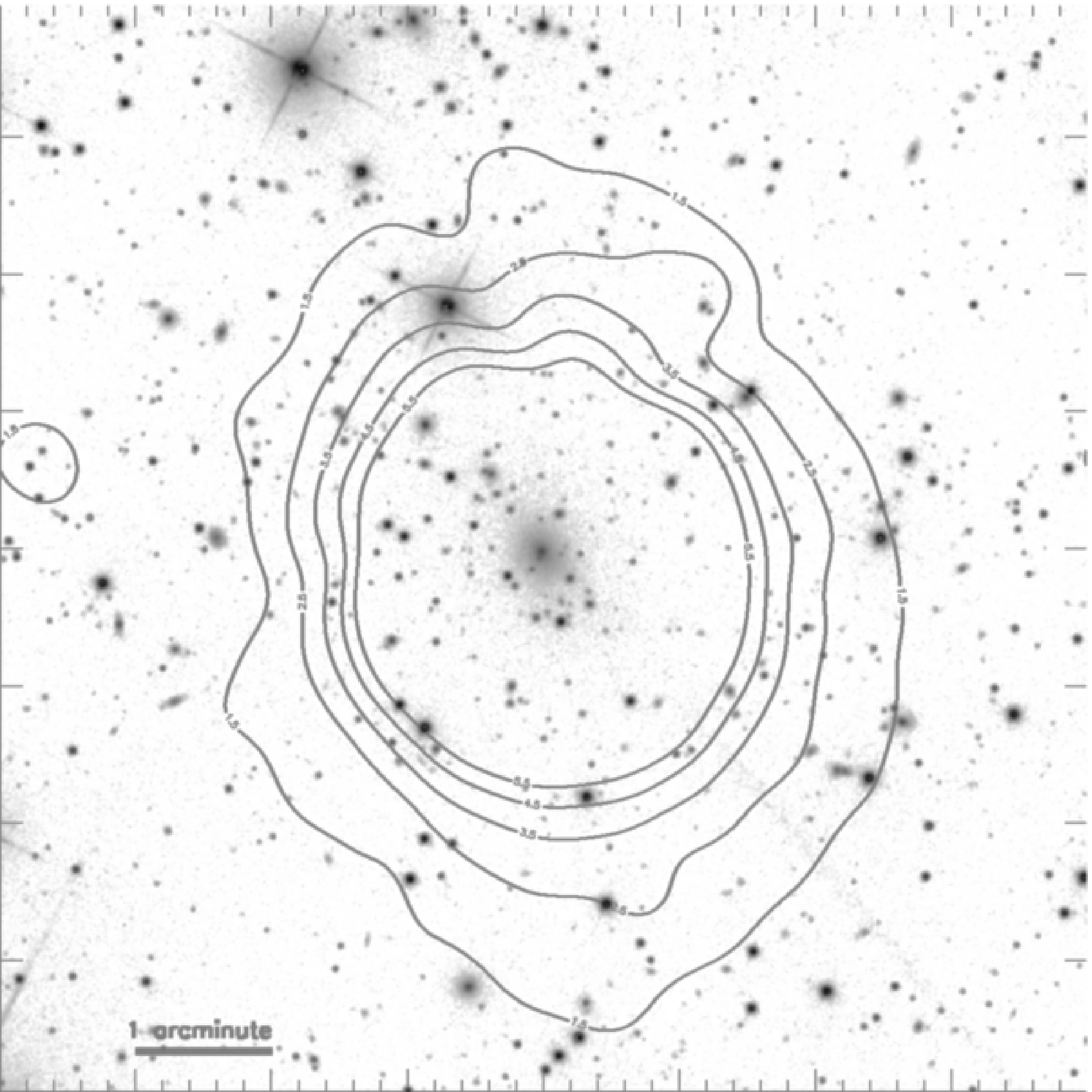}{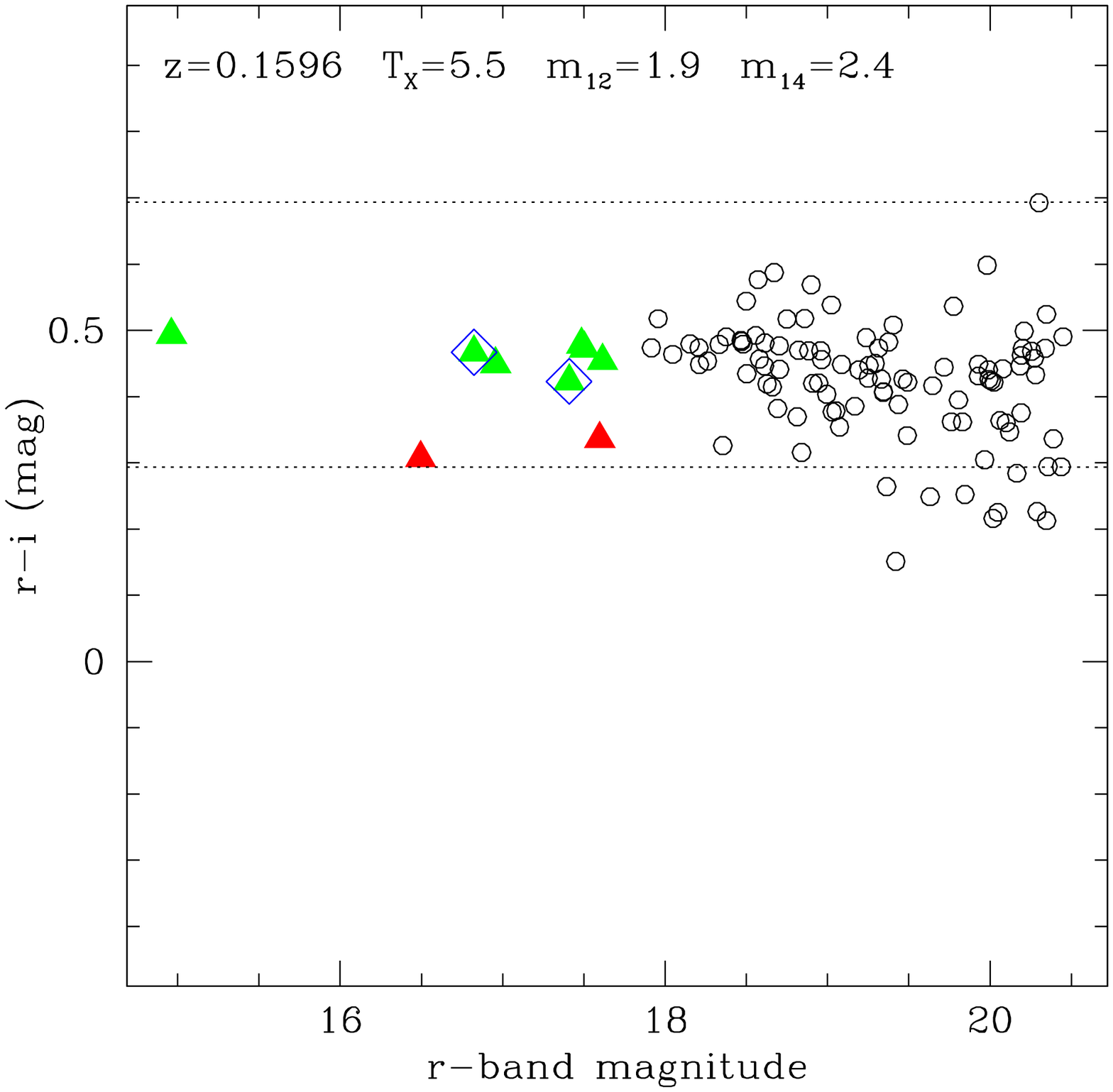}
\caption{$z=0.1596$ system located at 17:20:10.0 +26:37:24.7.}
\label{j172010}
\end{figure}

This $z=0.1596$ system is located at 17:20:10.0 +26:37:24.7 (Figure
\ref{j172010}). The CMD shows an obvious RS, which all spectroscopic
members lie on, and the magnitude gap is 2.45 based on SDSS
spectroscopic data. This is close enough to 2.5 that we accept it as
an FS. The two rejected galaxies are 2400 km s$^{-1}$ and 20000 km
s$^{-1}$ away from the FG. Reducing the spectroscopic redshift cut to
1000 km s$^{-1}$ would increase the magnitude gap by 0.4. The X-ray
temperature of the system is $T_X=5.5$ keV and the X-ray emission peak
lies $\sim 20.2$ kpc from the FG. The system is located near the edge
of the SDSS footprint but at a distance of $\sim 17R_{200}$ both
$\Delta m_{14}$ and $L_\mathrm{tot}$ are unaffected. The X-ray source
is extended with $R_{200}=1.54$ Mpc or $\sim 30R_{90}$. This system
was an \textit{XMM} target and was classified as an FS in
\citet{santos07}. It is also known as SDSS-C4 3072. The system has a
velocity dispersion of 768 km s$^{-1}$ based on 31 galaxies.


\begin{thebibliography}{}
\bibitem[Aguerri et al.(2011)]{aguerri11} Aguerri, J.~A.~L., Girardi,
  M., Boschin, W., et al.\ 2011, \aap, 527, A143
\bibitem[Arnaud et al.(2005)]{arnaud05} Arnaud, M., Pointecouteau, E.,
  \& Pratt, G.~W.\ 2005, \aap, 441, 893
\bibitem[Barnes(1989)]{barnes89} Barnes, J.~E.\ 1989, \nat, 338, 123 
\bibitem[Beers et al.(1990)]{beers90} Beers, T.~C., Flynn, K., \&
  Gebhardt, K.\ 1990, \aj, 100, 32
\bibitem[Bernardi et al.(2007)]{bernardi07} Bernardi, M., Hyde, J.~B.,
  Sheth, R.~K., Miller, C.~J., \& Nichol, R.~C.\ 2007, \aj, 133, 1741 
\bibitem[Bilir et al.(2005)]{bilir05} Bilir, S., Karaali, S., \&
  Tun{\c c}el, S.\ 2005, Astron.\ Nachr., 326, 321
\bibitem[Blanton \& Roweis(2007)]{blanton07} Blanton, M.~R., \&
  Roweis, S.\ 2007, \aj, 133, 734
\bibitem[Brinchmann et al.(2004)]{brinchmann04} Brinchmann, J.,
  Charlot, S., White, S.~D.~M., et al.\ 2004, \mnras, 351, 1151
\bibitem[Brough et al.(2005)]{brough05} Brough, S., Collins, C.~A.,
  Burke, D.~J., Lynam, P.~D., \& Mann, R.~G.\ 2005, \mnras, 364, 1354
\bibitem[Brough et al.(2002)]{brough02} Brough, S., Collins, C.~A.,
  Burke, D.~J., Mann, R.~G., \& Lynam, P.~D.\ 2002, \mnras, 329, L53
\bibitem[Brough et al.(2007)]{brough07} Brough, S., Proctor, R.,
  Forbes, D.~A., et al.\ 2007, \mnras, 378, 1507
\bibitem[Brough et al.(2011)]{brough11} Brough, S., Tran, K.-V.,
  Sharp, R.~G., von der Linden, A., \& Couch, W.~J.\ 2011, \mnras,
  414, L80
\bibitem[Bruzual \& Charlot(1993)]{bruzual93} Bruzual, A., G., \&
  Charlot, S.\ 1993, \apj, 405, 538
\bibitem[Bruzual \& Charlot(2003)]{bruzual03} Bruzual, G., \& Charlot,
  S.\ 2003, \mnras, 344, 1000
\bibitem[Cash(1979)]{cash79} Cash, W.\ 1979, \apj, 228, 939
\bibitem[Cavaliere \& Fusco-Femiano(1976)]{cavaliere76}Cavaliere, A.,
  \& Fusco-Femiano, R., 1976, A\&A, 49, 137
\bibitem[Charlot \& Fall(2000)]{charlot00} Charlot, S., \& Fall,
  S.~M.\ 2000, \apj, 539, 718
\bibitem[Charlot et al.(2002)]{charlot02} Charlot, S., Kauffmann, G.,
  Longhetti, M., et al.\ 2002, \mnras, 330, 876
\bibitem[Charlot \& Longhetti(2001)]{charlot01} Charlot, S., \&
  Longhetti, M.\ 2001, \mnras, 323, 887
\bibitem[Cid Fernandes et al.(2004)]{cidfernandes04} Cid Fernandes,
  R., Gu, Q., Melnick, J., et al.\ 2004, \mnras, 355, 273
\bibitem[Cid Fernandes et al.(2005)]{cidfernandes05} Cid Fernandes,
  R., Mateus, A., Sodr{\'e}, L., Stasi{\'n}ska, G., \& Gomes, J.~M.\
  2005, \mnras, 358, 363
\bibitem[Collins et al.(2009)]{collins09} Collins, C.~A., Stott,
  J.~P., Hilton, M., 2009, \nat, 458, 603
\bibitem[Conroy et al.(2007)]{conroy07} Conroy, C., Wechsler, R.~H.,
  \& Kravtsov, A.~V.\ 2007, \apj, 668, 826
\bibitem[Csabai et al.(2007)]{csabai07} Csabai, I., Dobos, L.,
  Trencs{\'e}ni, et al.\ 2007, Astron.\ Nachr., 328, 852
\bibitem[Cui et al.(2011)]{cui11} Cui, W., Springel, V., Yang, X., De
  Lucia, G., \& Borgani, S.\ 2011, \mnras, 416, 2997
\bibitem[Cypriano et al.(2006)]{cypriano06} Cypriano, E.~S., Mendes de
  Oliveira, C.~L., \& Sodr{\'e}, L., Jr.\ 2006, \aj, 132, 514
\bibitem[Dariush et al.(2007)]{dariush07} Dariush, A., Khosroshahi,
  H.~G., Ponman, T.~J., et al.\ 2007, \mnras, 382, 433
\bibitem[Dariush et al.(2010)]{dariush10} Dariush, A.~A.,
  Raychaudhury, S., Ponman, T.~J., et al.\ 2010, \mnras, 405, 1873 
\bibitem[De Lucia et al.(2006)]{delucia06} De Lucia, G., Springel, V.,
  White, S.~D.~M., Croton, D., \& Kauffmann, G.\ 2006, \mnras, 366,
  499
\bibitem[De Lucia \& Blaizot(2007)]{delucia07} De Lucia, G., \&
  Blaizot, J.\ 2007, \mnras, 375, 2
\bibitem[D{\'{\i}}az-Gim{\'e}nez et al.(2008)]{diaz-gimenez08}
  D{\'{\i}}az-Gim{\'e}nez, E., Muriel, H., \& Mendes de Oliveira, C.\
  2008, \aap, 490, 965
\bibitem[D'Onghia et al.(2005)]{d'onghia05} D'Onghia, E.,
  Sommer-Larsen, J., Romeo, A.~D., et al.\ 2005, \apjl, 630, L109
\bibitem[Dubinski(1998)]{dubinski98} Dubinski, J.\ 1998, \apj, 502,
  141
\bibitem[Fakhouri et al.(2010)]{fakhouri10} Fakhouri, O., Ma, C.-P.,
  \& Boylan-Kolchin, M.\ 2010, \mnras, 406, 2267
\bibitem[Feldmeier et al.(2002)]{feldmeier02} Feldmeier, J.~J., Mihos,
  J.~C., Morrison, H.~L., Rodney, S.~A., \& Harding, P.\ 2002, \apj,
  575, 779
\bibitem[Ferland(1996)]{ferland96} Ferland, G.~J.\ 1996, University of
  Kentucky Internal Report
\bibitem[Gastaldello et al.(2008)]{gastaldello08} Gastaldello, F.,
  Trevese, D., Vagnetti, F., \& Fusco-Femiano, R.\ 2008, \apj, 673,
  176
\bibitem[Herberich et al. (2010)]{herberich10} Herberich, E.,
Sikorski, J., \& Hothorn, T., 2010, A Robust Procedure for Comparing
Multiple Means under Heteroscedasticity in Unbalanced Designs, ed.\
F.\ Rapallo (Public Library of Science) PLoS ONE 5(3): e9788,
doi:10.1371/journal.pone.0009788
\bibitem[Johnston et al.(2007)]{johnston07} Johnston, D.~E., Sheldon,
  E.~S., Wechsler, R.~H., et al.\ 2007, arXiv:0709.1159
\bibitem[Jones et al.(2000)]{jones00} Jones, L.~R., Ponman, T.~J., \&
  Forbes, D.~A.\ 2000, \mnras, 312, 139
\bibitem[Jones et al.(2003)]{jones03} Jones, L.~R., Ponman, T.~J.,
Horton, et al.\ 2003, \mnras, 343, 627
\bibitem[Kauffmann et al.(2003)]{kauffmann03} Kauffmann, G., Heckman,
  T.~M., White, S.~D.~M., et al.\ 2003, \mnras, 341, 33
\bibitem[Khosroshahi et al.(2004)]{khosroshahi04} Khosroshahi, H.~G., 
Jones, L.~R., \& Ponman, T.~J.\ 2004, \mnras, 349, 1240
\bibitem[Khosroshahi et al.(2006a)]{khosroshahi06a} Khosroshahi, H.~G.,
Maughan, B.~J., Ponman, T.~J., \& Jones, L.~R.\ 2006a, \mnras, 369,
1211
\bibitem[Khosroshahi et al.(2006b)]{khosroshahi06b} Khosroshahi, H.~G., 
Ponman, T.~J., \& Jones, L.~R.\ 2006b, \mnras, 372, L68
\bibitem[Khosroshahi et al.(2007)]{khosroshahi07} Khosroshahi, H.~G., 
Ponman, T.~J., \& Jones, L.~R.\ 2007, \mnras, 377, 595 
\bibitem[Koester et al.(2007a)]{koester07a} Koester, B.~P., McKay,
  T.~A., Annis, J., et al.\ 2007, \apj, 660, 221
\bibitem[La Barbera et al.(2009)]{labarbera09} La Barbera, F., de
  Carvalho, R.~R., de la Rosa, et al.\ 2009, \aj, 137, 3942
\bibitem[Lauer et al.(2007)]{lauer07} Lauer, T.~R., Gebhardt, K.,
  Faber, S.~M., et al.\ 2007, \apj, 664, 226
\bibitem[Lavery \& Henry(1998)]{lavery98} Lavery, R.~J., \& Henry,
  J.~P.\ 1998, BAAS, 30, 864
\bibitem[Lin \& Mohr(2004)]{lin04} Lin, Y.-T., \& Mohr, J.~J.\ 2004,
  \apj, 617, 879
\bibitem[Lloyd-Davies et al.(2011)]{lloyddavies11} Lloyd-Davies,
E.~J., Romer, A.~K., Mehrtens, N., et al.\ 2011, \mnras, 418, 14 (LD11)
\bibitem[Loh \& Strauss(2006)]{loh06} Loh, Y.-S., \& Strauss, M.~A.\
  2006, \mnras, 366, 373 
\bibitem[Lotz et al.(2011)]{lotz11} Lotz, J.~M., Jonsson, P., Cox,
  T.~J., et al.\ 2011, \apj, 742, 103
\bibitem[Mantz et al.(2010)]{mantz10} Mantz, A., Allen, S.~W.,
  Ebeling, H., Rapetti, D., \& Drlica-Wagner, A.\ 2010, \mnras, 406,
  1773
\bibitem[Maughan(2007)]{maughan07} Maughan, B.~J.\ 2007, \apj, 668,
  772
\bibitem[Mehrtens et al.(2011)]{mehrtens11} Mehrtens, N., Romer,
  A.~K., Lloyd-Davies, E.~J., et al.\ 2011, arXiv:1106.3056 (M11)
\bibitem[Mendes de Oliveira et al.(2009)]{mendesdeoliveira09} Mendes
  de Oliveira, C.~L., Cypriano, E.~S., Dupke, R.~A., \& Sodr{\'e},
  L.\ 2009, \aj, 138, 502
\bibitem[Mendes de Oliveira et al.(2006)]{mendesdeoliveira06} Mendes
  de Oliveira, C.~L., Cypriano, E.~S., \& Sodr{\'e}, L., Jr.\ 2006,
  \aj, 131, 158
\bibitem[M{\'e}ndez-Abreu et al.(2012)]{mendezabreu12}
  M{\'e}ndez-Abreu, J., Aguerri, J.~A.~L., Barrena, R., et al.\ 2012,
  \aap, 537, A25
\bibitem[Miller et al.(2005)]{miller05} Miller, C.~J., Nichol, Robert
  C., Reichart, D., et al.\ 2005, \aj, 130, 968
\bibitem[Miller et al.(2011)]{miller11} Miller, E.~D., Rykoff, E.,
  Dupke, R., et al.\ 2012, ApJ, 747, 94
\bibitem[Milosavljevi{\'c} et al.(2006)]{milosavljevic06}
  Milosavljevi{\'c}, M., Miller, C.~J., Furlanetto, S.~R., \& Cooray,
  A.\ 2006, \apjl, 637, L9 
\bibitem[Mulchaey \& Zabludoff(1999)]{mulchaey99} Mulchaey,
  J.~S., \& Zabludoff, A.~I.\ 1999, \apj, 514, 133 
\bibitem[Murante et al.(2004)]{murante04} Murante, G., Arnaboldi, M.,
  Gerhard, O., et al.\ 2004, \apjl, 607, L83
\bibitem[Nagamine \& Loeb(2003)]{nagamine03} Nagamine, K., \& Loeb,
  A.\ 2003, New Astron., 8, 439
\bibitem[Osmond \& Ponman(2004)]{osmond04} Osmond, J.~P.~F., \&
  Ponman, T.~J.\ 2004, \mnras, 350, 1511
\bibitem[Patton \& Atfield(2008)]{patton08} Patton, D.~R., \& Atfield,
  J.~E.\ 2008, \apj, 685, 235
\bibitem[Pipino et al.(2011)]{pipino11} Pipino, A., Szabo, T.,
  Pierpaoli, E., MacKenzie, S.~M., \& Dong, F.\ 2011, \mnras, 417,
  2817
\bibitem[Ponman et al.(1994)]{ponman94} Ponman, T.~J., Allan, D.~J.,
  Jones, L.~R., et al.\ 1994, \nat, 369, 462
\bibitem[Popesso et al.(2005)]{popesso05} Popesso, P., B{\"o}hringer,
  H., Romaniello, M., \& Voges, W.\ 2005, \aap, 433, 415 
\bibitem[Proctor et al.(2011)]{proctor11} Proctor, R.~N., de Oliveira,
  C.~M., Dupke, R., et al.\ 2011, \mnras, 418, 2054
\bibitem[Quillen et al.(2008)]{quillen08} Quillen, A.~C., Zufelt, 
N., Park, J., et al.\ 2008, \apjs, 176, 39
\bibitem[Richards et al.(2009)]{richards09} Richards, J.~W., Freeman,
  P.~E., Lee, A.~B., \& Schafer, C.~M.\ 2009, \mnras, 399, 1044
\bibitem[Richardson (2011)]{richardson11} Richardson, A.\ 2011, in
  International Statistical review, Multiple Comparisons Using R, ed.\
  F.\ Bretz, T.\ Hothorn, \& P.\ Westfall, 79, 297, (Wiley Online
  Library)
\bibitem[Romer et al.(2001)]{romer01} Romer, A.~K., Viana, P.~T.~P.,
  Liddle, A.~R., \& Mann, R.~G.\ 2001, \apj, 547, 594 
\bibitem[Rudick et al.(2009)]{rudick09} Rudick, C.~S., Mihos, J.~C.,
  Frey, L.~H., \& McBride, C.~K.\ 2009, \apj, 699, 1518
\bibitem[Rudick et al.(2006)]{rudick06} Rudick, C.~S., Mihos, J.~C.,
  \& McBride, C.\ 2006, \apj, 648, 936
\bibitem[Rudick et al.(2011)]{rudick11} Rudick, C.~S., Mihos, J.~C.,
  \& McBride, C.~K.\ 2011, \apj, 732, 48
\bibitem[Ruszkowski \& Springel(2009)]{ruszkowski09} Ruszkowski, M.,
  \& Springel, V.\ 2009, \apj, 696, 1094
\bibitem[Salim et al.(2007)]{salim07} Salim, S., Rich, R.~M., et al.\
  2007, \apjs, 173, 267
\bibitem[Santos et al.(2007)]{santos07} Santos, W.~A., Mendes de
  Oliveira, C., \& Sodr{\'e}, L., Jr.\ 2007, \aj, 134, 1551
\bibitem[Schirmer et al.(2010)]{schirmer10} Schirmer, M., Suyu, S.,
  Schrabback, T., et al.\ 2010, \aap, 514, A60 
\bibitem[Shaffer (1995)]{shaffer95} Shaffer, J.P., 1995, Annual Review
    of Psychology, vol 46, pg 561
\bibitem[Smith et al.(2005)]{smith05} Smith, G.~P., Kneib, 
J.-P., Smail, I., et al.\ 2005, \mnras, 359, 417 
\bibitem[Sommer-Larsen(2006)]{sommerlarsen06} Sommer-Larsen, J.\ 2006,
  \mnras, 369, 958
\bibitem[Soltan \& Henry(1983)]{soltan83} Soltan, A., \& Henry, J.~P.\
  1983, \apj, 271, 442
\bibitem[Stott et al.(2010)]{stott10} Stott, J.~P., et al.\ 2010,
  \apj, 718, 23
\bibitem[Stott et al.(2008)]{stott08} Stott, J.~P., Edge, A.~C.,
  Smith, G.~P., Swinbank, A.~M., \& Ebeling, H.\ 2008, \mnras, 384,
  1502
\bibitem[Tavasoli et al.(2011)]{tavasoli11} Tavasoli, S., Khosroshahi,
  H.~G., Koohpaee, A., Rahmani, H., \& Ghanbari, J.\ 2011, \pasp, 123,
  1 
\bibitem[Tremonti et al.(2004)]{tremonti04} Tremonti, C.~A., Heckman,
  T.~M., Kauffmann, G., et al.\ 2004, \apj, 613, 898
\bibitem[Voevodkin et al.(2010)]{voevodkin10} Voevodkin, A., Borozdin,
  K., Heitmann, K., et al.\ 2010, \apj, 708, 1376 
\bibitem[Voevodkin et al.(2008)]{voevodkin08} Voevodkin, A., Miller,
  C.~J., Borozdin, et al.\ 2008, \apj, 684, 204
\bibitem[von Benda-Beckmann et al.(2008)]{vonbendabeckmann08} von
  Benda-Beckmann, A.~M., D'Onghia, E., Gottl{\"o}ber, S., et al.\
  2008, \mnras, 386, 2345 
\bibitem[von der Linden et al.(2007)]{vonderlinden07} von der Linden,
  A., Best, P.~N., Kauffmann, G., \& White, S.~D.~M.\ 2007, \mnras,
  379, 867
\bibitem[Whiley et al.(2008)]{whiley08} Whiley, I.~M.,
  Arag\'{o}n-Salamanca, A., De Lucia, G., et al.\ 2008, \mnras, 387,
  1253
\bibitem[White \& Rees(1978)]{white78} White, S.~D.~M., \& Rees,
  M.~J.\ 1978, \mnras, 183, 341
\bibitem[Wu et al.(1999)]{wu99} Wu, X.-P., Xue, Y.-J., \& Fang, L.-Z.\
  1999, \apj, 524, 22 
\bibitem[York et al.(2000)]{york00} York, D.~G., Adelman, J.,
  Anderson, J.~E., Jr., et al.\ 2000, \aj, 120, 1579
\bibitem[Zibetti et al.(2009)]{zibetti09} Zibetti, S., Pierini, D., \&
  Pratt, G.~W.\ 2009, \mnras, 392, 525
\end{thebibliography}
\end{document}